\newif\ifemulate
\emulatetrue

\ifemulate
	\documentclass[iop]{emulateapj-rtx4}
	\slugcomment{ApJ in press}
	\lefthead{Rudnick et al.}
	\righthead{CO detections in a $z=1.62$ cluster}
\else
	\documentclass[12pt, preprint]{aastex}
\fi

\usepackage{graphicx}
\usepackage{lscape}
\usepackage{natbib}
\usepackage{amsmath}
\newcommand\clg{XMM-LSS~J02182-05102}

\newcommand\ha{${\rm H\alpha}$}

\newcommand\ang{${\rm \AA}$}

\newcommand\mlstar{${ M_{\star}}/{\rm L}$}

\newcommand{\mlstarlamav}[1]{$\langle {\cal M_{\star}}/{\rm L}_{V}\rangle$}

\newcommand\msol{${\cal M_{\odot}}$}
\newcommand\lsol{$L_{\odot}$}

\newcommand\mgas{${\cal M}_{gas}$}
\newcommand\tcon{$t_{\rm con}$}
\newcommand\siggas{$\Sigma_{mol}$}
\newcommand\mustar{$\mu_\star$}
\newcommand\sigsfr{$\Sigma_{SFR}$}
\newcommand\mhtwo{${\cal M}_{\rm mol}$}

\newcommand{\mstar}{${\cal M_{\star}}$}

\newcommand\lir{$L_{IR}$}

\def\gtsima{$\; \buildrel > \over \sim \;$}
\def\gsim{\lower.7ex\hbox{\gtsima}}
\def\ltsima{$\; \buildrel < \over \sim \;$}
\def\lsim{\lower.7ex\hbox{\ltsima}}

\renewcommand\micron{$\mu$m}

\newcommand{\spitzer}{\textit{Spitzer}}

\newcommand{\lco}{$L^\prime_{CO}$}

\newcommand\co[2]{CO(#1--#2)}
\newcommand{\aco}{$\alpha_{\rm CO}$}

\def\apj{ApJ}	
\def\aj{AJ}	
\def\mnras{MNRAS}	
\def\aap{A\&A}
\def\apjs{ApJS}

\def\nat{Nature}

\def\apjl{ApJ}
	
\def\araa{ARA\&A}

\begin{document}

\title{Deep CO(1-0) Observations of $z=1.62$ Cluster Galaxies with Substantial Molecular Gas Reservoirs and Normal Star Formation Efficiencies}

\author{Gregory Rudnick\altaffilmark{1,2,3}, 
  Jacqueline Hodge\altaffilmark{2,4,5,6}, 
  Fabian Walter\altaffilmark{2},
  Ivelina Momcheva\altaffilmark{7}, 
  Kim-Vy Tran\altaffilmark{8},
  Casey Papovich\altaffilmark{8},
  Elisabete da Cunha\altaffilmark{2,10},
  Roberto Decarli\altaffilmark{2},
  Amelie Saintonge\altaffilmark{11},
  Christopher Willmer\altaffilmark{9},
  Jennifer Lotz\altaffilmark{12},
  Lindley Lentati\altaffilmark{13}}

\altaffiltext{1}{The University of Kansas, Department of Physics and Astronomy, Malott room 1082, 1251 Wescoe Hall Drive, Lawrence, KS, 66045, USA; \texttt{grudnick@ku.edu}}
\altaffiltext{2}{The Max-Planck-Institute for Astronomy, K\"onigstuhl 17, Heidelberg, 69120, Germany}
\altaffiltext{3}{Alexander von Humboldt Fellow}
\altaffiltext{4}{The National Radio Astronomy Observatory, 520 Edgemont Road
Charlottesville, VA 22903-2475, USA}
\altaffiltext{5}{Jansky Fellow}
\altaffiltext{6}{Leiden Observatory, Niels Bohrweg 2, 2333 CA Leiden, Netherlands}
\altaffiltext{7}{Astronomy Department, Yale University, P.O. Box 208101,New Haven, CT 06520-8101 USA}
\altaffiltext{8}{George P. and Cynthia Woods Mitchell Institute for Fundamental Physics and Astronomy, and Department of Physics and Astronomy, Texas A\&M University, College Station, TX 77843-4242, USA}
\altaffiltext{9}{Steward Observatory, University of Arizona, 933 N. Cherry Avenue, Tucson, AZ 85721, USA}
\altaffiltext{10}{Research School of Astronomy and Astrophysics, Australian National University, ACT 2611, Canberra, Australia}
\altaffiltext{11}{Astrophysics Group, Department of Physics and Astronomy, University College London, 3rd Floor, 132 Hampstead Road, London, NW1 2PS, United Kingdom}
\altaffiltext{12}{Space Telescope Science Institute, 3700 San Martin Drive, Baltimore, MD 21218, USA}
\altaffiltext{13}{Kavli Institute for Cosmology, c/o Institute of Astronomy, Madingley Road Cambridge CB3 0HA, Uk}

\begin{abstract}
We present an extremely deep
\co{1}{0} observation of a confirmed $z=1.62$ galaxy cluster.  We
detect two spectroscopically confirmed cluster members in \co{1}{0} with $S/N>5$.  Both galaxies
have log(\mstar/\msol)$>11$ and are gas rich, with \mhtwo/(\mstar+\mhtwo)$\sim 0.17-0.45$.  One of these
galaxies lies on the star formation rate (SFR)-\mstar\ sequence while the other lies an order of magnitude below.  We compare the cluster galaxies to other SFR-selected galaxies with CO measurements and find that they have
 CO luminosities consistent with expectations given their infrared luminosities.  We also find that they have comparable gas fractions and star formation efficiencies (SFE) to what is expected from published field galaxy scaling relations.   
The galaxies are compact in their stellar light distribution, at the extreme end for all high redshift star-forming galaxies.  However, their SFE is consistent with other field galaxies at comparable compactness.  This is similar to two other sources selected in a blind CO survey of the HDF-N.  Despite living in a highly quenched proto-cluster core, the molecular gas properties of these two galaxies, one of which may be in the processes of quenching, appear entirely consistent with field scaling relations between the molecular gas content, stellar mass, star formation rate, and redshift.  We speculate that these cluster galaxies cannot have any further substantive gas accretion if they are to become members of the dominant passive population in $z<1$ clusters.  
\end{abstract}


\section{Introduction}
\label{Sec:intro}

\subsection{The Evolution of Massive Galaxies}
Understanding the regulation and demise of star formation in the most massive (log(\mstar/\msol)$\gtrsim 11$) galaxies is a dominant theme of galaxy evolution studies.  An important epoch for understanding the evolution in this population is $1<z<2$.  This epoch was witness to one of the largest increases in the number and mass density of massive galaxies and by $z\sim 1$ roughly 50\% of  log(\mstar/\msol)$> 11$ galaxies were in place \citep[e.g.][]{Dickinson03,Rudnick03,Fontana03,Rudnick06,Fontana06,Pozzetti07,Marchesini09, Ilbert10, vandokkum10}.  

Large surveys of representative volumes in the local Universe, such as SDSS, have determined that the massive galaxy population has uniformly very low star formation rates (SFRs) and old stellar ages while lower mass galaxies are highly star-forming \citep[e.g.][]{Strateva01, Blanton03, Kauffmann03}.  Since discovering this "bimodality", a persistent question has been what caused the massive galaxies to cease their star formation and what has maintained their low levels of star formation, even in the presence of modest gas reservoirs \citep{Davis11}.  A piece of this puzzle was uncovered by \citet{Bell04}, who found that the mass density of passive galaxies has been increasing since $z\sim 1$.  This was confirmed by later studies \citep{Brown07, Arnouts07, Faber07} and eventually extended out to $z>2$ \citep{Ilbert10,Nicol11,Brammer11,Muzzin13,Ilbert13}.  These latter studies also highlighted the $1<z<2$ epoch as critical to understanding the transformation of massive galaxies, as it is the first time when the number and mass density of massive galaxies was dominated by those that are passive.

Immediately prior to becoming passive, these galaxies clearly must have been star forming galaxies and an emergent field in recent years has been the study of how star formation is supplied and regulated in these progenitors of the passive population.  We now know that the SFRs of most star forming galaxies are tightly correlated with their stellar mass, the so-called "main sequence" of star formation or \mstar-SFR relation \citep{Brinchmann04,Noeske07,Daddi07,Pannella09}.  This sequence is in place out to at least $z\sim2$ and increases its zeropoint towards higher redshift \citep{Elbaz11,Wuyts11,Karim11,Whitaker12} with the SFR of star forming galaxies increasing with redshift at a fixed stellar mass.  One result of these findings was a shift in our understanding of the driving forces behind the large SFRs typically observed at high redshift.  Locally galaxies with very high SFRs, usually characterized as being Ultra Luminous Infrared Galaxies (ULIRGs) with \lir$>10^{12}$\lsol, reside uniformly in major galaxy mergers \citep{Sanders96}.  In contrast, although the galaxies on the \mstar-SFR sequence at $z>1$ have much higher absolute SFRs than locally, their star formation likely proceeds in scaled up versions of extended galactic disks with similar dust temperature distributions as local galaxies on the \mstar-SFR sequence \citep{Papovich07}, although with significantly higher SFRs and SFR surface densities \citep{Elbaz11}.  

\subsection{Gas accretion as the driver of the \mstar-SFR relation}

Much effort has gone into understanding the origin of the tight \mstar-SFR relation.  A key result has been that the SFRs of galaxies on the \mstar-SFR sequence should be governed by the accretion of gas from the intergalactic medium (IGM.)  Such a scenario predicts that the SFRs should be roughly proportional to both the gas accretion and outflow rates, with galaxies having a relatively small SFR per unit gas mass, or star formation efficiency \citep[SFE;][]{Dutton10}.  This scenario is consistent with the results of hydrodynamical simulations, which show that massive galaxies at high redshift should receive substantial accretion from the IGM \citep{Keres09,Dekel09}.  In the presence of a Kennicutt-Schmidt like star formation law that links gas surface density and SFR surface density \citep{Kennicutt98b,Leroy08,Bigiel08}, large gas fractions from ample accretion would fuel correspondingly intense star formation.

Clearly, understanding how massive galaxies regulate their star formation and eventually shut it down requires a characterization of the gas contents of galaxies at $z>1$.  This is mostly accomplished via observations of the $^{12}$CO molecule, which can be converted to a molecular hydrogen gas mass via a conversion factor termed \aco\ \citep[see][for a review]{Bolatto13}.  The past 5 years have witnessed a rapid improvement in the study of gas at high redshift enabled mostly by observations of CO in distant galaxies using the improved capabilities of the Plateau de Bure Interferometer (now renamed NOEMA.)  These observations have been carried out on small samples of individual galaxies on the \mstar-SFR sequence and as part of the IRAM Plateau de Bure high-z blue sequence \co{3}{2} survey (PHIBSS) \citep{Aravena10,Tacconi10,Daddi10a,Daddi10b,Genzel10,Magdis12,Tacconi13,Carilli13}.  These studies have shown that normal star-forming galaxies at $1<z<3$ have very high gas fractions, $f_{gas}\equiv$\mhtwo/(\mstar+\mhtwo)$\sim 0.5$ and form stars with a relatively low SFE, similar to galaxies on the \mstar-SFR sequence locally.  In limited cases where the gas excitation has been measured, it appears to have moderate values similar to the Milky-Way \citep{Dannerbauer09}, although it may be that a higher excitation dense gas phase exists that is missing in normal local star-forming galaxies \citep{Daddi15}.  Additionally, in one case where the molecular gas could be directly spatially resolved, it appears that it is significantly extended in a turbulent Toomre unstable disk \citep{Genzel13}.  This again reinforces the view that very high star formation rates are being driven by spatially extended large gas reservoirs.   

A natural outcome of the large SFRs are short gas consumption timescales with galaxies on the \mstar-SFR sequence using up their gas in $\sim 0.7$~Gyr \citep{Tacconi13}.  The uniformly short consumption timescale seen in PHIBSS for high redshift star-forming galaxies argues for a replenishment of their gas supplies by accretion, in concordance with the predictions of simulations.  Recently, \citet{Genzel15} measured gas contents for galaxies below the \mstar-SFR sequence and has shown that the gas masses and SFRs decrease towards lower specific star formation rates (sSFR) such that the gas consumption timescale (\tcon$\equiv$\mhtwo/SFR) scales as $(1+z)^{-0.3}\times {\rm (sSFR/sSFR_{MS})}^{-0.5}$.  Hence a prediction of these observations is that galaxies move below the \mstar-SFR sequence because they are running out of gas.  

Despite the incredible advances afforded by these studies, they have several limitations.  First, they did not select galaxies primarily by their CO luminosity.    In PHIBSS, which will form the main comparison sample for this paper, galaxies at $z=1-1.5$ were selected to have high \mstar\ and SFR, such that the expected CO luminosity would make a detection likely.  Similarly, galaxies at $z=2-2.5$ were targeted based on the presence of H$\alpha$ emission from a parent sample of "BX/MD" galaxies chosen by their rest-frame UV colors \citep{Steidel04,Erb06}.  Given the time intensive nature of high-z CO observations, done one galaxy at a time, this preselection made sense for the early statistical studies.  However, it may present a limited view of the galaxy population and may be biased against galaxies with abnormally low SFEs (or high \mhtwo/SFR.).  

Second, most of the previous studies have relied on higher excitation lines of CO, for example PHIBSS relied exclusively on the \co{3}{2} rotational transition.  These lines are brighter than lower order transitions but most molecules do not lie in these excited states, thus necessitating an excitation correction.  As shown in \citet{Carilli13}, there is a large range in excitation values for color-selected galaxies at $z>1$, corresponding to a factor of $\sim 9$ range in $S_{(3-2)}/S_{(1-0)}$ ratio and hence in the line luminosities, although the $\nu^{-2}$ dependence of the conversion from line flux to CO line luminosity (\lco) reduces means that the variation in luminosities will be significantly less than the variation in the line fluxes.  In addition, \citet{Narayanan14} predict that the Spectral Line Energy Distribution (SLED) of star-forming galaxies varies strongly with the physical characteristics of the gas.  In \citet{Tacconi13}, however, the assumption is made of a constant ratio $L^\prime_{CO(3-2)}/L^\prime_{CO(1-0)}$, which may hide some of the intrinsic variations in excitation, and hence in $L^\prime_{CO(1-0)}$ and the molecular gas mass.

Finally, nearly all prior CO observations of distant galaxies have targeted galaxies with no pre-selection on environment and only a handful of surveys have purposefully targeted dense environments such as protoclusters \citep{Carilli11,Aravena12,Hodge13,Chapman15}.  This leaves wide open the potential effect of environment on the gas contents of distant galaxies, specifically those that will turn into the massive passive population that dominates clusters at $z<1$ \citep{Poggianti06,Muzzin12,vanderburg13}.  

\subsection{Studying the gas in distant cluster galaxies}
By modeling the evolution of the star-forming fraction in clusters at $0.4<z<0.8$, \citet{Poggianti06} proposed a model in which the massive passive cluster galaxy population at $z\sim 0.6$ have their star formation quenched during the epoch of cluster formation at $z>1$.  In the past 5 years, direct lookback observations of $z>1$ clusters may be observing this process in action.   We now know that clusters at high redshift possess a mix of massive star forming and massive passive galaxies \citep{Tran10,Fassbender11,Rudnick12,Strazzullo13,Tanaka13b,Fassbender14,Santos14,Ma15,Santos15} and that the fraction of star forming galaxies in clusters starts dropping at $z\sim1.5$  \citep{Brodwin13,Alberts14} and continues dropping to $z=0$ \citep{Saintonge08,Finn10}.  This drop in the SFRs of massive cluster galaxies that enter cluster environments is predicted by the models, which show that they should be decoupled from their IGM umbilical cords and hence their gas supply, with the SFR subsequently decreasing \citep{Keres09,Dekel09}.  To test whether this cutoff of gas accretion plays an important role in the evolution of massive cluster galaxies at early times it is necessary to directly observe the gas in dense environments.  

We have constructed an observational program to address these shortcomings.  We targeted a $z=1.62$ cluster \citep{Papovich10,Tanaka10} in the UKIRT Infrared Deep Sky Survey (UKIDSS) Ultra Deep Survey (UDS) field with the Karl G. Jansky Very Large Array (VLA) to observe the \co{1}{0} line.  The observations presented in this paper constitute the deepest \co{1}{0} exposure every undertaken with the VLA.  We use \co{1}{0} as it traces the bulk of the CO and does not suffer from the uncertain excitation corrections required to go from higher CO transitions to the ground state.  Our observations also constitute one of a very small but growing number of blind CO surveys \citep{Decarli14,Chapman15} and is one of the only ones targeting a distant cluster.  Additionally, the dense concentration of galaxies in cluster cores may make them good locations for high efficiency targeting of multiple galaxies within a single primary beam.  

In this paper we describe two galaxies securely detected in \co{1}{0} from our integration on this cluster.  These two galaxies show evidence for significant molecular gas reservoirs, with star formation efficiencies (SFE) and gas consumption timescales similar to those for field galaxies.  This paper presents the evidence for these conclusions and discusses the implications when these galaxies and other blindly detected CO emitters are viewed in the context of the bulk of existing gas measurements of $z>1$ normal star-forming galaxies.

The paper is organized as follows.  In \S\ref{Sec:obs} we discuss the data and observations, including the supporting ground-based and HST data and the derivation of SFRs, \mstar, and rest-frame optical sizes.  In \S\ref{Sec:results} we discuss our results, including the detection of \co{1}{0} in the two galaxies, the comparison of the CO and total infrared luminosities and their counterparts \mhtwo\ and SFR, and the gas fraction.  In \S\ref{Sec:discussion} we discuss our results and the implications for the SFE, the stability of the gas, the gas consumption timescales, and the future of gas accretion in these sources.  We present caveats to our analysis in \S\ref{Sec:caveats} and summarize in \S\ref{Sec:summ}.

Throughout we assume ``concordance'' $\Lambda$-dominated cosmology
with
$\Omega_\mathrm{M}=0.3,~\Omega_{\Lambda}=0.7,~\mathrm{and~H_o}=70~{\rm
  h_{70}~km~s^{-1}~Mpc^{-1}}$ unless explicitly stated otherwise.  All
magnitudes are quoted in the AB system.  

\section{Data and Observations}
\label{Sec:obs}

\subsection{A $z=1.62$ galaxy cluster}

\begin{figure*}
\includegraphics[scale=0.6,angle=90]{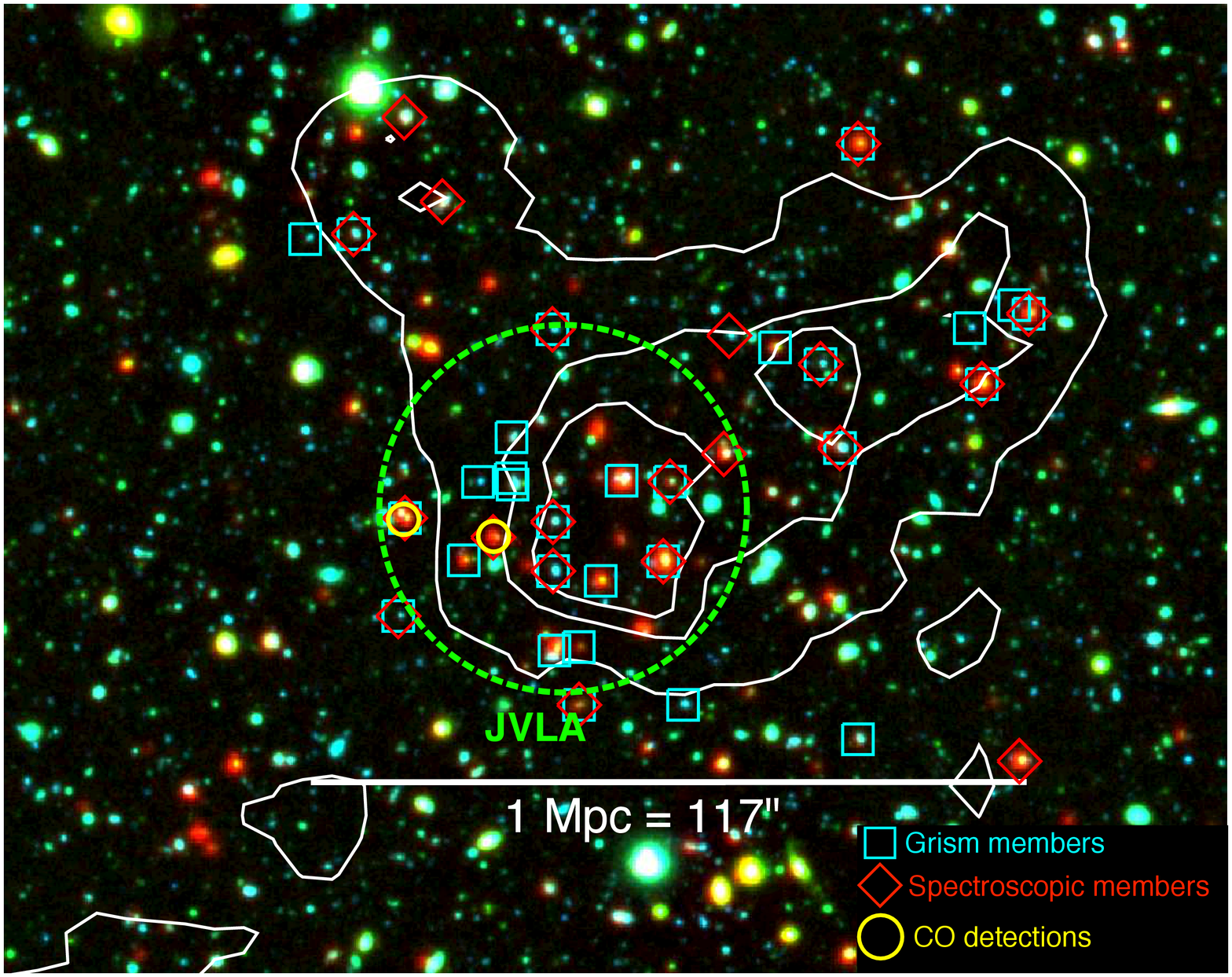}
\caption {A $Bi[4.5\mu$m] image of \clg.  The contours denote regions
  with 5, 10, and 15$\sigma$ above the mean density of galaxies with
  $1.5<z_{phot}<1.7$ from the UKIDSS UDS $K$-selected catalog presented in \citet{Papovich10}.  The green dashed circle illustrates our
  pointing of the VLA, with the size of the circle corresponding to
  the FWHM of the beam at 43.913~GHz.  The yellow circles indicate the
  two \co{1}{0} detections.  The red diamonds mark all
  spectroscopically confirmed members and the cyan squares mark all member as determined by their grism redshifts
  \citep[][Momcheva et al. in prep.]{Papovich10,Tanaka10,Tran15}.}
\label{Fig:fov}
\end{figure*}

Our VLA observations targeted the forming cluster \clg\footnote{Also referred to as IRC0218 or CLG J0218-0510 in the literature.} at $z=1.6233$ \citep{Papovich10,Tanaka10,Tran15}.  This cluster was selected in the UKIDSS UDS as an overdensity of sources with red IRAC [3.6]-[4.5] colors.  As shown in \citet{Papovich08}, this simple color selection, coupled with a requirement that galaxies are faint in the observed optical, is a reliable method for isolating galaxies at $z>1.3$ regardless of their rest-frame color.  Details of the selection and confirmation are given in \citet{Papovich08}, \citet{Papovich10}, and \citet{Tanaka10}.  The cluster was also marginally detected in x-rays at the 2.3$\sigma$ level \citep{Pierre12}.  The cluster is shown in Figure~\ref{Fig:fov}.  This cluster consists of a 20$\sigma$ overdensity of galaxies compared to the mean number density at this epoch and is the most significant overdensity in the UDS at high redshift.  

\subsection{Multi-wavelength imaging and spectroscopy}

This cluster has been imaged at $BRizJK[3.6][4.5][5.6][8.0]$ as part of the UKIDSS UDS survey and the initial cluster identification and spectroscopic selection used photometry and photometric redshifts from \citet{Williams09}.  The cluster was subsequently observed by CANDELS \citep{Koekemoer11,Grogin11}, 3D-HST \citep{Brammer12}, and our own Cycle-19 HST program \citep{Papovich12} and for this paper we use the V4.2 publicly available $uBVV_{606W}RiI_{814W}zJJ_{125W}$ $HH_{140W}H_{160W}K[3.6][4.5][5.6][8.0]$ 3D-HST  catalog \citep{Skelton14}.

\clg\ was observed at 24\micron\ with the \spitzer/MIPS instrument as part of SpUDS\footnote{\texttt{http://ssc.spitzer.caltech.edu/spitzermission/\\observingprograms/legacy/spuds/}} and these observations and the source catalog were presented in \citet{Tran10}.  The MIPS photometry was performed by detecting sources independently in the MIPS catalog and matching them with a 1\arcsec\ search radius against the F160W-selected photometric catalog.  The cluster was also observed with the SPIRE and PACS instruments on \textit{Herschel} at 100, 160, 250, 350, and 500\micron, as presented in \citet{Santos14}.   

This cluster has been the subject of an extended ground-based spectroscopic campaign.  Our ground-based spectroscopy comes from Magellan/IMACS \citep{Papovich10}, Subaru/MOIRCS \citep{Tanaka10}, Magellan/MMIRS (Momcheva et al. in prep), and Keck/LRIS+MOSFIRE \citep{Tran15}.  In addition, this cluster was observed with HST/WFC3 using both the G141 and G102 grisms.  The G141 observations were taken as part of 3D-HST \citep{Brammer12,Momcheva16} and the G102 observations were taken as part of our Cycle 19 program \citep[PI: Papovich;][]{Lee-Brown17}.  Grism redshifts were determined by using a modified version of the EAZY code \citep{Brammer08} run on the combination of the \citet{Skelton14} photometry and either the G141 grism or G102 grism (Momcheva et al. in prep.)  In the case where both redshifts were extracted, we took the average of the two.  For those cases, the median difference was -0.008 and the biweight scatter was 0.005.  For regions of the VLA beam where we have greater than 50\% peak sensitivity, we have eight spectroscopically confirmed members, and an additional four whose membership is based on their grism redshifts.  We also have three non-members whose grism redshifts would put CO in the observable range.

\subsection{VLA data}

The VLA pointing (Figure~\ref{Fig:fov}) was chosen to coincide with the peak of the photometric and spectroscopic redshift members with MIPS detections from \citet{Tran10}.  We observed the cluster in the Q-band at a central observed frequency of 43.913 GHz (6.8mm), corresponding to the rest-frame frequency of \co{1}{0} at 115.271GHz redshifted to the cluster redshift of $z=1.625$.  We used the full 2GHz bandwidth, which at this frequency probes \co{1}{0} over the range $1.546<z<1.666$.  This is well in excess of the formal 250 km~s$^{-1}$ velocity dispersion of this unrelaxed forming cluster.  The full width half power (FWHP) size of the primary beam is 60\arcsec\ at $\nu_{obs}=43.913$GHz.  The FWHM of the synthesized beam was $\approx 1\farcs5$at this frequency.

Observations were obtained in 2011, 2013, 2014, and 2015.  The 60 hours of 2011 observations were conducted in shared risk mode in the D configuration.  Much of our 2011 observations were taken between September 20, 2011 and December 3, 2011 and were subject to the documented "1 second problem"\footnote{\texttt{http://www.vla.nrao.edu/astro/archive/issues/\#1009}}, during which only 1 second of each 3 second scan was read out.  This caused an effective factor of 3 loss in the exposure time for these scheduling blocks (SB).   The 45 hours of observations in 2013 were conducted in the D configuration (25h) and the DnC configuration (20h).  The total amount of on source time, including the loss of the exposure time due to the 1s problem was 45.5 hours. The RMS of our maps around the central observed frequency following 2013 was $26\mu$Jy in 44MHz channels, compared to the $19\mu$Jy that we expected from the exposure time calculator (ETC).  Using our two sets of observations we determined that the ETC is overoptimistic in terms of its sensitivity by a factor of $\sim 3$ in the Q-band.  A further 96h of observations were proposed and accepted to bring us up to our originally proposed sensitivity.  These were mostly completed in early 2015,  and the resultant RMS was 21$\mu$Jy in 44MHz channels, close to our final value.  The failure to reach our final values is likely because we were forced to use short SB lengths (see below) to facilitate scheduling, which resulted in significantly larger overheads.

The observations were conducted in SBs with lengths of 1.5, 2.5, 4, or 5 hours.  We observed 3C48 as our flux calibrator for all observations and targeted it once every SB.  In each scan we first observed a phase calibrator that was near on the sky to our target and then observed on target for $\approx 4$~minutes.  We observed a pointing source (J0239-0234) once at the beginning of every SB and again repeatedly during our scan loops.

Observations were reduced with the Common Astronomy Software Applications (CASA).  Visibilities with bad RMS were flagged and removed from the analysis.  An image with 4MHz resolution was constructed from the sum of all observations.  Channels near the edge of each subband were flagged not included in any line fits or derived properties.

\begin{figure*}
\vspace{-0cm}
\centering
 \begin{minipage}[b]{0.49\textwidth}
   \centering
  \includegraphics[width=\textwidth,angle=90,scale=0.7]{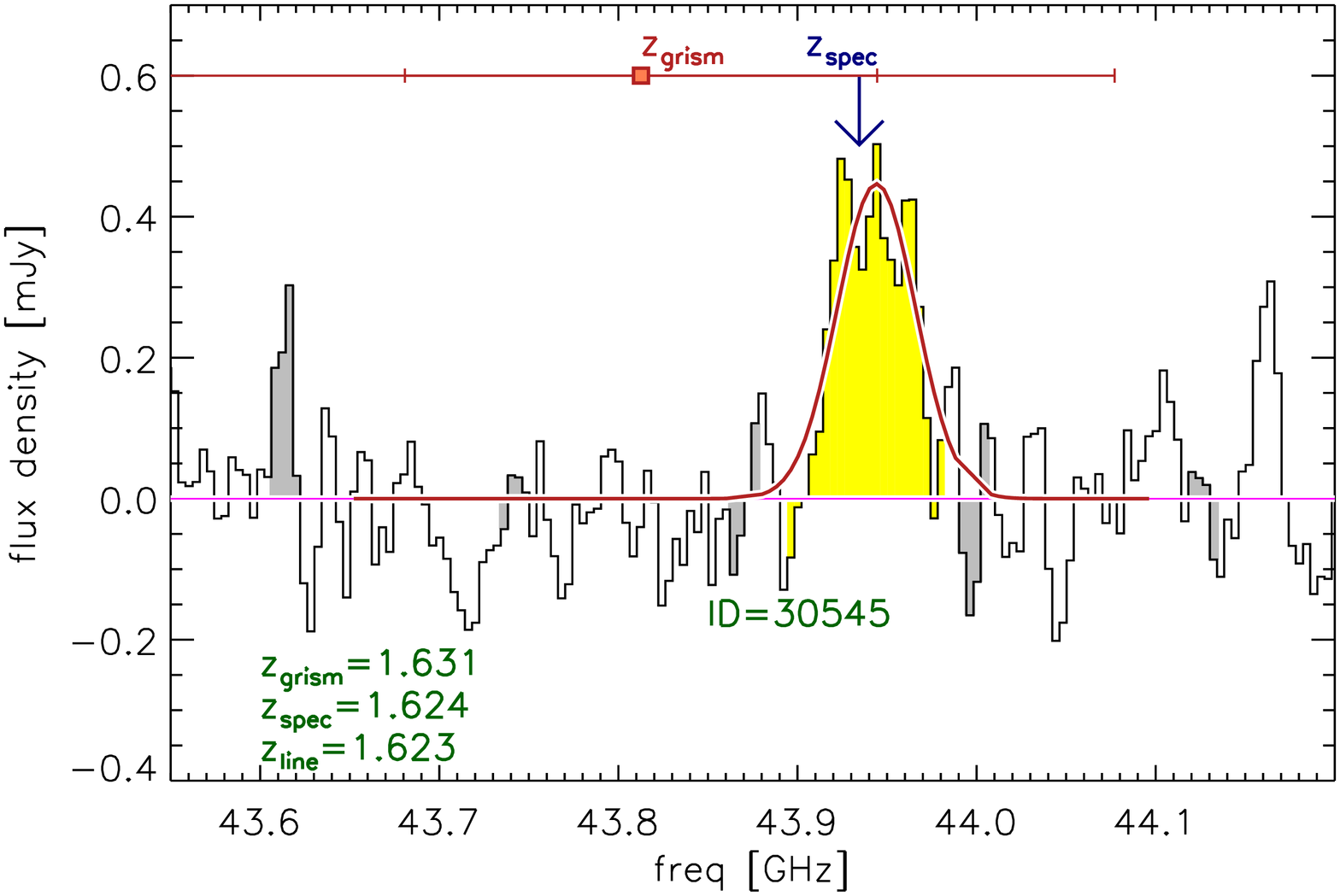}
\end{minipage}
\begin{minipage}[b]{0.49\textwidth}
  \centering
  \includegraphics[width=\textwidth,angle=90,scale=0.7]{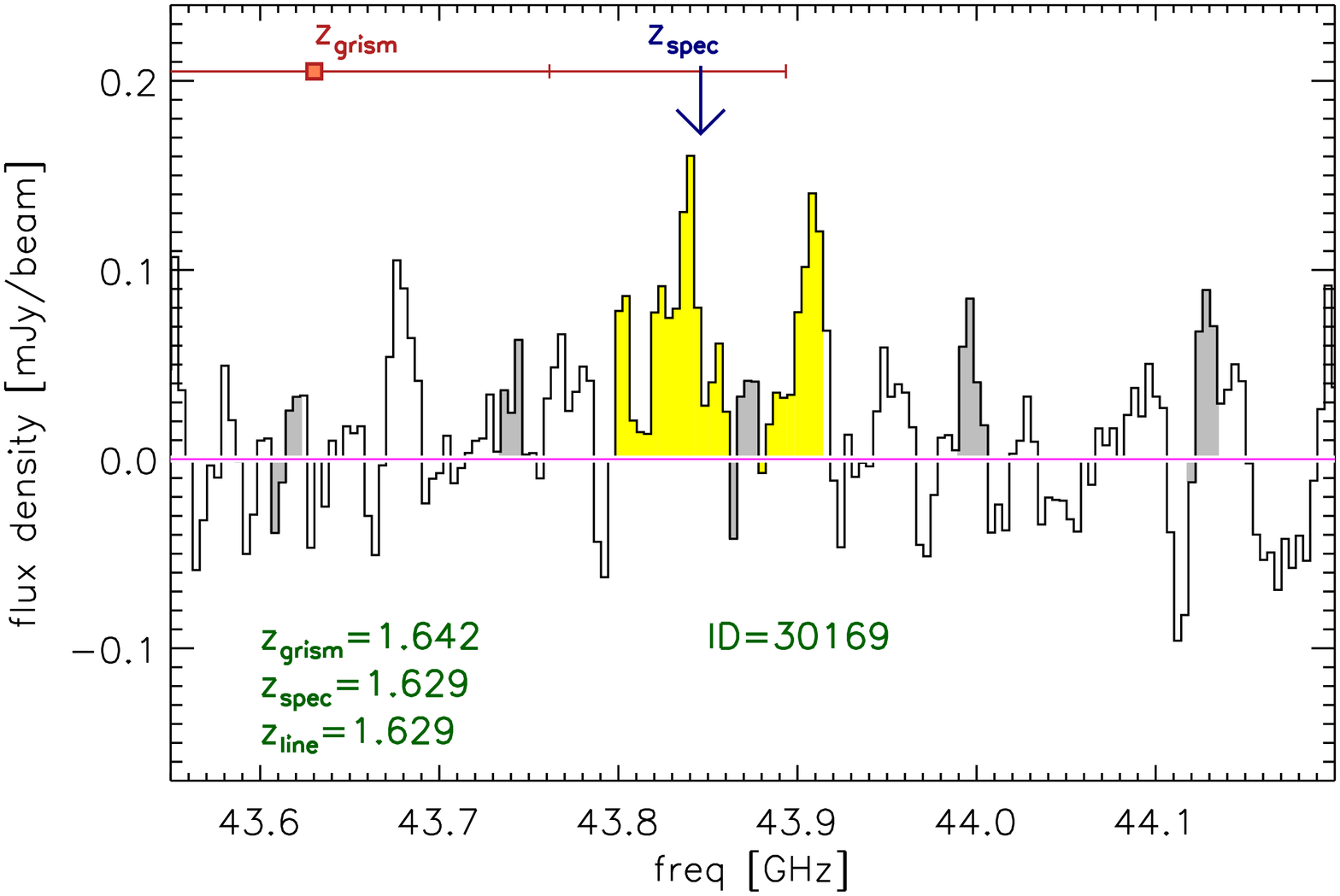}
\end{minipage}
\caption{Two detections of \co{1}{0} in star-forming cluster galaxies from our VLA data, shown at 4MHz resolution smoothed by 8MHz.    
The symbols at the top of each panel indicate the spectroscopic and HST/WFC3 grism redshift \citep[][Momcheva et al. in prep.]{Papovich10,Tanaka10,Tran15}.   The yellow regions correspond to the frequencies over which we collapsed the images to estimate the S/N and derive the contours shown in Figure~\ref{Fig:co_im}.  The gray portions of the spectra correspond to bad channels.  For both sources we compute the line center using a Gaussian fit.  For 30545 we show the Gaussian fit to the data but omit it from 30169 given the irregular velocity structure. The error bars on the grism redshifts are the 68 and 95\% confidence intervals on the redshift.  
}
\label{Fig:co_spec}
\end{figure*}

\begin{figure*}
\vspace{-0cm}
\centering
 \begin{minipage}[b]{0.49\textwidth}
   \centering
  \includegraphics[width=\textwidth]{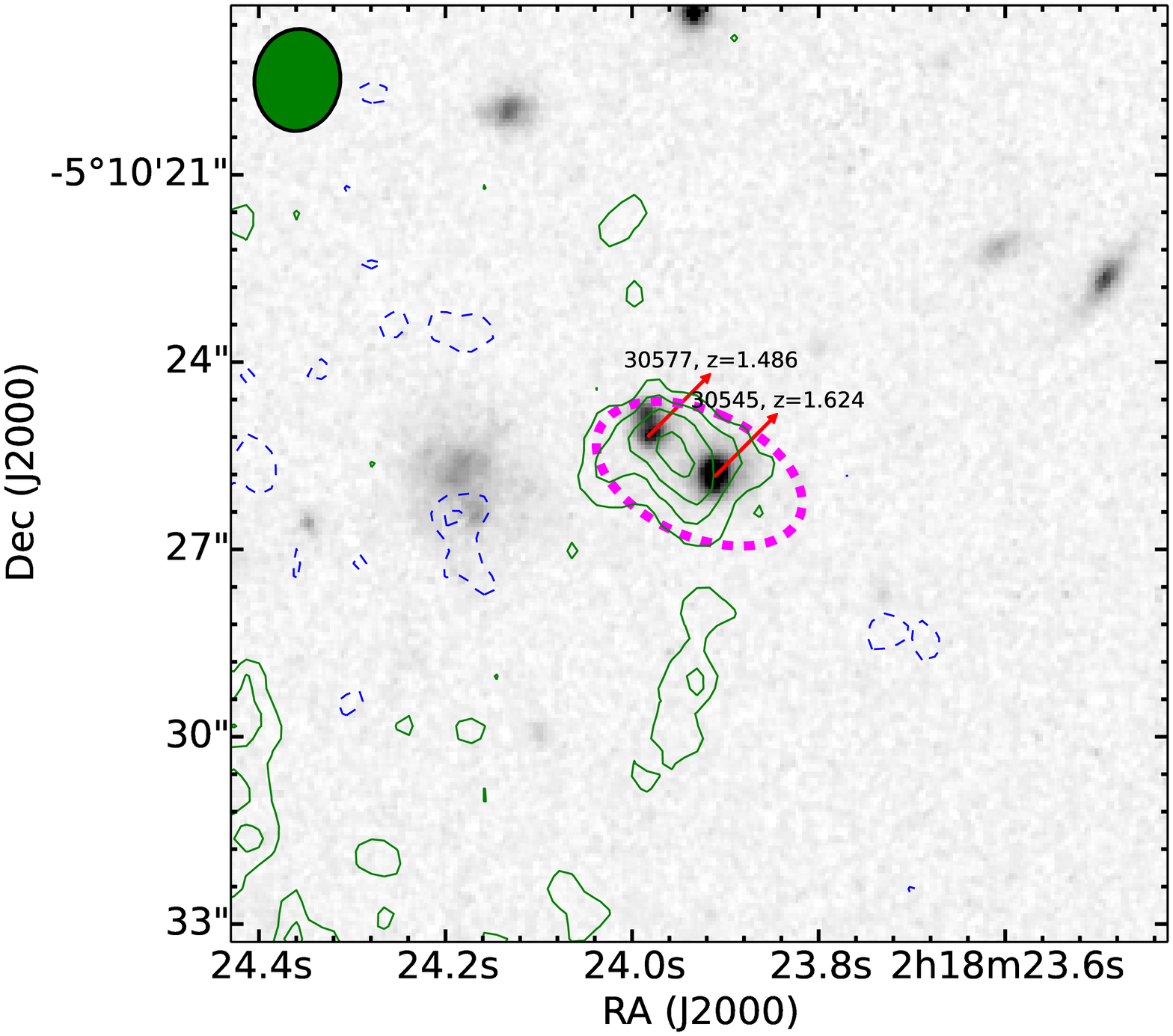}
\end{minipage}
\begin{minipage}[b]{0.49\textwidth}
  \centering
  \includegraphics[width=\textwidth]{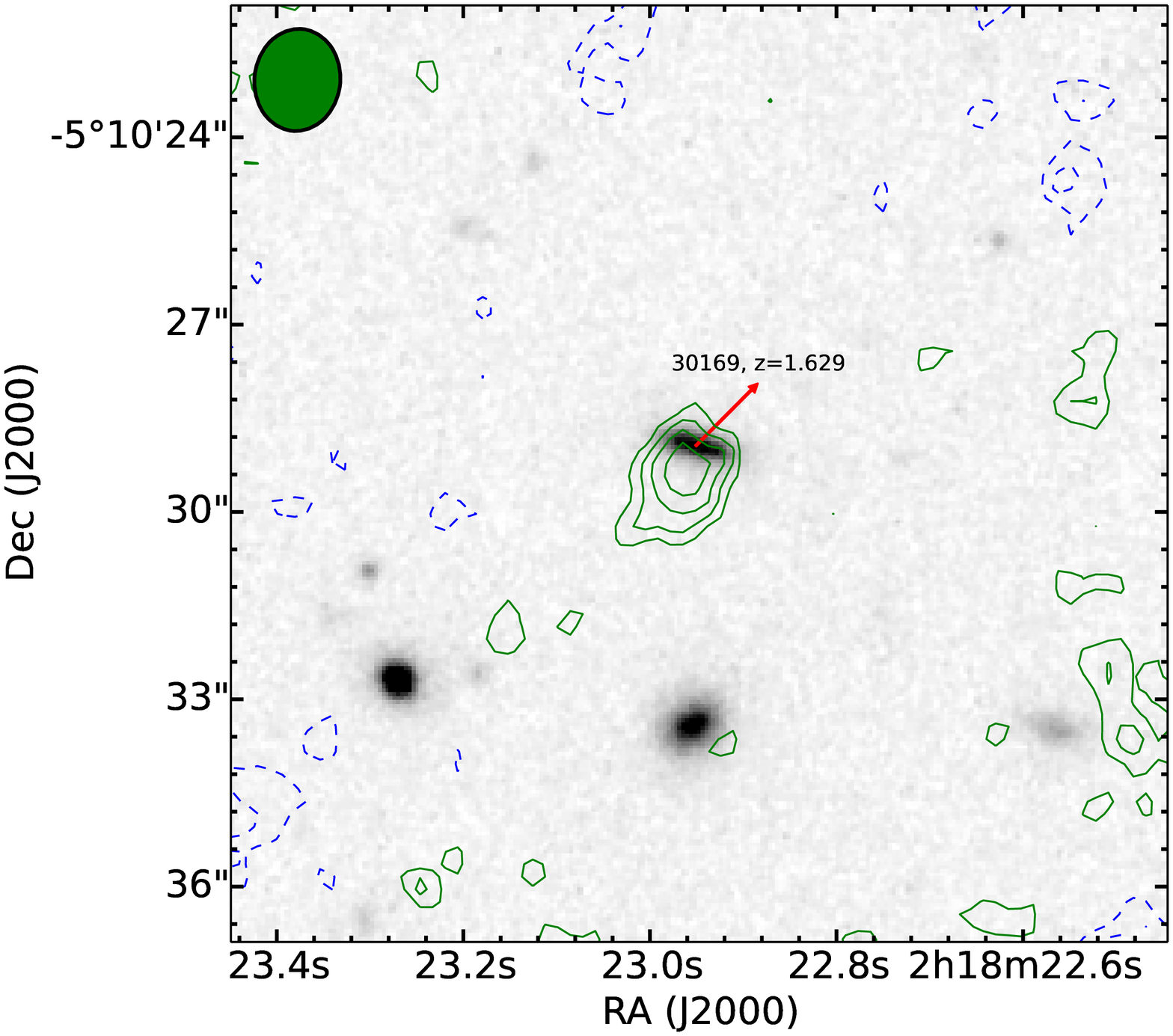}
\end{minipage}
\vspace{1cm}
\caption{CO contours overlaid on top of F160W HST/WFC3 images of our two detections.  We show 2, 3, 4, and 5-$\sigma$ contours as computed from the  collapsed and cleaned CO images.  Solid green contours are positive and blue dashed contours are negative.  Our synthesized beam is indicated in the upper left-hand corner. We also mark the redshift of sources near the CO source with red arrows.  The source to the NE of 30545 is at a different redshift and is unlikely to contribute to the extended CO.  The magenta ellipse in the left panel represents the aperture over which the CO flux is measured for that object.}
\label{Fig:co_im}
\end{figure*}

With these data we detect two sources in CO.  We show the CO spectra in Figure~\ref{Fig:co_spec}, the HST images and CO contours in Figure~\ref{Fig:co_im}, and will describe them in \S\ref{Sec:COdet}.  

\subsection{SFRs, Stellar Masses, and Sizes}
\label{sec:sfr_mstar}

\begin{figure}
\epsscale{1.10}
\plotone{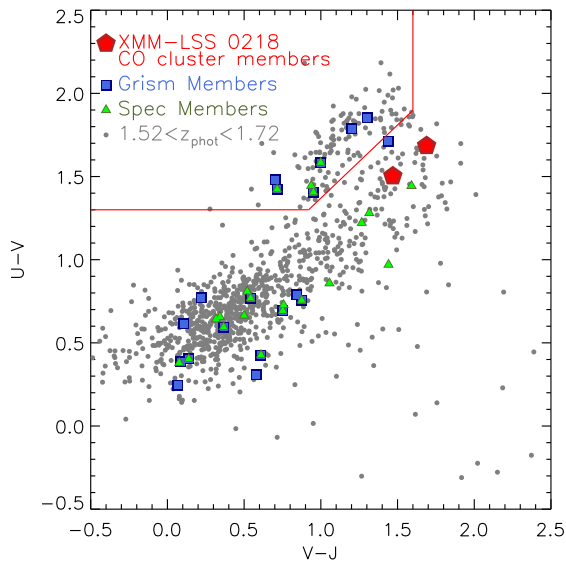}
\caption {The optical/NIR colors of our CO sources (see \S\ref{Sec:COdet}) compared to those of spectroscopic and grism members as well as objects with photometric redshifts close to the cluster redshift and $J<24.5$.  The red region marks the division between passive (upper left) and star-forming (lower right) as determined from \citet{Williams09}. Our two CO sources are consistent with being dust-obscured star-forming galaxies.}
\label{Fig:uvj}
\end{figure}

\begin{figure*}
\epsscale{1.10}
\plottwo{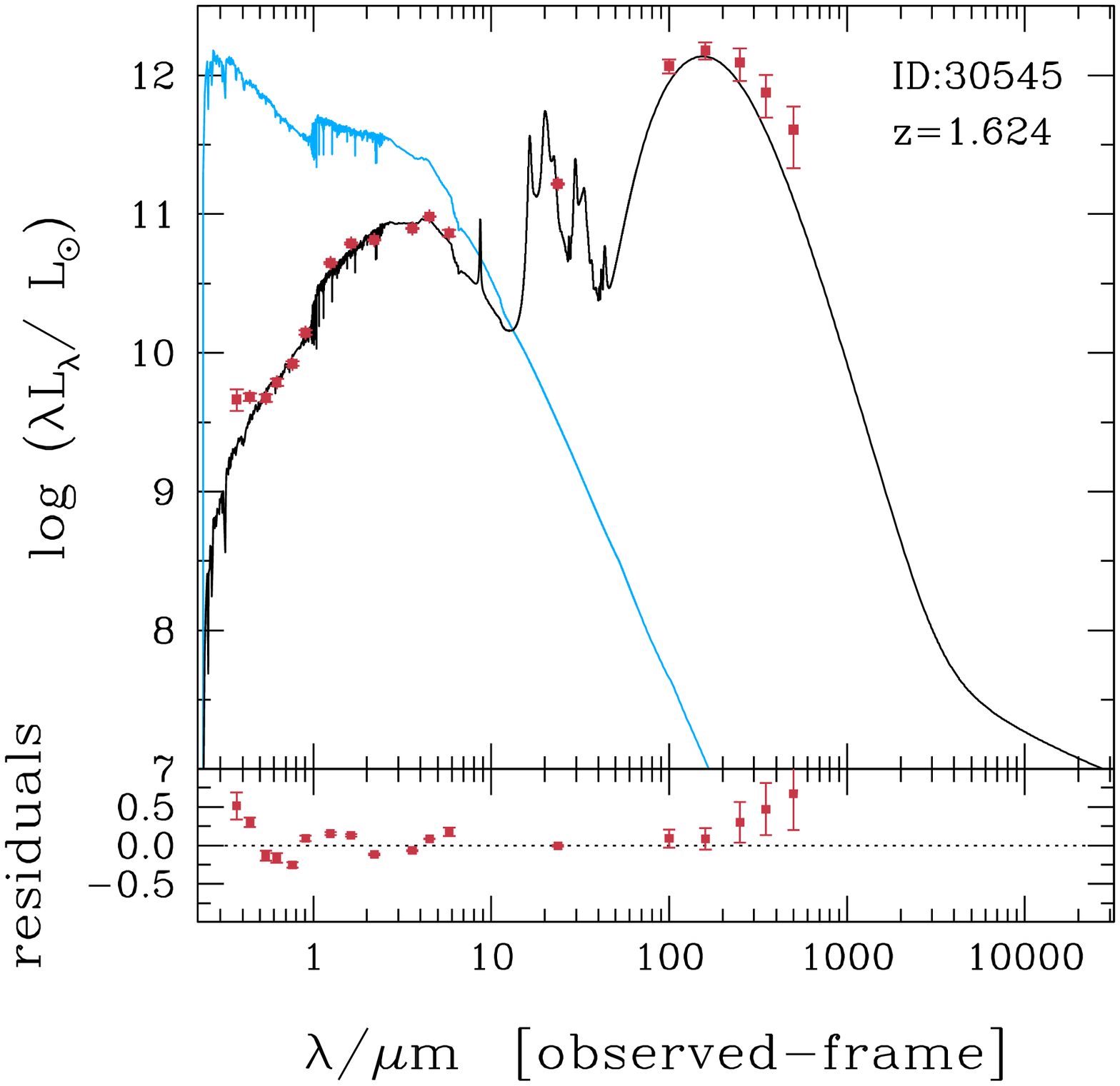}{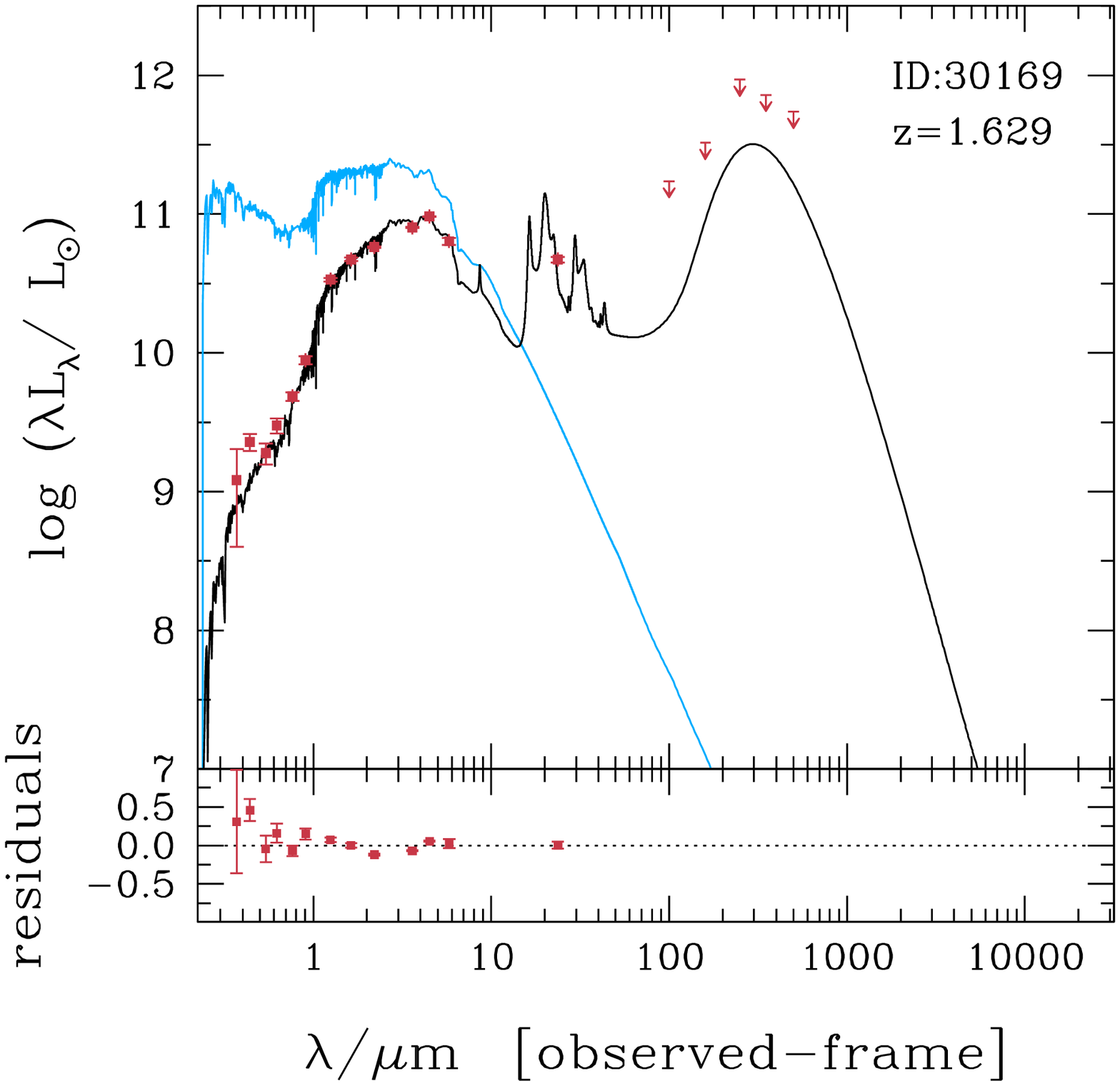}
\caption {The SEDs and model fits for our two CO detected galaxies.
  The fits were performed with the MAGPHYS package \citep{dacunha08}.
  The red points are the data with uncertainties.  3-$\sigma$ upper limits for \textit{Herschel} are shown as downward arrows.   \textit{Top panel:}
  The blue curve represents the unattenuated stellar continuum.  The
  black curve shows the attenuated stars and the dust
  emission. \textit{Bottom panel:} The residuals from the SED fit.  
  }
\label{Fig:sed}
\end{figure*}

Both of the CO detected galaxies (see \S\ref{Sec:COdet}) are detected at 24\micron\ and only the brightest (30545) with \textit{Herschel}.  In Figure~\ref{Fig:uvj} we show the position of these galaxies in the rest-frame $U-V$ vs. $V-J$ space pioneered by \citet{Wuyts07} and \citet{Williams09} to separate dusty star-forming from passive galaxies.  Our two sources have colors consistent with dusty star-forming objects.  We quantify the star formation rates (SFR) and stellar masses (\mstar) using the "HIGHz" extension of the MAGPHYS SED modeling software \citep{dacunha08,dacunha15}\footnote{\texttt{http://www.iap.fr/magphys/magphys/MAGPHYS.html}} assuming a \citet{Chabrier03} initial mass function (IMF).  MAGPHYS uses the physically motivated \citet{Charlot00} dust model to account for the light absorbed in the rest-frame UV through NIR and self-consistently requires that this absorbed energy is output in the mid-to-far infrared.  This code has been tested on simulated isolated galaxies and major mergers and been shown to correctly retrieve \mstar, SFRs, and \lir\ of the simulated objects \citep{Hayward15}.  It was also shown in \citet{dacunha13} that MAGPHYS, when used to fit $U-K$ photometry, can accurately predict the \lir\ derived for the same galaxies from \textit{Herschel} measurements.  The SED fits are shown in Figure~\ref{Fig:sed} and the derived parameters are given in Table~\ref{stellpops_tab}.  Despite the formally small uncertainties in the fitting provided by the exquisite data, we acknowledge that there are unaccounted for systematic errors in the stellar population models and the derived parameters.  We therefore assume a minimum error of 0.15~dex for the \mstar, \lir, and SFR measures.  For our two CO-detected objects (see \S\ref{Sec:COdet}), 30169 has an \lir$=2.9\times10^{11}$\lsol\ and object 30545 has \lir$=1.7\times10^{12}$\lsol.  

The Herschel fluxes for 30545 are not fit very well by the SED, although they are within 1-1.5$\sigma$ of the model fit.  To assess the effect of this on the derived SFR for 30545 we attempted to fit the SED with three different variations: 1) We only fit the data longword of $\lambda_{jobs}=3$\micron, 2) we relaxed the energy balance constraint, such that the absorbed optical light did not need to exactly equal that emitted in the IR, and 3) we increased the weight of the Herschel bands so that they contributed more to the fit.  In case 1 and 2 the SED fit the Herschel flux perfectly, although at the expense of fitting the rest-frame UV.  In all three cases the SFR remained within 0.05~dex of the original value.  We are therefore confident that the small mismatch between the model and data in the FIR is not influencing our \lir\ or SFR values.  We also note that the two bluest points for 30545 are significantly deviant from the best-fit model.  To assess the impact of this mismatch we forced the photometry to fit the UV-optical data for 30545 but found that this gave an entirely unacceptable (and low) fit to the Herschel and 24\micron\ data.  This is because the low $A_V$ required by the models to match the UV data resulted in too low IR emission.  We suspect that this is potentially because of an abnormal dust distribution or because a contribution from the x-ray AGN that is in this source but makes a small contribution to the IR flux (see below).  Given that the energy output for 30545 is clearly dominated by the IR emission, the small disagreements in the rest-frame UV do not affect our derived SFR or \lir.

\begin{figure}
\epsscale{1.20}
\plotone{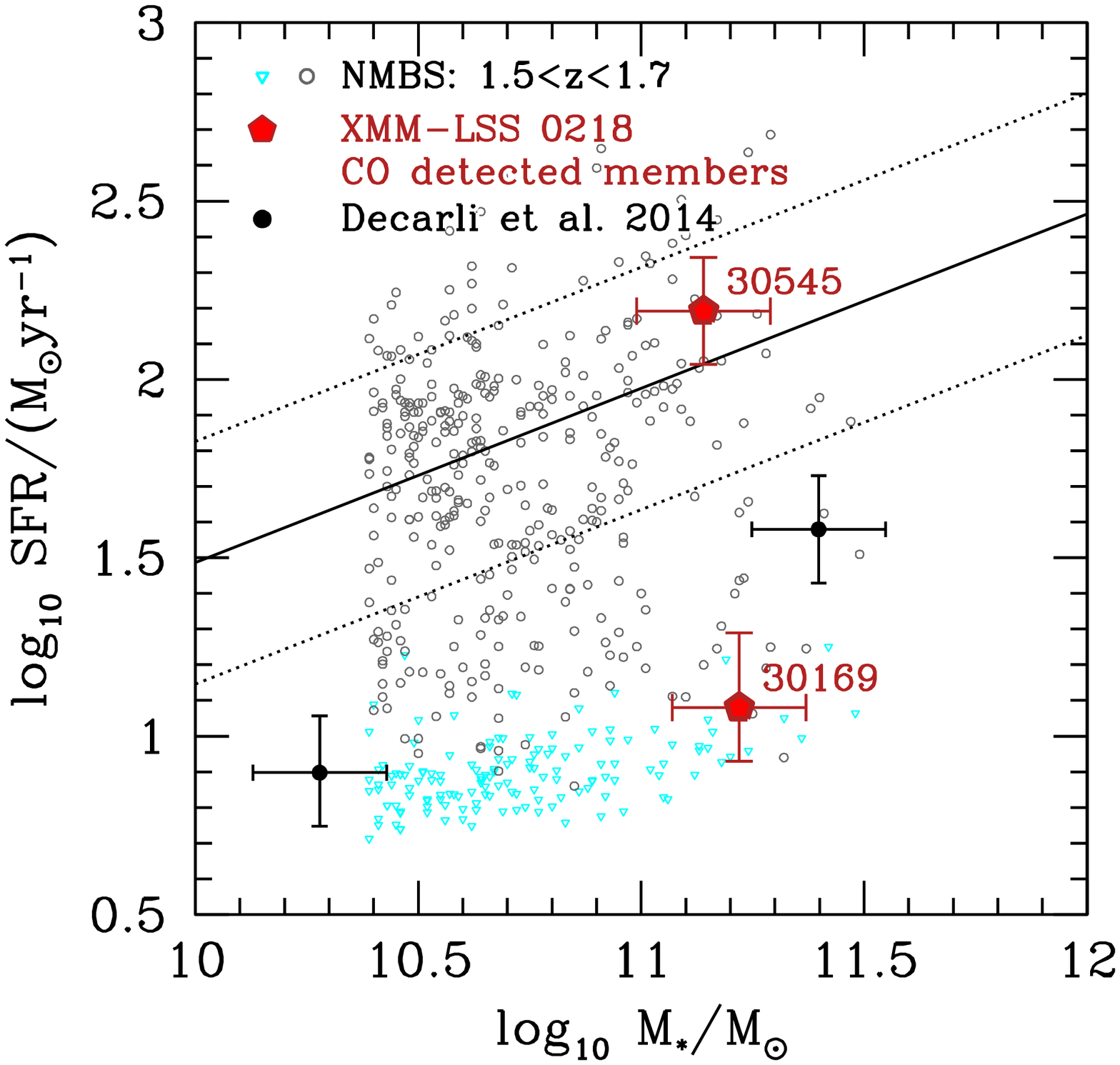}
\caption {The \mstar\ and SFRs for our two sources compared to those from the NEWFIRM Medium Band Survey \citep[NMBS;][]{Whitaker12}.  The SFRs from NMBS were computed using a combination of UV+IR.  Galaxies with IR detections are shown as dark gray circles.  Those not detected in the IR are indicated as 1$\sigma$ upper limits with cyan triangles.  The two CO detected sources have their SFRs measured from their full rest-frame UV through FIR SEDs (Figure~\ref{Fig:sed}).  On of our CO-detected cluster members is on the SF sequence but 30169 has a measured SFR that is an order of magnitude lower than that of the sequence.  In black we also plot two sources from \citet{Decarli14}, that were detected in a blind CO survey of the HDF-N.}
\label{Fig:mstar_sfr}
\end{figure}

In Figure~\ref{Fig:mstar_sfr} we plot the location of our two CO-detected galaxies in the \mstar\ vs. SFR plane.  Both objects have \mstar$\sim1.5\times10^{11}$\msol.  Object 30545 has SFR$=155$\msol$/yr$ and object 30169 has SFR$=12$\msol$/yr$.  Object 30545 lies on the \mstar-SFR relation for star-forming galaxies while object 30169, which is also star-forming, lies well below the sequence.  Object 30545 hosts an x-ray AGN and has moderately broad H$\alpha$ emission but the IR SED from MAGPHYS does not indicate an especially hot dust component, with $T_{dust}=45$K.  \citet{Santos14} determined the AGN contribution to \lir\ and concluded that an AGN could only contribute $\sim 4\%$ to the luminosity.  Note that any AGN contribution would lower the SFR inferred from the SED, moving this object even further below the \mstar-SFR sequence.  Despite its ample infrared luminosity, object 30169 is roughly an order of magnitude below the \mstar-SFR sequence.  
The best fit unattenuated stellar SED for 30169 also has a significant contribution from evolved stars as is evidenced by the strong 4000\AA\ break (Figure~\ref{Fig:sed}) and much of the \lir\ in this context may reflect the SFR averaged over the past $\sim100$Myr and not the instantaneous SFR.  We note that the Main Sequence from \citet{Whitaker12} that we plot in Figure~\ref{Fig:mstar_sfr} is within 0.15~dex of the more recent determination by \citet{Tomczak16}.

We use the rest-frame optical major axis effective radii for our objects as measured using CANDELS HST imaging \citep{vanderwel12}.  As object 30169 appears to be a disk, the semi-major $r_{1/2}$ is appropriate as it is inclination independent.  Object 30545 has an axis ratio of 0.75 and so the semi-major $r_{1/2}$ will not differ significantly from the circularized effective radius.  Object 30169 has $r_{1/2}=4.1$~kpc and 30545 has $r_{1/2}=1.93$~kpc (Table~\ref{stellpops_tab}).  These sizes correspond to 0.5 and 0.2\arcsec\ respectively and given our synthesized beam of 1\farcs5 we do not expect to resolve the CO if it has a similar radial extent as the stars.  

\ifemulate
	\begin{deluxetable*}{ccccccc}
\else
	\begin{deluxetable}{ccccccc}
\fi
\tablecaption{Stellar Population Parameters of CO-detected Galaxies}
\tablewidth{0pt}
\tablehead{\colhead{ID} & \colhead{log(\mstar/\msol)\tablenotemark{a}} & \colhead{SFR\tablenotemark{a}} & \colhead{log(L$_{IR}/$\lsol)\tablenotemark{a}} & \colhead{$r_{1/2}$\tablenotemark{b}} & \colhead{$n$\tablenotemark{c}} & \colhead{$q$\tablenotemark{d}}\\
\colhead{} & \colhead{} & \colhead{[\msol~yr$^{-1}$]} & \colhead{} & \colhead{[kpc]} & \colhead{} & \colhead{}}
\startdata
30169 & $11.22^{+0.15}_{-0.15}$ & $12.0^{+7.5}_{-3.5}$ & $11.46^{+0.15}_{-0.15}$ & $4.15\pm0.17$ & $0.6\pm0.1$ & $0.23\pm0.03$\\\\
30545\tablenotemark{e} & $11.14^{+0.15}_{-0.15}$ & $155.6^{+64.2}_{-45.4}$ & $12.23^{+0.15}_{-0.15}$ & $1.93\pm0.15$ & $2.7\pm0.4$ & $0.76\pm0.05$
\enddata
\label{stellpops_tab}
\tablenotetext{a}{Computed from the MAGPHYS \citep{dacunha08} fits to the full SED from the $u$-band through the \textit{Herschel} SPIRE bands at 500\micron.  We assign a minimum 0.15~dex uncertainty to all quantities.}
\tablenotetext{b}{The effective radius for a \citet{Sersic68} fit to the F160W HST/WFC3 imaging from \citet{vanderwel12}.}
\tablenotetext{c}{The \citet{Sersic68} index of the fit to the F160W HST/WFC3 imaging from \citet{vanderwel12}.}
\tablenotetext{d}{The minor-to-major axis ration of the fit to the F160W HST/WFC3 imaging from \citet{vanderwel12}.}
\tablenotetext{e}{The observed optical and NIR photometry for this source are well separated from the neighbor 30577.  It is possible that the MIPS 24\micron\ and Herschel fluxes may include contributions from 30545 and the neighbor 30577.  As the SFR is dominated by the FIR emission for the Herschel source, if it is blended we should still be measuring the total SFR corresponding to the CO detection.}
\ifemulate
	\end{deluxetable*}
\else
	\end{deluxetable}
\fi

\section{Results}
\label{Sec:results}

\subsection{\co{1}{0} detections of two star-forming galaxies}
\label{Sec:COdet}

We searched the data cube both blindly and at the location of each of our sources, using the available redshift information, i.e. $z_{spec}$, $z_{grism}$, or $z_{phot}$.  We securely detect a line in two cluster members, which we associate with \co{1}{0} (Figure~\ref{Fig:co_spec}).  From now on we refer to the objects by their closest match in the 3D-HST catalog \citep[see below;][]{Skelton14}, namely 30169 and 30545.
For each line we collapsed the image cube around the detection and slightly recentered the extraction pixel at the peak of the flux distribution.  We then collapsed the image again over the full extent of the visible line in the new 1D spectrum, shown in yellow in Figure~\ref{Fig:co_spec}.   
This frequency range was $\Delta\nu=43.7946-43.9168$GHz for 30169 and $\Delta\nu=43.8959-43.9869$GHz for 30545.  We cleaned these collapsed images using the clean task and cleaned down to 1.5$\sigma$ using a tight clean box around the source.  The cleaned images are shown in Figure~\ref{Fig:co_im}.  We determine the $S/N$ of these lines by comparing the flux at the peak of the collapsed clean source to the rms computed between 2 and 8 arc seconds from the source, i.e. an area with a similar primary beam correction.  The $S/N$ of the lines thus computed is 4.9 and 7.1 for 30169 and 30545 respectively.  

Fitting the profile of 30169 within CASA shows it to be consistent with a point source.  For that reason we extracted the spectra at the peak of the emission, as appropriate for an unresolved source.  We made images collapsed around the frequency 
As we will discuss below, the emission for 30545 is likely extended and we measured the flux in an elliptical aperture shown in Figure~\ref{Fig:co_im}.  We fit each spectrum with a Gaussian line profile to 30545 using the MPFITPEAK routine in IDL.  30169 is clearly non-Gaussian in nature and therefore we directly integrate the line.  To estimate the noise spectrum we compute the RMS of each channel in the annulus described above.  For 30545 we correct this noise spectrum to account for the multiple beams covering our aperture.  The redshift of the lines are $z_{line}=1.624\pm0.0006$ for 30545 and $z_{line}=1.629\pm0.001$ for 30169.  The ID numbers correspond to the sources from the 3D-HST catalog that are most closely matched in spatial and redshift coordinates to the CO line flux.  In Figure~\ref{Fig:co_im} we shown contours at the 2, 3, 4, and 5$\sigma$ level.  We now discuss the optical counterparts to the CO emission.

The location of the CO emission for 30169 is within 0\farcs3 of the position of the CANDELS NIR source, which corresponds to 2.5 kpc at the redshift of this galaxy.  We explored whether the two peaks in the spectra seen in Figure~\ref{Fig:co_spec} have different positions and thus contribute to the small offset of the CO source from the NIR source.  We collapsed the image around each peak and found the source to be in both maps and to be in the same location.   We therefore conclude that the CO emission from this galaxy is slightly offset from stellar light.  30169 has a grism redshift that agrees at the 95\% level with the CO redshift. Object 30169 has an H$\alpha$ redshift from observations with MOSFIRE \citep{Tran15}.  The spectroscopic redshifts of 1.629 agree perfectly with the CO redshift of 1.629 for 30545 and 30169 respectively.  We therefore unambiguously identify the CO emission with object 30169.

The source in the collapsed and cleaned CO map peaks half-way between 30545 and the source 30577 to the NE of 30545 (Figure~\ref{Fig:co_im}).  In our F160W data there is a possibility that there is some diffuse emission between the two sources but only at the faintest levels and it is not clear if it just represents the individual extended emission from each optical source. The CO emission may also be slightly extended and we use the \textit{imfit} task in CASA to estimate the intrinsic size of this source.  The source is resolved and has an intrinsic size of 2\farcs1$\times$0\farcs9 although with significant uncertainties.  30545 has an optical redshift from Magellan/IMACS \citep{Papovich12} and an H$\alpha$ redshift from observations with MOSFIRE \citep{Tran15}.  The spectroscopic redshift of 1.624 agrees perfectly with the CO redshift of 1.623.  Source 30577 has no spectroscopic redshift but we computed an improved grism redshift by jointly fitting the \citet{Skelton14} photometry, 3D-HST G141 data and our G102 data \citep{Lee-Brown17}.  The resulting redshift has a peak at $z=1.486$.  There is, however, a less likely second probability peak at $z=1.6$. There are no strong emission lines in the grism but a weak line is identified as H$\beta$ at $z=1.486$.  This weak line is not fit well at $z=1.6$.  We estimate the likelihood that this source is contributing the CO emission by integrating the grism $P(z)$ over the redshift range allowed by the full extend of the CO line ($z=1.620-1.626$).  This results in only a 1.4\% probability of being at that redshift, indicating that it is very unlikely that 30577 lies at the redshift of the CO line. 

Taking these arguments into account, we identify the CO line with 30545 for two reasons.  First, there is a perfect match between the spectroscopic redshift of 30545 and the CO line redshift and the grism redshift makes it highly unlikely that 30577 is at the correct redshift.  Note that the grism redshift for 30545 agrees very well with the spectroscopic redshift.  Second, the 24\micron\ detection and the 3.6\micron\ source  is more closely associated with 30545 and this increases the likelihood that both the Herschel flux and CO flux are coming from this object.  Nonetheless, the moderate $S/N$ and poor resolution of our CO data prevent us from being conclusive about the proper counterpart for this line.   We will require higher $S/N$ and higher resolution CO data with ALMA and a spectroscopic redshift for 30577 to definitively determine the counterpart.  There is a precedent for large offsets between CO emission and the rest-frame optical emission in high redshift intensely star-forming galaxies that may result from highly non uniform obscuration \cite[e.g.][]{Chapman05,Capak08,Riechers10,Hodge12} and such a large offset as seen in 30545 may therefore be physically plausible.  For now we assume that the stellar mass, SFR, \lir, and \lco\ all come from 30545.  As the SFR is clearly dominated by the FIR, assuming that it all comes from the same source or from a blend will not alter the total SFR of the system.  If the CO line is a blend of the two sources, then the main parameter that will be affected is the stellar mass.  However, 30545 has a stellar mass more than a factor of four more than 30577, implying that including 30577 will change the stellar mass by less than 25\%.

The velocity width of 30545 is FWHM$=351\pm12$km/s.  The line for 30169 is clearly non-gaussian and the window over which we collapse the CO image corresponds to 836 km/s.  It is not clear from our analysis if these velocities reflect purely dynamical motions or also include a large contribution from turbulence or molecular outflows.  It is possible that 30169 shows signs of a double horned profile but the data are currently too shallow to say this definitively.  Spatially resolved and higher signal-to-noise data may help us address that issue and for the remainder of the analysis we assume that the velocity widths for 30545 are dominated by dynamics, while we will be unable to use the velocity width for 30169.  We note that the MOSFIRE spectra also reveal broad H$\alpha$ for both galaxies, which is consistent with the broad CO line widths. 

The integrated flux for the lines from the Gaussian fit are $S_{CO}dv=0.19\pm 0.013$ and $0.05\pm0.02$~Jy~km/s for 30545 and 30169 respectively, both corrected for the primary beam sensitivity.    We give the CO line properties in Table~\ref{copar_tab}.

\subsection{Continuum detections}

We constrain the continuum level at a rest-frame frequency of 44.25~GHz by performing a weighted average of the spectra over the full 2~GHz bandwidth at the location of the two sources, masking out bad channels and the location of the emission lines.  We find no detection for 30169 with a 3$\sigma$ upper limit of 0.011~mJy.  We find a 3$\sigma$ detection of 30545 with $S_{44GHz}=0.015\pm0.005$~mJy.  We consider the implication of these detections in \S\ref{Sec:cont_mgas}.

\subsection{Comparison of the IR luminosity and CO luminosity}

We derive the CO luminosity \lco\ from the CO line flux using equation 3 from \citet{Solomon05}
\begin{equation}
L^\prime_{CO}=3.25\times 10^7 S_{CO}~dv~\nu_{\rm obs}^{-2}~D_L^2~(1+z)^{-3}
\end{equation}
and give the \lco\ in Table~\ref{copar_tab}.

\begin{figure*}
\plottwo{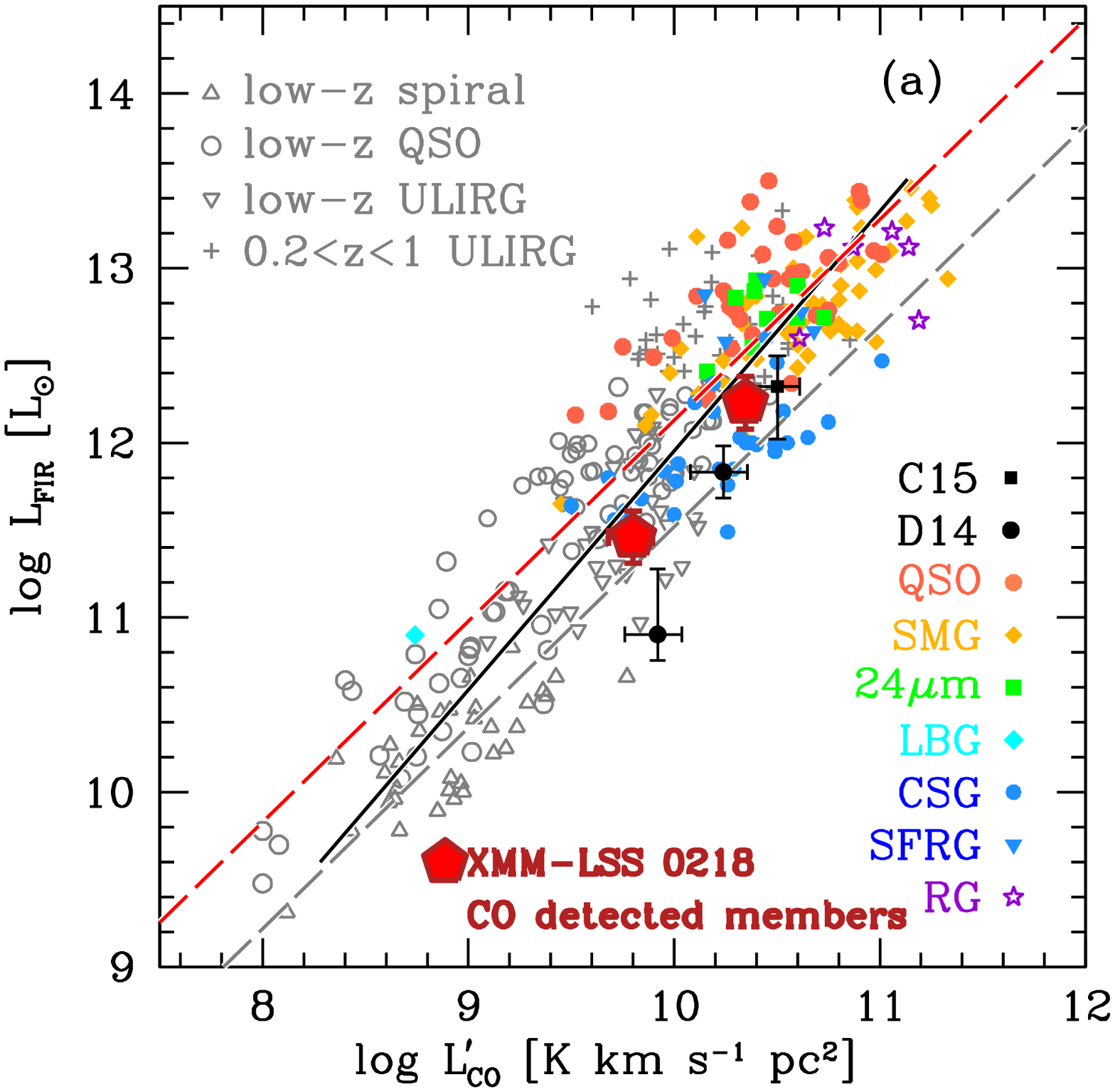}{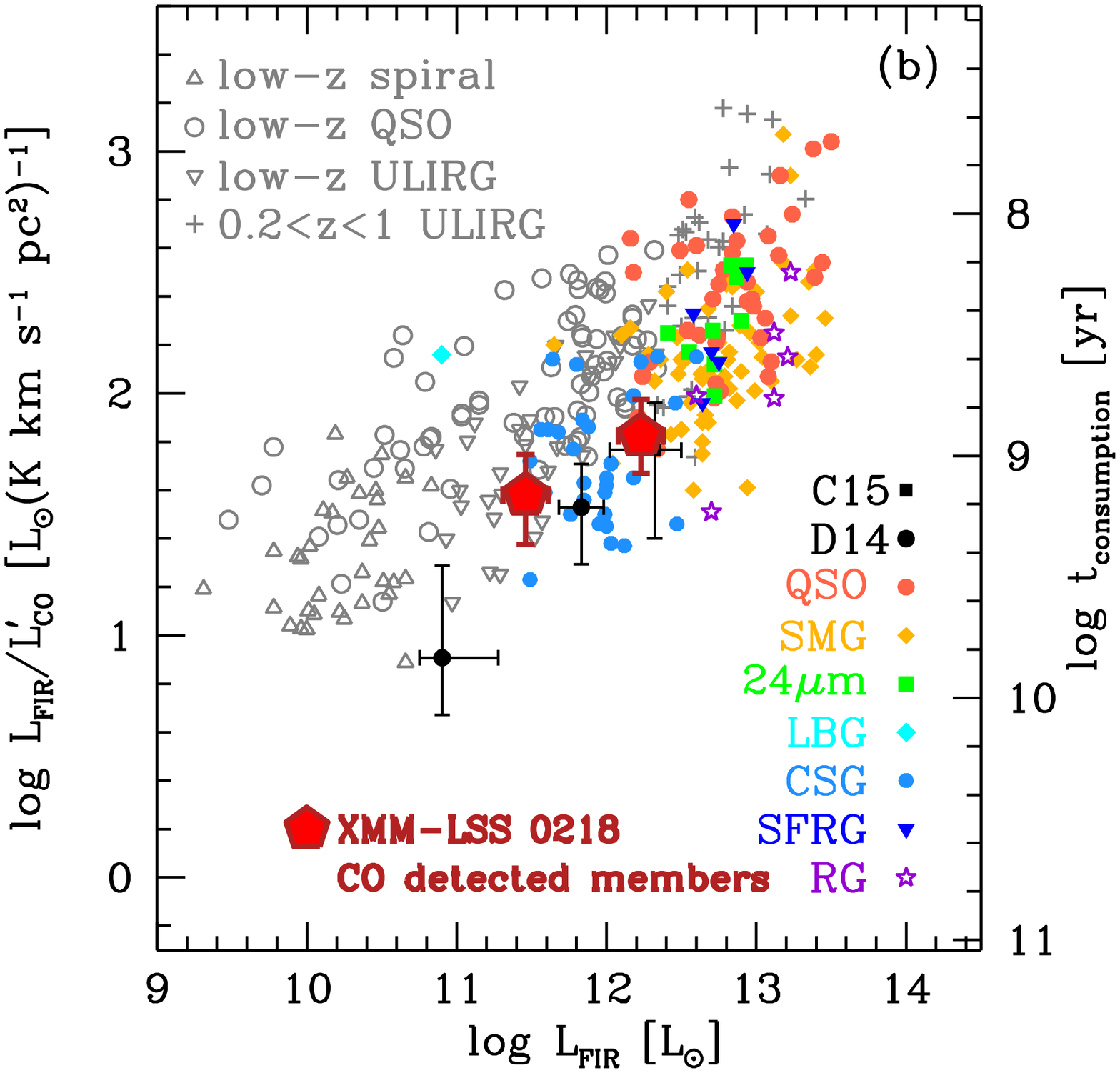}
\caption {\textit{Left Panel:} A comparison of the infrared luminosities and CO luminosities of our two CO detected cluster members at $z=1.625$ (large filled Pentagons) with a sample of star-forming galaxies and QSOs over a wide range of redshift taken from \citet{Carilli13} and which includes various local galaxies as well as all systems detected in CO at $z>1$ as of 2013.   In addition, we show two galaxies from D14 that were detected in a blind CO survey of the HDF-N and one from \citet[][C15]{Chapman15} that was detected in a blind survey of a proto-cluster at $z=2.3$.  \lir\ is a proxy for the SFR and the \lco\ is a proxy for the gas mass, modulo \aco.  The solid line is a fit to all data points, which gives a slope of $1.35\pm0.04$. The dashed lines indicate the best fits for the main sequence galaxies (gray) and starburst galaxies (red) derived by \citet{Genzel10} and \citet{Daddi10a}.  \textit{Right Panel:}  We compare the ratio of \lir$/$\lco\ to \lir\ for the same galaxies as shown in the left-hand panel.  \lir$/$\lco\ is a proxy for SFR/\mhtwo\ or the star formation efficiency.  On the right axis we plot the consumption timescale.  Our two cluster members are forming stars with typical SFE and have \tcon\ similar to other gas-rich galaxies at their \lir.   The legend abbreviations in both plots  stand for: QSO -- quasi-stellar objects; SMG -- submillimeter galaxies; 24\micron -- sources selected by 24\micron\ flux; LBG -- Lyman Break galaxies; CSG -- rest-frame UV color-selected "BM/BX" galaxies; SFRG -- star-forming radio galaxies; RG -- radio galaxies. }
\label{Fig:lir_lco}
\end{figure*}

In Figure~\ref{Fig:lir_lco} we compare the \lir\ and \lco\ of our galaxies to nearly all systems detected in CO at $z>1$ as of 2013 \citep[from][]{Carilli13}, as well as the two blind CO detections from \citet[][hereafter D14]{Decarli14} and the one blind detection from \citet[][C15]{Chapman15}.   The parameters for the D14 and C15 galaxies are shown in Table~\ref{comp_tab}.  Excitation corrections have been applied to all higher CO transitions but as we are using the \co{1}{0} line for our two galaxies, the excitation corrections there are minimal.  Our two CO detected cluster galaxies have an \lco\ that is within the range seen for field galaxies of comparable \lir\ at this epoch.

We interpret the \lco\ as \mhtwo\ after applying the conversion factor \aco, which we will discuss in \S\ref{subsec:mgas}.  We can also interpret \lir\ as the SFR, which is likely appropriate for galaxies of this \lir\ and is indicated by the MAGPHYS SED fits.  With that interpretation it would appear that our sources have typical SFRs for their \mhtwo.  We phrase the SFR$/$\mhtwo\ as the SFE, which implies that our two gas rich star-forming galaxies are converting their molecular gas to stars at a similar rate as galaxies that are targeted for CO observations based on their SFRs.   This is shown in the right-hand panel of Figure~\ref{Fig:lir_lco}.  We note that the other blind CO detections from D14 and \citet{Chapman15} are also consistent with the general locus of SFR-selected galaxies, indicating that blind CO surveys may not be selecting galaxies that are preferentially overluminous in CO.  

\subsection{Constraints from Stacking}

We attempt a stacking analysis of the CO data centered on all the galaxies and those in the star-forming region of $UVJ$ space (Figure~\ref{Fig:uvj}), excluding the two directly detected objects.  We extracted a spectrum at the pixel corresponding to the location of the NIR source in the 3D-HST catalog.  For each class of objects we make separate stacks for galaxies with spectroscopic redshifts and for galaxies with spectroscopic or grism redshifts.  The stacks have between 4 and 13 galaxies.  To estimate the flux in the stack we sum over an interval corresponding to the 1-sigma accuracy for each redshift determination, 340~km/s for spectroscopic redshifts and 1000~km/s for grism redshifts, added in quadrature with the 275~km/s that corresponds to the intrinsic width of the galaxy.  We detect no flux in any of the stacked spectra and the 3$\sigma$ upper limit on \lco\ is $5.14\times10^{10}$[K~km~s$^{-1}$~pc$^2$], which is higher than nearly any \lco\ shown in Figure~\ref{Fig:lir_lco}.  Therefore the stacking result places no useful constraints.

The lack of a detection in the stack may be driven primarily by the low numbers of spectroscopic members and by the non-negligible redshift errors in the grism data.  This cluster is also highly quenched in its core \citep{Lee-Brown17}, which further limits the number of star-forming galaxies eligible for a stack.

\subsection{\mhtwo, \mstar, and gas fractions}
\label{subsec:mgas}

We convert our \lco\ measurements to total molecular gas masses via ${\cal M}_{H_2}= L^\prime_{CO} \alpha_{CO}$, where we use a Galactic $\alpha_{CO}= 4.36~{\cal M_{\odot}} {\rm (K~km~s^{-1} pc ^{2})^{-1}}$ \citep[e.g.][]{Genzel15}.  This conversion factor includes the 36\% correction for Helium, which means that our gas masses reflect both the Helium and molecular hydrogen contents of galaxies. We give the gas mass in Table~\ref{copar_tab}. There is mounting evidence that a Galactic conversion factor is appropriate for galaxies on or below then local SFR-\mstar\ sequence (MS) and possibly even at higher redshift\citep[e.g.][]{Bolatto13}, although with significant variation.  Much of this variation in \aco\ stems from a metallicity dependence \citep[e.g.][]{Sandstrom13,Bolatto13} yet our galaxies both are massive and likely have near-solar metallicities, as do similarly massive star-forming galaxies in this cluster \citep{Tran15}.   In \S\ref{Sec:aco} we discuss in detail our justification for our choice of \aco\ and how our results depend on this choice.

\begin{figure}
\vspace{-0cm}
\centering
\includegraphics[scale=0.4,angle=0]{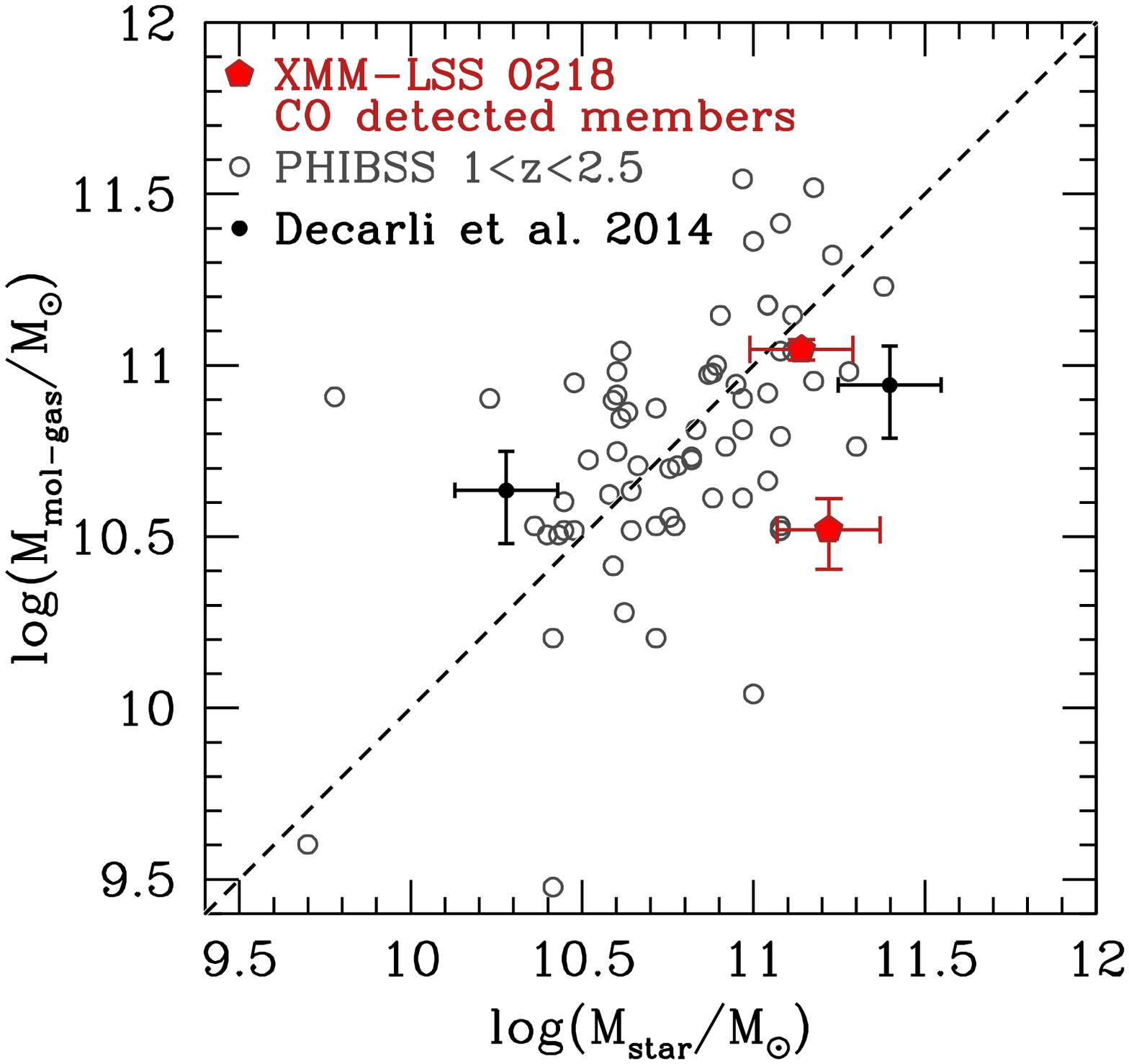}
\caption{\mstar\ and \mhtwo\ for our two galaxies detected in \co{1}{0}.  \mhtwo\ was estimated from \lco\ using a Galactic \aco, which is consistent with our dynamical constraints from the CO line width and the rest-frame optical size.  The dashed line is the one-to-one relation.  Compared to galaxies from the PHIBSS sample \citep{Tacconi13}, our two galaxies are at the high end of the range of \mstar\ and have gas fractions of \mhtwo$/$\mstar$=0.2-0.8$ and \mhtwo$/($\mhtwo$+$\mstar$)=0.17-0.45$, which are comparable to or lower than PHIBSS galaxies.  In addition, we show two galaxies from \citet{Decarli14} that were detected in a blind CO survey of the HDF-N.}
\label{Fig:mgas_mstar}
\end{figure}

We compare our stellar and gas masses to those for other star-forming galaxies on and near the SFR-\mstar\ relation in Figure~\ref{Fig:mgas_mstar}.  We find that our two CO-detected galaxies are at the massive end of the galaxies from PHIBSS in stellar mass but have typical to low molecular gas masses.  The gas fractions are \mhtwo/\mstar=0.2-0.8 or $f_{gas}\equiv$\mhtwo/(\mstar+\mhtwo$)=0.17-0.45$.  This is not unusual for vigorously star-forming galaxies at this epoch, as log (\mstar$/$\msol)$\approx 11$ galaxies from \citet{Tacconi13} have $f_{gas}\approx 0.4$.    Nonetheless, one of our galaxies is forming stars a factor of $\sim 10$ below the levels of galaxies of similar mass that lie on the SFR-\mstar\ sequence yet still has substantial amounts of molecular gas.  We address the low SFRs in the presence of the measured gas fractions in subsequent sections.  As a comparison we also show two galaxies from D14 that were detected in a blind CO scan of the HDF-N with PdBI.  

\ifemulate
	\begin{deluxetable*}{ccccccc}
\else
	\begin{deluxetable}{ccccccc}
\fi
\tablecaption{CO line properties}
\tablewidth{0pt}
\tablehead{\colhead{ID\tablenotemark{a}} & \colhead{$z_{CO}$\tablenotemark{b}} & \colhead{$S/N$\tablenotemark{c}} & \colhead{$S_{CO}~dv$\tablenotemark{b}} & \colhead{$\Delta v_{CO}$\tablenotemark{d}} & \colhead{\lco\tablenotemark{b}} & \colhead{log(\mhtwo/\msol)\tablenotemark{e}} \\\\
\colhead{} & \colhead{} & \colhead{} & \colhead{[Jy~km~s$^{-1}$]} & \colhead{[km~s$^{-1}$]} & \colhead{[K~km~s$^{-1}$~pc$^2$]} & \colhead{}}
\startdata
30169 & $1.629\pm0.001$ & 4.9 & $0.06\pm0.01$ & $836$ & $0.76\pm0.18\times10^{10}$ & 10.52$^{+0.09}_{-0.12}$\\\\
30545 & $1.624\pm0.0006$ & 7.1 & $0.19\pm0.013$ & $351\pm12$ & $2.55\pm0.18\times10^{10}$ & 11.05$^{+0.03}_{-0.03}$
\enddata
\label{copar_tab}
\tablenotetext{a}{ID is from 3D-HST catalog of \citet{Skelton14}.}
\tablenotetext{b}{For 30169, this was computed from the direct sum over the line weighted by the inverse variance, as the line is clearly non-Gaussian.  For 30545 it was computed from a Gaussian fit to the line profiles from Figure~\ref{Fig:co_spec}.  Nonetheless, the $S_{CO}~dv$ value is the same to within 10\% if using the Gaussian fit or if directly summing over the line.}
\tablenotetext{c}{S/N is computed from the cleaned collapsed image, using the peak flux density and the rms computed in an annulus around the source.}
\tablenotetext{d}{For 30169, this is the full velocity width of the line that was used to integrate the flux.   For 30545 it was computed from the Gaussian fit and corresponds to the FWHM.}
\tablenotetext{e}{Computed assuming \aco$=4.36$.}
\ifemulate
	\end{deluxetable*}
\else
	\end{deluxetable}
\fi

\ifemulate
	\begin{deluxetable*}{lllllllll}
\else
	\begin{deluxetable}{ccccccccc}
\fi
\tablecaption{Comparison sample properties}
\tablewidth{0pt}
\tablehead{\colhead{ID} & \colhead{$z$} & \colhead{source} & \colhead{log(L$_{IR}/$\lsol)} & \colhead{SFR} & \colhead{log(\mstar/\msol)} & \colhead{\lco} & \colhead{$r_{1/2}$} & \colhead{q} \\\\
\colhead{} & \colhead{} & \colhead{} & \colhead{} & \colhead{[\msol~yr$^{-1}$]} & \colhead{} &  \colhead{[K~km~s$^{-1}$~pc$^2$]}  & \colhead{kpc} & \colhead{} }
\startdata
03 & 1.7844 & D14 & $11.83^{+0.04}_{-0.01}$ & $ 38.0^{+8.0}_{-1.0}$ & 11.40 & $ 2.01\pm0.60\times10^{10}$ & $ 0.23\pm0.00$ & $ 0.75\pm0.01$\\
19 & 2.0474 & D14 & $10.90^{+0.38}_{-0.06}$ & $ 7.9^{+3.5}_{-1.4}$ & 10.28 & $ 0.99\pm0.30\times10^{10}$ & $ 0.14\pm0.00$ & $ 0.58\pm0.02$\\
 DRG55 & 2.296 & C15 & $12.32$ & $ 210$ & $\sim 11$ & $ 3.6\pm1.0\times10^{10}$ &  & 
\enddata
\label{comp_tab}
\ifemulate
	\end{deluxetable*}
\else
	\end{deluxetable}
\fi

\section{Discussion}
\label{Sec:discussion}

We have presented our two CO-detected galaxies that reside in a $z=1.625$ cluster and have shown that these galaxies are massive (log (\mstar$/$\msol)$\approx 11$) and gas-rich (log (\mhtwo$/$\msol)$\approx 10.5-11.05$) and are forming stars at values similar to those seen for comparably massive and gas-rich galaxies.  In the following section we discuss the SFEs and the implications these have for the future of these cluster galaxies.  

\subsection{Star Formation Efficiencies}
\label{Sec:lowsfe}

\begin{figure*}
\plottwo{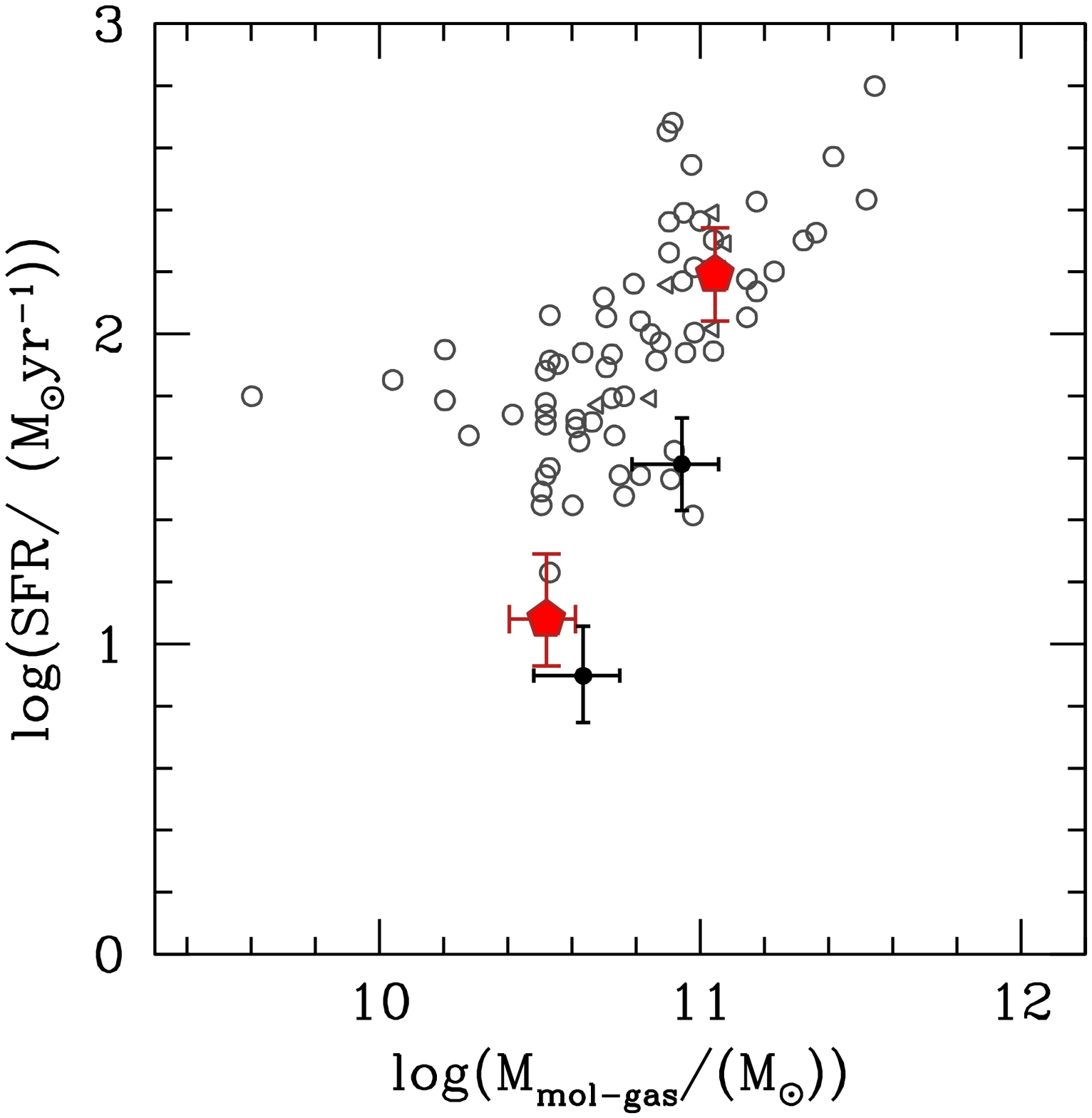}{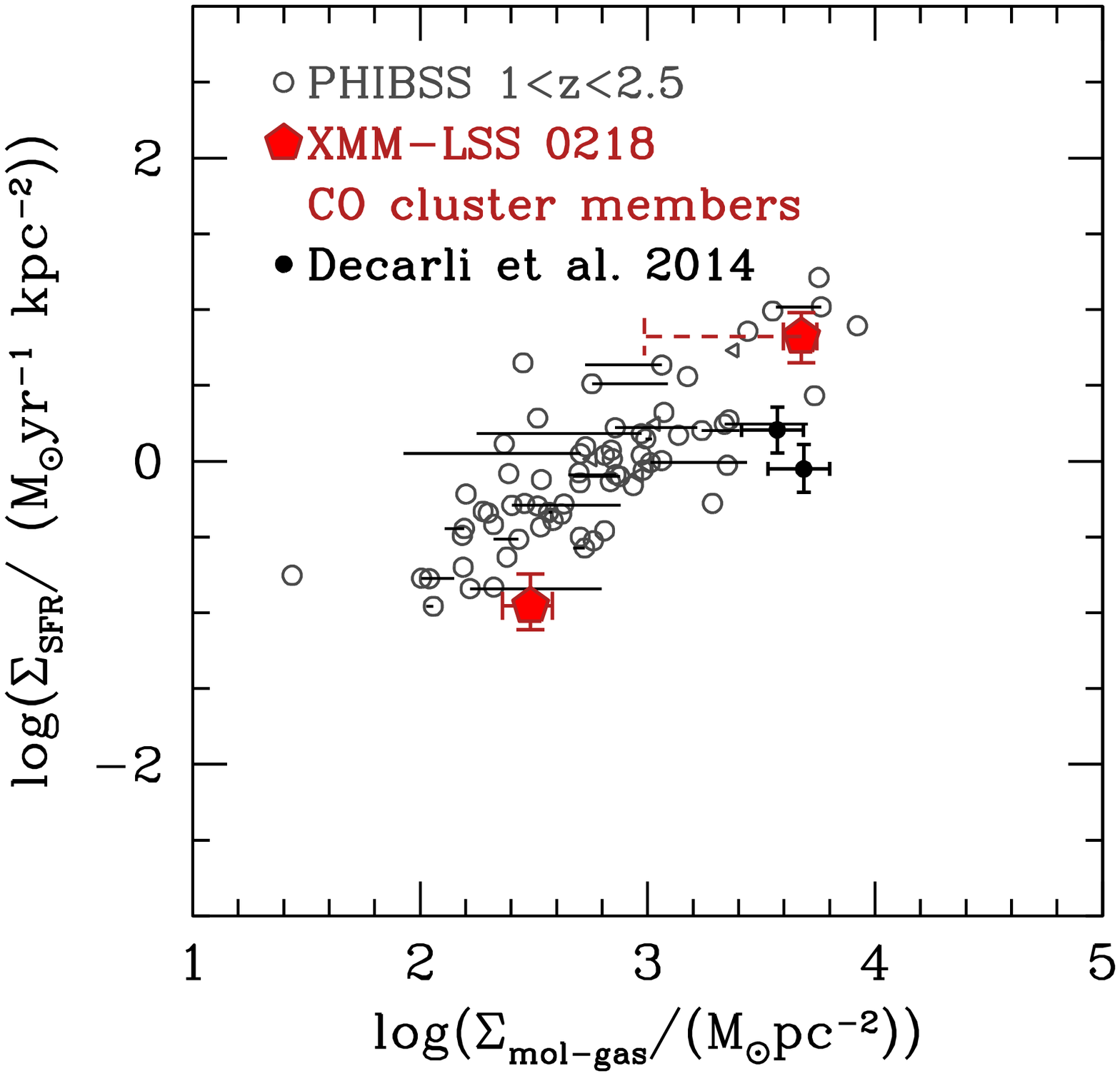}
\caption{\textit{Left Panel:} The SFR and molecular gas mass  for our two CO detected cluster members, galaxies from PHIBSS \citep{Tacconi13} and the two blind detections from \citet{Decarli14}.  We convert \lco\ to gas mass using a Milky Way \aco.  30169 has a SFR lower than the PHIBSS sources while 30545 is consistent with the distribution of PHIBSS sources in \mhtwo\ and SFR.  \textit{Right Panel:}  The surface density of star formation and molecular gas for the same sources.  In this diagram the star formation efficiency decreases down and to the right.  30169 has a \sigsfr\ less than nearly all of the PHIBSS points while 30545 is at the upper end of the distribution and is consistent with the PHIBSS distribution.  The dashed error bar on the upper red point (30545) shows how the SFR surface density would change if integrated over the extend of the resolved CO line instead of over the stellar disk.  If this is appropriate then the SFE for 30545 would be higher than that of galaxies in the PHIBSS sample.  A more precise comparison will require actual gas size measurements for our sources.  Note that the same size is used for both the SFR and gas surface density for all measurements and this may partly explain the strong correlation between the two parameters in the right-hand panel.  The horizontal black lines on the \citet{Tacconi13} points show how the surface densities change for those sources that have direct CO size measurements.}
\label{Fig:sigsfr_sigh2}
\end{figure*}

As shown in Figure~\ref{Fig:lir_lco}, our two sources have typical \lir\ for their CO luminosity.  We interpret this as a normal SFE, where SFE$\equiv$SFR$/$\mhtwo.  That is, our two galaxies are forming stars at typical rates for their gas masses.  We show this in another way in the left-hand panel of Figure~\ref{Fig:sigsfr_sigh2}, in which we plot the total SFR vs. \mhtwo, which also shows that our galaxies lie within the locus of the PHIBSS sources.  To gain further insight we plot the surface density of molecular gas (\siggas$\equiv$\mhtwo$/(2\pi r_{1/2}^2$) vs. that star formation (\sigsfr$\equiv$SFR$/(2\pi r_{1/2}^2$) in Figure~\ref{Fig:sigsfr_sigh2}.  Lacking a spatially resolved measure of the SFR or \mhtwo\ we adopt the rest-frame optical half-light radius as the relevant spatial scale for the SFR and gas.  This differs somewhat from \citet{Daddi10a} and \citet{Tacconi13}, who use the rest-frame UV half-light radius.  However, \citet{vanderwel12} provide a fitting function for the wavelength dependence of $r_{1/2}$ in CANDELS galaxies at similar redshifts and correcting our F160W sizes to those measured with F814W would result in a 0.1~dex increase in the sizes and only a $40\%$ (0.2~dex) change in our surface densities.   We note that changing the size to account for systematic differences between the rest-frame optical and UV sizes will affect the \siggas\ and \sigsfr\ in the same way and so will move objects parallel to the locus of PHIBSS galaxies.  An additional source of error would clearly be if the CO size is systematically different from the size of the rest-frame UV or optical light.  This may be true for 30545 as we have measured the gas to be marginally extended (\S\ref{Sec:COdet}).  The dashed error bar for this sources indicates how the gas surface density would change if we use the 2\farcs1$\times$0\farcs9 size, but note that this size is uncertain given the low resolution of our data.  We assume going forward that the sizes are the same \citep{Daddi10a,Tacconi13} but will need high resolution CO imaging to test this assumption.  Under the assumption that the gas and star formation have the same spatial distribution - the same assumption made for the PHIBSS galaxies - this therefore implies that these two CO detections may have  lower \sigsfr\ than galaxies with equivalently high \siggas\ or conversely that they may be forming stars with a somewhat smaller spatially resolved SFE.


We further examine how our galaxies compare to the global star-forming population at their redshift by comparing them to the scaling relations for \mhtwo/\mstar\ and \tcon\ from \citet{Genzel15}.  That paper uses a large sample of galaxies with SFR, \mstar, and \mhtwo\ measurements spanning a large range in redshift ($0<z<3$).  They found that \mhtwo/\mstar\ and \tcon\ followed scaling relations with separable dependencies on redshift, \mstar, and distance with respect to the \mstar-SFR sequence.  The sense of the trends is such that galaxies below the \mstar-SFR sequence at a fixed redshift and stellar mass have lower SFRs and lower gas fractions than those on the sequence.  This results in galaxies below the \mstar-SFR sequence having higher \tcon\ (or lower SFE) than those on the sequence.  

\begin{figure*}
\plottwo{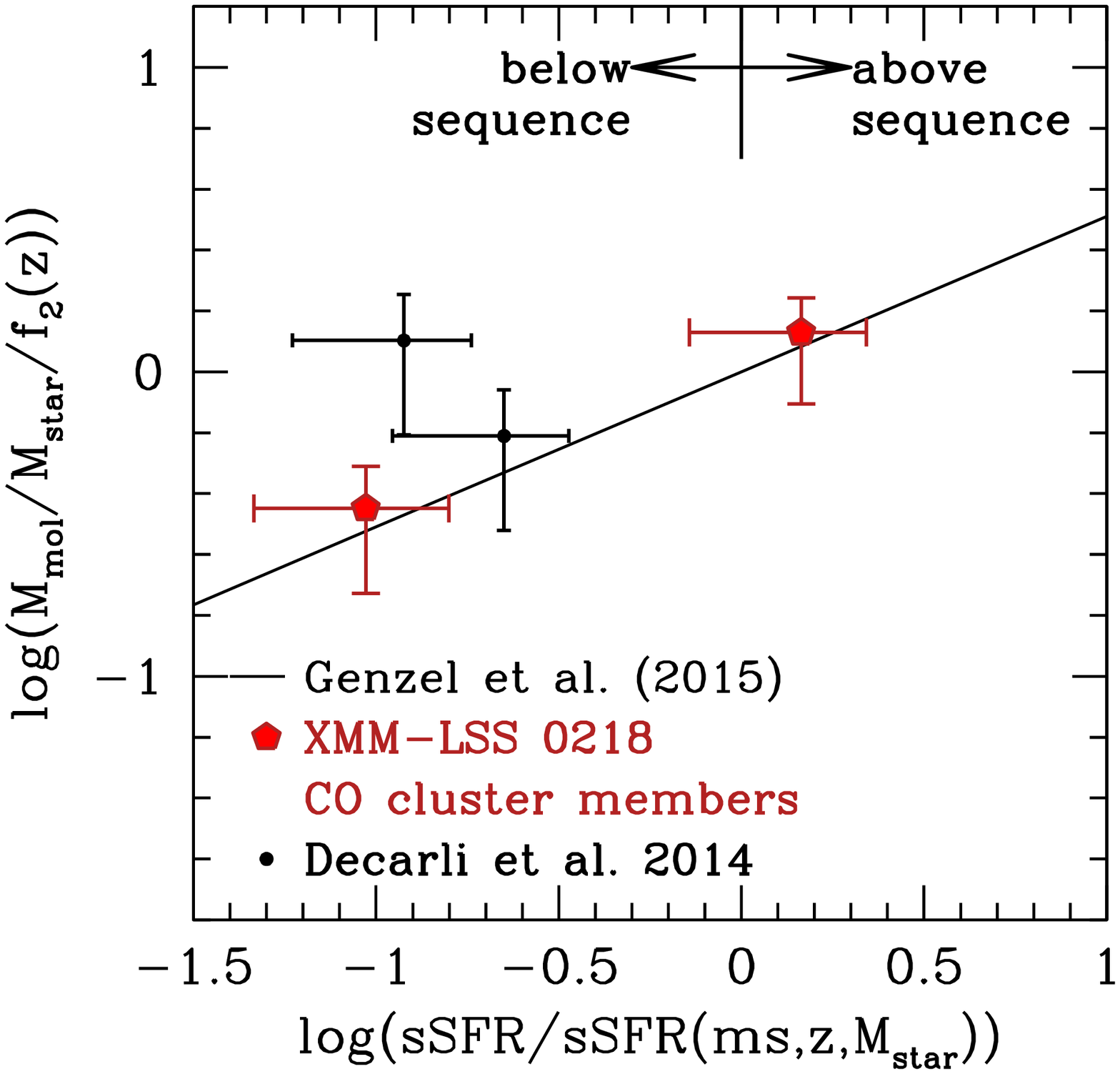}{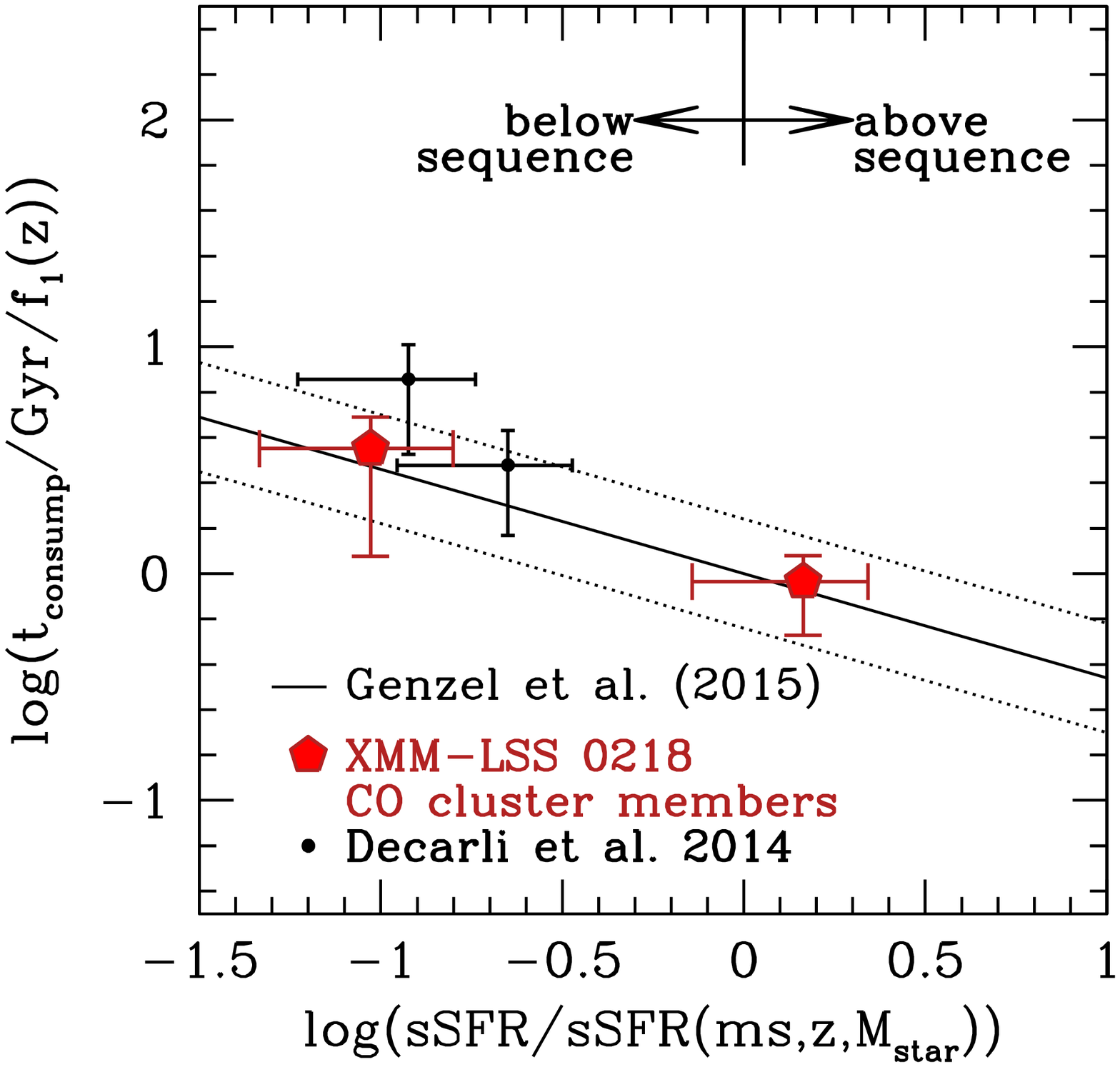}
\caption {A comparison of our galaxies with the scaling relations from \citet{Genzel15}.  In both panels, the x-axis is the distance from the main sequence, which is parameterized, as in \citet{Genzel15}, using the relation from \citet{Whitaker12}. Both scaling relations and galaxies have had the stellar mass and redshift dependence removed (see text.) \textit{Left Panel:}  The molecular gas fraction of our galaxies and those of \citet{Decarli14}.  The four galaxies have gas fractions consistent with the scaling relations.  \textit{Right panel:} The consumption timescales for the same four galaxies. The dotted lines indicate the 0.24~dex scatter around the scaling relation from \citet{Genzel15}.  All four galaxies have \tcon\ consistent with the scaling relations.}
\label{Fig:scale_rel}
\end{figure*}

We plot our galaxy on those scaling relations in Figure~\ref{Fig:scale_rel}.  The scaling relations depend weakly on stellar mass and we have removed this dependence from \citet{Genzel15} (using the formula from their Table 3 and 4) and the redshift dependence of the scaling law using the fitting functions $f_1(z)=10^{-0.04-0.165 \times log(1+z)}$ and $f_2(z)=10^{-1.23+2.71 \times log(1+z)}$.  We have also normalized our galaxies with respect to the \mstar-SFR sequence at the redshift and stellar mass of each galaxy such that each galaxy's specific SFR is given with respect to the main sequence. Our two cluster galaxies and the two sources from \citet{Decarli14} are consistent with the \citet{Genzel15} scaling relations for field galaxies at $z<3$.  In the context of these relations, the interpretation of the low gas content for 30169 is consistent with its low SFR, although we note that there are no galaxies at $z>0.6$ in the PHIBSS2 sample with such low SFRs.  Hence the scaling relations are not calibrated at such low SFRs.  It is therefore interesting that our galaxies nonetheless agree so well with the scaling relation prediction.

\begin{figure}
\vspace{-0cm}
\centering
\includegraphics[scale=0.4,angle=0]{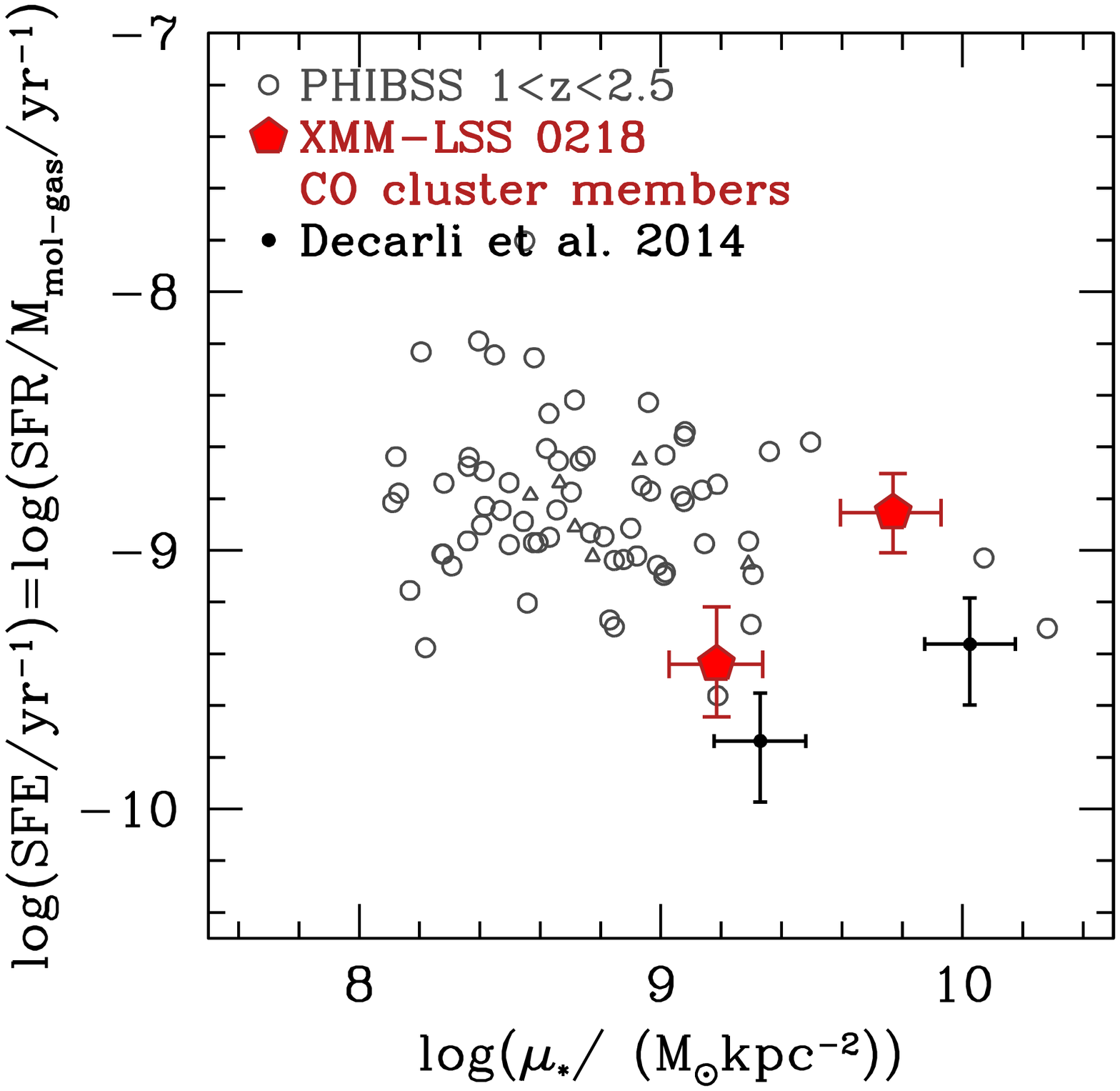}
\caption{The Star Formation Efficiency (SFE$\equiv$SFR$/$\mhtwo) vs the stellar surface mass density for our two CO detected cluster members, galaxies from PHIBSS \citep{Tacconi13} and the two blind detections from \citet{Decarli14}.  
We convert \lco\ to gas mass using a Milky Way \aco, which is consistent with the dynamical constraints from the CO line width and the rest-frame optical size.  Our galaxies and the two other blind detections are at the extreme high end of surface mass density. }
\label{Fig:sigmstar_sfe}
\end{figure}

To further place our sources in the context of larger field galaxy surveys we compare how their SFE relates to their central surface mass density.  We first calculate the stellar mass surface density within the half-light radius as \mustar$\equiv$\mstar$/(2\pi r_{1/2}^2)$, assuming that half the stellar mass is contained within $r_{1/2}$.  We therefore have assumed that the H-band light traces the stellar mass for our galaxies and the two blind detections and that there are no significant color gradients.  We also use the rest-frame UV size as a proxy for the stellar mass size for the PHIBSS galaxies.  As described above, the difference in the size in the rest-frame NUV and optical is only 0.1~dex and will not affect our results.   We plot SFE vs \mustar\ for our two galaxies, the points from \citet{Decarli14}, and the points from PHIBSS2 in Figure~\ref{Fig:sigmstar_sfe}.  We find that our sources and those from \citet{Decarli14} are at the extreme high end of \mustar\ for galaxies of nearly any SFE from PHIBSS.  We do not know what is driving this compact mass distribution, i.e. if our galaxies are dominated by  compact spheroids or disks.  We note that 30545 is round and compact, with $r_{1/2}=1.93$kpc, within the official criteria of the compact star-forming galaxies that might be progenitors of compact passive galaxies at these redshifts \citep[e.g.]{Stefanon13}.  Galaxy 30169 is a larger object with $r_{1/2}=4.1$kpc.  It looks like an edge on disk, although we lack observations of sufficient resolution and sensitivity to kinematically confirm bulk rotation.  There is a slight color gradient in this object, however, such that the center is slightly redder than the outskirts (Figure~\ref{Fig:co_spec}).  Correcting the light profile for this color (and hence \mlstar) gradient will presumably make the stellar mass more concentrated than the H-band light and will increase the implied effective stellar mass concentration.  Further blind CO studies will be needed to understand if SFR-selected samples are biased to lower stellar mass surface density compared to CO-selected samples.  This might be the case as there is a trend at these redshifts between SFR and size, such that SF galaxies tend to be more extended \citep{Toft07}.

Finally, we must consider that 30169 would require 13~Gyr to form its stellar mass at a constant SFR, which is clearly longer than the age of the Universe at this epoch.  Therefore the SFR must have been much higher prior to the epoch of observation and since declined.  If we are catching this object in the process of quenching, during which it is depleting its molecular gas reservoir, then this process may occur in a way that keeps galaxies on the \citet{Genzel15} scaling relations.  


To make a more accurate analysis of the stability of the gas and its physical characteristics we will need spatially resolved CO with ALMA or PdBI/NOEMA and potentially higher spatial resolution stellar mass and SFR maps with HST and eventually, JWST.  Ultimately, we will require spatially resolved excitation maps of our galaxies to understand how the physical conditions of the gas vary across their surface.

\subsubsection{Continuum-based \mgas}
\label{Sec:cont_mgas}

We use our continuum detection at a rest-frame frequency of 44.25~GHz to obtain an alternate measurement of the gas mass using the scaling between thermal dust emission and the gas mass described in \citet{Scoville16}.  We use their equation 16 (corrected using published erratum) 
\begin{multline}
M_{mol}=1.78~S_{\nu_{obs}}[{\rm mJy}]~(1+z)^{-4.8}~\left(\frac{\nu_{850\mu m}}{\nu_{obs}}\right)^{3.8}\\
\times(d_{L}[Gpc])^2~\left(\frac{6.7\times 10^{19}}{\alpha_{850}}\right) \frac{\Gamma_0}{\Gamma_{RJ}}~10^{10}\mathcal{M_\odot}
\end{multline}
Where $S_{\nu_{obs}}$ is the continuum flux, $d_L$ is the luminosity distance, $\alpha_{850}$ is a conversion from the 850\micron\ luminosity to a molecular gas mass, and $\Gamma_0$ and $\Gamma_{RJ}$ are the corrections for departure in the rest frame of the Planck function from Rayleigh-Jeans at a redshift of 0 and at the redshift of the source respectively.  We adopt the same value of $\alpha_{850}$ as \citet{Scoville16} of $6.7\times10^{19}$erg~s$^{-1}$Hz$^{-1}$\msol$^{-1}$.  $\Gamma_0=0.7$ and $\Gamma_{RJ}$ is given by the equation
\begin{equation}
\Gamma_{RJ}(T_d,\nu_{obs},z)=\frac{h\nu_{obs}(1+z)/kT_d}{e^{h\nu_{obs}(1+z)/kT_d}-1}
\end{equation}
where $T_d$ is the mass-weighted dust temperature (see \citet{Scoville16} for a discussion of the differences between mass-weighted and luminosity weighted dust temperatures.)  We adopt $T_d=25$K as in \citet{Scoville16}.   

Using the above formalism we derive a dust-based gas mass of log(\mgas/\msol)$=11.90^{+0.11}_{-0.17}$.  This is 2.7$\sigma$ above the gas mass derived from the CO emission.  Such a difference is at the limit of what is expected by comparisons between these two methods for larger samples
of galaxies \citep{Genzel15} but may be compatible within the significant uncertainties in our dust-based gas mass. Reconciling the difference between these two estimates is at face
value not trivial as it would require increasing \aco\
significantly above our adopted value.  In addition, the conversion
from dust emission to a CO gas mass is relatively insensitive to the
dust temperature.  On the other hand, our dust-based gas mass measurement relies on a factor of eight extrapolation in frequency from that used in \citet{Scoville16}, which is a source of significant uncertainty.  Galaxy 30545 also hosts an x-ray AGN that contributes a small amount to the IR SED and may also cause the dust-based gas mass estimate to be uncertain.

Given these uncertainties we do not know the origin of the gas mass discrepancy but
note that if the true gas mass were more consistent with the
continuum-based value that this galaxy would have a gas fraction and
depletion time significantly higher than galaxies of similar stellar
mass, SFR, and redshift.

\subsection{The relative role of environmental affects and CO selection on the gas contents of cluster galaxies}

As we have shown in the previous sections, our cluster galaxies have molecular gas contents very similar to those of field galaxies.  We now explore how conditions in the cluster environment and selection effects related to our CO selection may be playing a role in determining our observed gas fractions.  

First, there have been multiple studies that indicate that this proto-cluster is a merger rich environment \citep{Papovich12, Rudnick12, Lotz13}.  Due to its low velocity dispersion \citep{Tran15}, significant substructure, and high density of galaxies, \clg\ is an  environment conducive to mergers.  \citet{Lotz13} directly measured a merger rate 3--10 times higher than for massive galaxies in the field at $z\sim1.6$ and noted that most close pairs in the cluster implied minor mergers (${\cal M}_{\rm primary}/{\cal M}_{\rm satellite} \geq 4$).  Likewise \citet{Rudnick12} concluded that the average passive galaxy in the cluster must undergo 3--4 (mostly minor) mergers by $z\sim 0.6$ to explain the evolution in the red sequence luminosity function.  Finally, \citet{Papovich12} found that minor mergers were a potential explanation for the small size differences in passive galaxies between the cluster and the field.  Neither of our sources appears to be undergoing a merger in the deep CANDELS imaging but if mergers affected the molecular gas contents and SFRs in the past then they must have done so in a way that moves galaxies along the scaling relations.

An additional potential effect of the environment could be stripping effects, both of the cold ISM and of the accretion flows that feed galaxies.  The stripping of the cold gas is unlikely for these galaxies given the low cluster velocity dispersion.  However, the galaxies may be decoupled from the accretion flows by weak hydrodynamic effects and by tidal forces.  Indeed, \citet{vandeVoort17} showed using the EAGLE simulation that massive satellite galaxies at $z\sim 2$ in halos with a similar mass to ours do undergo a modest reduction in the amount of accreted gas, although their statistics are very poor for galaxies with log~(\mstar$/{\cal M}_\odot)\approx 11$.  In \S\ref{sec:accretion} we discuss how a cutoff in accretion may be used to understand the future evolution of our sources.  If accretion has been shut off by being in the cluster environment then the consumption of the existing gas must occur in such a way as to keep the galaxies on the field scaling relations.

Although we targeted a cluster with a large number of star-forming members, on a galaxy-by-galaxy basis, our survey did not target individual galaxies based on their position relative to the SFR-\mstar\ sequence.  Our survey may also be considered a pseudo-blind survey.   The survey of D14 is explicitly blind.  Naively, one would expect that a blind CO detection would yield sources that are over-luminous in CO compared to those selected by some other property, e.g. SFR.  However, our galaxies and those in the HDF-N are completely consistent with the scaling relation predictions.  Nonetheless, the galaxy sample is small and further blind studies will be needed to determine how blind and pointed surveys compare.  

%

\subsection{The lack of future gas accretion in massive cluster galaxies}
\label{sec:accretion}

Regardless of the cause of the low SFE, it remains true that our two cluster galaxies have high gas fractions and low SFEs.  We can phrase this in terms of the gas consumption timescale (\tcon$\equiv{\rm SFE}^{-1}=$\mhtwo/SFR), which we show on the right-hand axis of Figure~\ref{Fig:lir_lco}b.  \tcon\ for our sources is $2.8\pm1.4$~Gyr and $0.7\pm0.3$~Gyr for 30169 and 30545 respectively.\footnote{The right axis of Figure~\ref{Fig:lir_lco}b is calculated assuming \lir$\propto$SFR, which is not strictly true and is the source of the slight differences compared to the numbers in the text.}  This \tcon\ estimate is in some senses a lower limit as it assumes a constant SFR, whereas our massive galaxies likely have declining SFRs and a correspondingly longer \tcon.  On the other hand, our estimate ignores outflows, which have been found to be ubiquitous around star-forming galaxies at $z\sim1$ \citep{Weiner09,Erb12} and would drive \tcon\ down.  Acknowledging these uncertainties, we show in \S\ref{Sec:lowsfe} and Figure~\ref{Fig:scale_rel} that our two sources have systematically long \tcon\ compared to that expected from gas scaling relations \citep{Genzel15}.


The \tcon\ for our galaxies may have implications for their future gas accretion histories.  The short \tcon\ of star-forming galaxies at $z\sim1.5$ has been used to argue for the importance of accretion in powering the continued high SFRs of galaxies at these epochs \citep{Daddi10b,Genzel10,Tacconi10}.  To place constraints on the future gas accretion history of our two cluster galaxies, we attempt to identify their likely descendants. We know that the descendants of our two galaxies must be at least as massive and reside in the likely descendant halo of \clg.   \citet{Rudnick12} showed that the accretion history of \clg\ results in it likely being a log$({\cal M}_{clust}/{\cal M}_\odot)=14.1-14.35$ cluster at $z=1$.  Recently, \citet{Muldrew15} have shown that the mass of $z=0$ descendants of $z\sim2$ simulated protoclusters can be predicted with an 0.5~dex scatter, and that this scatter is reduced to 0.3~dex if one also knows the mass ratio of the primary and secondary clump in the protoclusters.    It is therefore likely that our cluster at high-z can be associated with a cluster  of comparable or higher mass, although it is hard to accurately determine the descendant mass without identifying other substructures and their masses. 

Galaxies at $z=1$ with log~(\mstar$/{\cal M}_\odot)\approx 11$ and in clusters of comparable mass to \clg\ from GCLASS \citep{Muzzin12} have a passive fraction of $\approx 0.75$.\footnote{Galaxies in GCLASS were determined to be passive via a lack of [\ion{O}{2}] at the EW$\sim 2$\AA\ level.  This corresponds to star formation rates of $\sim 5$\msol/yr for galaxies with log~(\mstar$/{\cal M}_\odot)\approx 11$.  \citet{Muzzin12} use deep stacked spectra of passive galaxies identified this way to estimate that $>90\%$ of them have sSFR$\lesssim 5\times10^{-11}$yr$^{-1}$.} \clg\ has a similarly high passive fraction of $1.0^{+0.0}_{-0.37}$ \citep{Lee-Brown17}.  Given that this cluster will accrete galaxies from the field, where the SF-fraction is higher \citep{Hatch16,Lee-Brown17}, the galaxies in the proto-cluster core at $z=1.62$ likely must consume most of their gas in the intervening 1.8 Gyr between $z=1.6$ and 1 in order to become part of this passive population.  This timespan is consistent with the \tcon\ for both 30545 and 30169.  While \tcon\ estimated from \mgas\ and SFR is uncertain due to the unknown SFH and the impact of outflows, if our \tcon\ estimate is correction it would imply that neither of our galaxies can tolerate any further gas accretion from the cosmic web at redshifts lower than $z=1.6$.  

This is not unexpected as galaxies in simulations lose their connection to their gas umbilical cords as they become satellite members \citep{Keres09,Dekel09}.  At lower redshift, this process may be analogous to the process of strangulation, in which a galaxy's gas supply is truncated and the galaxy uses up its remaining fuel \citep{Larson80,Balogh00,Bekki02}, usually on a few Gyr timescale.  We now know that star-forming galaxies can drive outflows with significant mass loading \citep[e.g.][]{Tremonti07,Weiner09,Tripp11,Martin12}.  Using the ubiquitous nature of outflows in star-forming galaxies, \citet{McGee14} made an interesting adjustment to the timescale for environmental quenching because the mass-loaded winds can cause the SFR rate to drop much quicker than the classic consumption timescale once gas accretion has shut off.  It is worth noting that the fast suppression of SFR in this model results partly from those authors assuming that the SFR remains unchanged until the gas is depleted.  It also assumes that the mass from winds is completely ejected from the halo whereas observations show that the wind in intermediate redshift star-forming galaxies often cannot completely escape from the galaxy and can populate the lower-redshift circumgalactic medium \citep{Rubin14}.  If winds are an important factor in quenching, however, such a high mass loading and short truncation timescale matches the evolution in the group and cluster galaxy passive fractions at $z<1$ \citep{McGee14,Balogh16} and may provide a way to significantly reduce gas consumption timescales. This process is not expected to be limited to dense clusters but may also be active at the group mass scale \citep{Kawata08}, similar to what is found in high-z forming clusters.  We think we may be seeing evidence of this cutoff of gas accretion playing a role in high redshift cluster galaxies.    

\section{Caveats}
\label{Sec:caveats}

Our conclusions suffer from a few potential uncertainties.  We outline these below and discuss their effect.

\subsection{\aco\ and the nature of high redshift star formation}
\label{Sec:aco}

The foremost uncertainty is the value for \aco, which determines the conversion of \lco\ to \mhtwo.  We adopt a Galactic value of \aco=4.36 and show that it is broadly consistent with our limited dynamical constraints, although an \aco\ that is lower by 50\% may be appropriate for 30545 to avoid having the baryonic mass for 30545 in excess of the dynamical mass (see \S\ref{subsec:mgas}).  If a ULIRG-like \aco=0.8 is more appropriate, it would reduce our gas masses by a factor of $\sim 5$ and make our SFEs more consistent with other star-forming galaxies at these redshifts. 

\aco\ has a strong dependence on the density and temperature of the molecular gas and the fraction of CO in diffuse as opposed to concentrated components.  We must therefore consider this dependence in the context of our choice of \aco.  For example, \aco\ in extreme local starburst galaxies with dense gas configurations is thought to be significantly lower than that for the Milky Way, with $\alpha_{\rm CO,ULIRG}\sim1$ \citep[e.g.][]{Scoville97, Downes98, Genzel15}. 

Our two galaxies have significantly higher SFRs than comparably massive galaxies locally.  In interpreting these SFRs in the context of \aco, we must note that the elevated SFRs are likely due to the increasing SFRs of all galaxies going back in time, as the SFR-\mstar\ sequence evolves to higher SFR \citep[e.g.]{Noeske07, Whitaker12}.  Additionally, there is mounting evidence that the characteristics of star-formation in galaxies on the SFR-\mstar\  sequence out to $z=2$ is more similar to that in local main-sequence galaxies than to that in local ULIRGs, despite the distant galaxies having absolute SFRs more similar to the latter.  For example, multiple authors have found that distant MS galaxies have IR SEDs comparable to local MS galaxies, and yet different from galaxies above the main sequence at low redshift that have similar absolute SFRs \citep[e.g.][]{Papovich07,Elbaz11}.  The interpretation of the SEDs is that the distant MS galaxies have a higher contribution from diffuse IR emission than their more luminous counterparts. 

MS galaxies at $z<3$ follow the same scaling relations of SFR and molecular gas content \citep{Genzel15} over a large redshift range.  Likewise, the CO emission \citep{Tacconi13} and \ha\ emission \citep{Shapiro08,ForsterSchreiber09,ForsterSchreiber11,Wisnioski15} in high-z MS galaxies is more extended than in local ULIRGS.   If MS galaxies at high redshift have gas with a spatial extent similar to the Milky Way, then it might also be that the CO emission is likely generated in molecular clouds with mean densities similar to those in the Milky Way \citep[$\langle n_H\rangle \sim 10^2 -10^3~{\rm cm}^{-3}$;][]{Dannerbauer09,Daddi10b}.   The implication is that a MW-like \aco, which relates to the physical state of the gas, might be appropriate for intensely star-forming (in an absolute sense) MS galaxies at high redshift, as their gas characteristics might be more similar to local MS galaxies than to local galaxies of comparable absolute SFRs.   Therefore it is reasonable to assume that high redshift galaxies on the MS with solar metallicity have \aco\ similar to the Milky Way, as
also assumed in the PHIBSS and PHIBSS 2 samples which form the backbone of our comparison samples.  We therefore feel justified in our choice of a Galactic \aco\ and add that it allows us to compare our gas masses to the PHIBSS galaxies that also assume a galactic \aco.  

%

As an independent check on our gas masses we compared our \mstar\ and \mhtwo\ estimates with the dynamical constraints from the CO line widths.  For our \aco\ to be valid, the total baryonic mass (\mhtwo$+$\mstar) cannot exceed the dynamical mass.  Object 30545 is compact and round and we estimate its dynamical mass from its rest-frame optical size and velocity dispersion $\sigma_v$ using the isotropic virial estimator from \citet{ForsterSchreiber09} and \citet{ForsterSchreiber06}, who derived it from \citet{Binney08}.  We believe that this estimator is valid for 30545 as there is no indication of a disk morphology.
\begin{equation}
{\cal M}_{dyn}=\frac{6.7 r_{1/2} \sigma_v^2}{ G}.
\end{equation}
The resultant ${\cal M}_{dyn}=6.7\pm0.5\times10^{10}$\msol.  This is less than the stellar mass, which may reflect systematic uncertainties in our stellar mass estimates (typically 0.3~dex) or an inappropriate dynamical mass estimator.  This taken together with the significant spatial displacement of the gas from the stars Figure~\ref{Fig:co_im} implies that our dynamical mass is likely very inaccurate and thus cannot be used to constrain \aco.  On the other hand, 30169 has an irregular velocity profile and it is impossible with the current data to constrain the dynamics.  We therefore decide to perform our remaining analysis using a Galactic \aco\ value to be consistent with other work being done on "typical" star-forming galaxies at this epoch, e.g. from PHIBSS \citep{Tacconi13}.  We hope that future improved constraints on the dynamical mass with spatially resolved measurements using ALMA will allow us to more strongly constrain \aco\ in the future. 

If a much lower \aco\ were appropriate, it would be difficult to justify in the context of previous work.  Most galaxies with an inferred low \aco\ sit significantly above the SFR-\mstar\ sequence and may be forming stars in a different, merger-dominated, mode \citep{Elbaz11,Kartaltepe12} where \aco\ may be lower because of the different physical conditions at high gas surface densities \citep{Daddi10a,Genzel10,Narayanan11b,Narayanan12,Bolatto13}.  \citet{Narayanan12} give a fitting formula to find \aco\ in terms of the CO surface brightness, but since we cannot measure this directly we are wary of applying this formula and instead prefer to adopt a constant Galactic \aco\ for all of our galaxies, which are forming stars at or below the value of the SFR-\mstar\ sequence.  If we were to apply a different \aco\ to our galaxies and to those of \citet{Decarli14} to reconcile their SFEs with the PHIBSS galaxies, its significantly lower value compared to the Galactic one would imply that compact galaxies and those below the SFR-\mstar\ sequence might have  different physical conditions of the molecular gas than galaxies on the \mstar-SFR sequence.  This could post a complication in the use of SFR-selected galaxies to construct scaling laws of gas content vs. gas consumption timescales.  This would be an interesting result in itself and would in turn emphasize the need for more blind CO surveys to probe the full range of star formation modes in the distant universe.

Another possible reason for low \aco\ is if the metallicity is significantly sub-solar \citep[e.g.][]{Bolatto13,Sandstrom13}.  However, \citet{Tran15} used MOSFIRE observations of rest-frame optical emission lines to find that the gas-phase metallicity of log~(\mstar$/{\cal M}_\odot)\approx 11$ star-forming members in \clg\ is close to solar.  This provides a further argument for Galactic \aco, which is appropriate for regions of solar metallicity.

\citet{Daddi15} found that typical star forming galaxies have complex SLEDs with a low and high excitation component.  It is not yet clear what impact this will have on the inferred \aco\ but clearly tells us something about the physical state of the gas.  Improving on the above uncertainties would require directly measuring the spatial extent of the CO and its excitation state, through multi-transition SLEDs, to see if the excitation of the gas in these compact galaxies is different from that in galaxies on the SFR-\mstar\ sequence.  Spatially unresolved studies can be conducted with the NOrthern Extended Millimeter Array (NOEMA) and spatially resolved excitation studies will be possible with deep ALMA observations in extended array configurations.

\subsection{Uncertainty in \lir\ and the SFR}

Increasing \lir\ by $\approx0.5~$dex would help to reconcile our galaxies with those from PHIBSS.  Object 30545 is detected in \textit{Herschel} bands and its SFR and \lir\ should be accurate.  Object 30169 lacks a \textit{Herschel} detection but the upper limit on the \textit{Herschel} fluxes provides a strong constraint on the SFR and LIR (Figure~\ref{Fig:sed}).  We therefore conclude that uncertainty in \lir\ is not an issue.

To assess our uncertainties in SFR we compare our MAGPHYS measure to that from LIR and the 2800\ang\ luminosity using the \citet{Wuyts11} scalings that were derived from \citet{Kennicutt98b}.  We find a SFR for 30169 and 30545 of 47 and 185 \msol$/$year, within the uncertainties on our SFR measure.  

\subsection{Sample size}

Our results are based primarily on two galaxies from our forming cluster, along with two galaxies from \citet{Decarli14}.  We find that objects detected in blind CO observations are gas rich, have low SFEs, and are spatially compact in the rest-frame optical compared to SFR selected objects.  Clearly the small sample size means that our results are preliminary.  We need to verify them by assembling a larger sample of CO-detected objects found in blind surveys to comparable or fainter line luminosities.  Once the full depth data for the present program is processed, we will see if any more CO detections become available.

We have also observed galaxies in a single high redshift cluster.  At lower redshift there is a well known cluster-to-cluster variation in galaxy properties, \citep[e.g.][]{Poggianti06,Rudnick09} and it may be that our finding regarding the cutting off of the gas supply is not indicative of processes affecting typical star-forming galaxies in forming clusters.  Indeed, recent observations of CO in 11 galaxies in 3 clusters at $z\sim1.6$ from \citet{Noble17} show evidence for elevated gas fractions with respect to the field. Making progress in this arena will require deep CO observations of multiple high redshift clusters and ground-based spectroscopy and grism spectroscopy to help improve membership and look for faint lines.  

\section{Summary}
\label{Sec:summ}

In this paper we present very deep \co{1}{0} observations with the VLA of a cluster of galaxies at $z=1.625$ and report the detection of two star-forming spectroscopically confirmed cluster members.  The redshift for one of the sources was found first from ground-based spectroscopy and the other was found first with our \co{1}{0} observations and later confirmed with ground-based and HST grism spectroscopy.  We use the \co{1}{0} line to trace the molecular gas in these galaxies and compare their \mhtwo\ to other physical properties such as their \mstar, SFR, rest-frame optical size, and \lir.  We summarize our findings as follows:

\begin{itemize}

\item Both CO-detected galaxies are massive (log~(\mstar$/{\cal M}_\odot)\approx 11$) and are forming stars, with one of them doing so at levels significantly below the SFR-\mstar\ sequence.  One of the objects is compact and hosts an x-ray AGN that is not energetically dominant in the IR.  The other appears to be a nearly edge-on disk with a slight color gradient such that it gets redder towards the center.

\item The galaxies are detected in \co{1}{0} with $S/N=4.9$ and 7.1  and have large line widths, indicating large dynamical masses.  Assuming a Galactic \aco\ we find that these galaxies have $f_{gas}\equiv$\mhtwo/(\mstar+\mhtwo$)=0.17-0.45$, within the gas fraction distribution for typical star-forming galaxies at $1<z<2.5$.

\item The CO luminosities for these two galaxies are within the range defined by galaxies with similar \lir.  Likewise, the $f_{gas}$, gas consumption timescales (\tcon), and star formation efficiencies (SFE)  are consistent with those of galaxies having similar redshift, \mstar, and SFR. The compatibility with field galaxies is also found when comparing the surface density of star formation to the surface density of the molecular gas.

\item The lower $f_{gas}$ and SFE for our galaxy with the lowest SFR is consistent with the predictions based on studies of local and high redshift star-forming galaxies, which find that the SFEs drop as galaxies move below the \mstar-SFR sequence and that this drop coincides with a drop in $f_{gas}$.  

\item Our galaxies have among the highest stellar surface mass densities of any CO-detected star-forming galaxies at the same epoch, comparable to that of compact quiescent galaxies found at similar redshifts. 

\item The gas consumption timescales for our galaxies are between 0.7 and 2.8~Gyr.  This is consistent with the depletion times of galaxies at these sSFRs taken from CO surveys that select galaxies by their SFR and rest-frame optical color.  However, our galaxies lie in a forming cluster and studies of the likely descendant clusters at $z<1$ indicate that our galaxies have a high probability of becoming passive in the intervening 1.8~Gyr between $z=1.62$ and 1.  If that is their destiny, then to become passive, and presumably gas poor, by $z=1$ means that our galaxies cannot tolerate any further gas accretion following the epoch in which we observe them.  This might indicate that galaxies in the forming cluster environment have been decoupled from their gas umbilical cords that connect them to the cosmic web and may be the early manifestation of the process known variously as starvation or strangulation.

\item We compare our two blindly detected cluster galaxies with three other blindly detected CO emitters and find that the SFEs, gas consumption timescales, and most surprisingly, the high surface stellar mass densities are similar to the galaxies found in our cluster.  This may indicate that deep blind CO surveys are sensitive to star formation in compact galaxies that is not common in SFR selected samples. 

\end{itemize}

While enticing, the conclusions presented here are based on only two galaxies in one forming cluster and a handful of other sources.  To understand if these findings are indicative of blind-CO detected sources in general at these redshifts will require a significant blind CO survey.  Likewise, to understand the role that environment may play in modulating the gas contents and SFEs of galaxies will require blind CO observations to be conducted in high redshift cluster environments.  To uncover the physical conditions of the gas will require spatially resolving it, understanding its excitation, and ultimately, spatially resolving the excitation.  These are ideal programs for future observations with ALMA and NOEMA (formerly PdBI).

\acknowledgements

This paper is based on data collected at VLA, which is operated by the National Radio Astronomy Observatory.  The National Radio Astronomy Observatory is a facility of the National Science Foundation operated under cooperative agreement by Associated Universities, Inc.  GHR thanks Alberto Bolatto, Andreas Burkert, Adam Leroy, Karin Sandstrom, Sharon Meidt, Arjen van der Wel, Reinhardt Genzel, and Linda Tacconi for useful discussions that improved this paper.  He is especially thankful to Allison Noble for finding a bug in one of the programs used in this paper.  GHR thanks the 3D-HST and CANDELS teams for their released catalogs and the 3D-HST team for their grism redshifts.  GHR acknowledges funding support from HST program HST-GO-12590.011-A, AR-14310.001and NSF AST grants 1211358 and 1517815, and an Alexander von Humboldt foundation fellowship for experienced researchers.  GHR acknowledges the excellent hospitality of the Max-Planck-Institute for Astronomy, the University of Hamburg Observatory, the Max-Planck-Institute for Extraterrestrial Physics, the International Space Sciences Institute, and the European Southern Observatory, where some of this research was conducted.  JH acknowledges support of the VIDI research programme with project number 639.042.611, which is (partly) financed by the Netherlands Organisation for Scientific Research (NWO).

\clearpage


\begin{thebibliography}{132}
\expandafter\ifx\csname natexlab\endcsname\relax\def\natexlab#1{#1}\fi

\bibitem[{{Alberts} {et~al.}(2014){Alberts}, {Pope}, {Brodwin}, {Atlee}, {Lin},
  {Dey}, {Eisenhardt}, {Gettings}, {Gonzalez}, {Jannuzi}, {Mancone},
  {Moustakas}, {Snyder}, {Stanford}, {Stern}, {Weiner}, \&
  {Zeimann}}]{Alberts14}
{Alberts}, S., {et~al.} 2014, \mnras, 437, 437

\bibitem[{{Aravena} {et~al.}(2010){Aravena}, {Carilli}, {Daddi}, {Wagg},
  {Walter}, {Riechers}, {Dannerbauer}, {Morrison}, {Stern}, \&
  {Krips}}]{Aravena10}
{Aravena}, M., {et~al.} 2010, \apj, 718, 177

\bibitem[{{Aravena} {et~al.}(2012){Aravena}, {Carilli}, {Salvato}, {Tanaka},
  {Lentati}, {Schinnerer}, {Walter}, {Riechers}, {Smolci{\'c}}, {Capak},
  {Aussel}, {Bertoldi}, {Chapman}, {Farrah}, {Finoguenov}, {Le Floc'h}, {Lutz},
  {Magdis}, {Oliver}, {Riguccini}, {Berta}, {Magnelli}, \& {Pozzi}}]{Aravena12}
---. 2012, \mnras, 426, 258

\bibitem[{{Arnouts} {et~al.}(2007){Arnouts}, {Walcher}, {Le F{\`e}vre},
  {Zamorani}, {Ilbert}, {Le Brun}, {Pozzetti}, {Bardelli}, {Tresse}, {Zucca},
  {Charlot}, {Lamareille}, {McCracken}, {Bolzonella}, {Iovino}, {Lonsdale},
  {Polletta}, {Surace}, {Bottini}, {Garilli}, {Maccagni}, {Picat},
  {Scaramella}, {Scodeggio}, {Vettolani}, {Zanichelli}, {Adami}, {Cappi},
  {Ciliegi}, {Contini}, {de la Torre}, {Foucaud}, {Franzetti}, {Gavignaud},
  {Guzzo}, {Marano}, {Marinoni}, {Mazure}, {Meneux}, {Merighi}, {Paltani},
  {Pell{\`o}}, {Pollo}, {Radovich}, {Temporin}, \& {Vergani}}]{Arnouts07}
{Arnouts}, S., {et~al.} 2007, \aap, 476, 137

\bibitem[{{Balogh} {et~al.}(2000){Balogh}, {Navarro}, \& {Morris}}]{Balogh00}
{Balogh}, M.~L., {Navarro}, J.~F., \& {Morris}, S.~L. 2000, ApJ, 540, 113

\bibitem[{{Balogh} {et~al.}(2016){Balogh}, {McGee}, {Mok}, {Muzzin}, {van der
  Burg}, {Bower}, {Finoguenov}, {Hoekstra}, {Lidman}, {Mulchaey}, {Noble},
  {Parker}, {Tanaka}, {Wilman}, {Webb}, {Wilson}, \& {Yee}}]{Balogh16}
{Balogh}, M.~L., {et~al.} 2016, \mnras, 456, 4364

\bibitem[{{Bekki} {et~al.}(2002){Bekki}, {Couch}, \& {Shioya}}]{Bekki02}
{Bekki}, K., {Couch}, W.~J., \& {Shioya}, Y. 2002, \apj, 577, 651

\bibitem[{{Bell} {et~al.}(2004){Bell}, {Wolf}, {Meisenheimer}, {Rix}, {Borch},
  {Dye}, {Kleinheinrich}, {Wisotzki}, \& {McIntosh}}]{Bell04}
{Bell}, E.~F., {et~al.} 2004, \apj, 608, 752

\bibitem[{{Bigiel} {et~al.}(2008){Bigiel}, {Leroy}, {Walter}, {Brinks}, {de
  Blok}, {Madore}, \& {Thornley}}]{Bigiel08}
{Bigiel}, F., {Leroy}, A., {Walter}, F., {Brinks}, E., {de Blok}, W.~J.~G.,
  {Madore}, B., \& {Thornley}, M.~D. 2008, \aj, 136, 2846

\bibitem[{{Binney} \& {Tremaine}(2008)}]{Binney08}
{Binney}, J., \& {Tremaine}, S. 2008, {Galactic Dynamics: Second Edition}
  (Princeton University Press)

\bibitem[{{Blanton} {et~al.}(2003){Blanton}, {Brinkmann}, {Csabai}, {Doi},
  {Eisenstein}, {Fukugita}, {Gunn}, {Hogg}, \& {Schlegel}}]{Blanton03}
{Blanton}, M.~R., {et~al.} 2003, \aj, 125, 2348

\bibitem[{Bolatto {et~al.}(2013)Bolatto, Wolfire, \& Leroy}]{Bolatto13}
Bolatto, A.~D., Wolfire, M., \& Leroy, A.~K. 2013, Annual Review of Astronomy
  {\&} Astrophysics, 51, 207

\bibitem[{{Brammer} {et~al.}(2008){Brammer}, {van Dokkum}, \&
  {Coppi}}]{Brammer08}
{Brammer}, G.~B., {van Dokkum}, P.~G., \& {Coppi}, P. 2008, \apj, 686, 1503

\bibitem[{{Brammer} {et~al.}(2011){Brammer}, {Whitaker}, {van Dokkum},
  {Marchesini}, {Franx}, {Kriek}, {Labb{\'e}}, {Lee}, {Muzzin}, {Quadri},
  {Rudnick}, \& {Williams}}]{Brammer11}
{Brammer}, G.~B., {et~al.} 2011, \apj, 739, 24

\bibitem[{{Brammer} {et~al.}(2012){Brammer}, {van Dokkum}, {Franx},
  {Fumagalli}, {Patel}, {Rix}, {Skelton}, {Kriek}, {Nelson}, {Schmidt},
  {Bezanson}, {da Cunha}, {Erb}, {Fan}, {F{\"o}rster Schreiber}, {Illingworth},
  {Labb{\'e}}, {Leja}, {Lundgren}, {Magee}, {Marchesini}, {McCarthy},
  {Momcheva}, {Muzzin}, {Quadri}, {Steidel}, {Tal}, {Wake}, {Whitaker}, \&
  {Williams}}]{Brammer12}
---. 2012, \apjs, 200, 13

\bibitem[{{Brinchmann} {et~al.}(2004){Brinchmann}, {Charlot}, {White},
  {Tremonti}, {Kauffmann}, {Heckman}, \& {Brinkmann}}]{Brinchmann04}
{Brinchmann}, J., {Charlot}, S., {White}, S.~D.~M., {Tremonti}, C.,
  {Kauffmann}, G., {Heckman}, T., \& {Brinkmann}, J. 2004, \mnras, 351, 1151

\bibitem[{{Brodwin} {et~al.}(2013){Brodwin}, {Stanford}, {Gonzalez}, {Zeimann},
  {Snyder}, {Mancone}, {Pope}, {Eisenhardt}, {Stern}, {Alberts}, {Ashby},
  {Brown}, {Chary}, {Dey}, {Galametz}, {Gettings}, {Jannuzi}, {Miller},
  {Moustakas}, \& {Moustakas}}]{Brodwin13}
{Brodwin}, M., {et~al.} 2013, \apj, 779, 138

\bibitem[{{Brown} {et~al.}(2007){Brown}, {Dey}, {Jannuzi}, {Brand}, {Benson},
  {Brodwin}, {Croton}, \& {Eisenhardt}}]{Brown07}
{Brown}, M.~J.~I., {Dey}, A., {Jannuzi}, B.~T., {Brand}, K., {Benson}, A.~J.,
  {Brodwin}, M., {Croton}, D.~J., \& {Eisenhardt}, P.~R. 2007, \apj, 654, 858

\bibitem[{{Capak} {et~al.}(2008){Capak}, {Carilli}, {Lee}, {Aldcroft},
  {Aussel}, {Schinnerer}, {Wilson}, {Yun}, {Blain}, {Giavalisco}, {Ilbert},
  {Kartaltepe}, {Lee}, {McCracken}, {Mobasher}, {Salvato}, {Sasaki}, {Scott},
  {Sheth}, {Shioya}, {Thompson}, {Elvis}, {Sanders}, {Scoville}, \&
  {Tanaguchi}}]{Capak08}
{Capak}, P., {et~al.} 2008, \apjl, 681, L53

\bibitem[{{Carilli} {et~al.}(2011){Carilli}, {Hodge}, {Walter}, {Riechers},
  {Daddi}, {Dannerbauer}, \& {Morrison}}]{Carilli11}
{Carilli}, C.~L., {Hodge}, J., {Walter}, F., {Riechers}, D., {Daddi}, E.,
  {Dannerbauer}, H., \& {Morrison}, G.~E. 2011, \apjl, 739, L33

\bibitem[{{Carilli} \& {Walter}(2013)}]{Carilli13}
{Carilli}, C.~L., \& {Walter}, F. 2013, \araa, 51, 105

\bibitem[{{Chabrier}(2003)}]{Chabrier03}
{Chabrier}, G. 2003, \apjl, 586, L133

\bibitem[{{Chapman} {et~al.}(2005){Chapman}, {Blain}, {Smail}, \&
  {Ivison}}]{Chapman05}
{Chapman}, S.~C., {Blain}, A.~W., {Smail}, I., \& {Ivison}, R.~J. 2005, \apj,
  622, 772

\bibitem[{{Chapman} {et~al.}(2015){Chapman}, {Bertoldi}, {Smail}, {Blain},
  {Geach}, {Gurwell}, {Ivison}, {Petitpas}, {Reddy}, \& {Steidel}}]{Chapman15}
{Chapman}, S.~C., {et~al.} 2015, \mnras, 449, L68

\bibitem[{{Charlot} \& {Fall}(2000)}]{Charlot00}
{Charlot}, S., \& {Fall}, S.~M. 2000, \apj, 539, 718

\bibitem[{{da Cunha} {et~al.}(2008){da Cunha}, {Charlot}, \&
  {Elbaz}}]{dacunha08}
{da Cunha}, E., {Charlot}, S., \& {Elbaz}, D. 2008, \mnras, 388, 1595

\bibitem[{{da Cunha} {et~al.}(2013){da Cunha}, {Walter}, {Decarli}, {Bertoldi},
  {Carilli}, {Daddi}, {Elbaz}, {Ivison}, {Maiolino}, {Riechers}, {Rix},
  {Sargent}, {Smail}, \& {Weiss}}]{dacunha13}
{da Cunha}, E., {et~al.} 2013, \apj, 765, 9

\bibitem[{{da Cunha} {et~al.}(2015){da Cunha}, {Walter}, {Smail}, {Swinbank},
  {Simpson}, {Decarli}, {Hodge}, {Weiss}, {van der Werf}, {Bertoldi},
  {Chapman}, {Cox}, {Danielson}, {Dannerbauer}, {Greve}, {Ivison}, {Karim}, \&
  {Thomson}}]{dacunha15}
---. 2015, \apj, 806, 110

\bibitem[{{Daddi} {et~al.}(2007){Daddi}, {Dickinson}, {Morrison}, {Chary},
  {Cimatti}, {Elbaz}, {Frayer}, {Renzini}, {Pope}, {Alexander}, {Bauer},
  {Giavalisco}, {Huynh}, {Kurk}, \& {Mignoli}}]{Daddi07}
{Daddi}, E., {et~al.} 2007, \apj, 670, 156

\bibitem[{{Daddi} {et~al.}(2010{\natexlab{a}}){Daddi}, {Elbaz}, {Walter},
  {Bournaud}, {Salmi}, {Carilli}, {Dannerbauer}, {Dickinson}, {Monaco}, \&
  {Riechers}}]{Daddi10a}
---. 2010{\natexlab{a}}, \apjl, 714, L118

\bibitem[{{Daddi} {et~al.}(2010{\natexlab{b}}){Daddi}, {Bournaud}, {Walter},
  {Dannerbauer}, {Carilli}, {Dickinson}, {Elbaz}, {Morrison}, {Riechers},
  {Onodera}, {Salmi}, {Krips}, \& {Stern}}]{Daddi10b}
---. 2010{\natexlab{b}}, \apj, 713, 686

\bibitem[{{Daddi} {et~al.}(2015){Daddi}, {Dannerbauer}, {Liu}, {Aravena},
  {Bournaud}, {Walter}, {Riechers}, {Magdis}, {Sargent}, {B{\'e}thermin},
  {Carilli}, {Cibinel}, {Dickinson}, {Elbaz}, {Gao}, {Gobat}, {Hodge}, \&
  {Krips}}]{Daddi15}
---. 2015, \aap, 577, A46

\bibitem[{{Dannerbauer} {et~al.}(2009){Dannerbauer}, {Daddi}, {Riechers},
  {Walter}, {Carilli}, {Dickinson}, {Elbaz}, \& {Morrison}}]{Dannerbauer09}
{Dannerbauer}, H., {Daddi}, E., {Riechers}, D.~A., {Walter}, F., {Carilli},
  C.~L., {Dickinson}, M., {Elbaz}, D., \& {Morrison}, G.~E. 2009, \apjl, 698,
  L178

\bibitem[{{Davis} {et~al.}(2011){Davis}, {Alatalo}, {Sarzi}, {Bureau}, {Young},
  {Blitz}, {Serra}, {Crocker}, {Krajnovi{\'c}}, {McDermid}, {Bois}, {Bournaud},
  {Cappellari}, {Davies}, {Duc}, {de Zeeuw}, {Emsellem}, {Khochfar},
  {Kuntschner}, {Lablanche}, {Morganti}, {Naab}, {Oosterloo}, {Scott}, \&
  {Weijmans}}]{Davis11}
{Davis}, T.~A., {et~al.} 2011, \mnras, 417, 882

\bibitem[{{Decarli} {et~al.}(2014){Decarli}, {Walter}, {Carilli}, {Riechers},
  {Cox}, {Neri}, {Aravena}, {Bell}, {Bertoldi}, {Colombo}, {Da Cunha}, {Daddi},
  {Dickinson}, {Downes}, {Ellis}, {Lentati}, {Maiolino}, {Menten}, {Rix},
  {Sargent}, {Stark}, {Weiner}, \& {Weiss}}]{Decarli14}
{Decarli}, R., {et~al.} 2014, \apj, 782, 78

\bibitem[{{Dekel} {et~al.}(2009){Dekel}, {Birnboim}, {Engel}, {Freundlich},
  {Goerdt}, {Mumcuoglu}, {Neistein}, {Pichon}, {Teyssier}, \&
  {Zinger}}]{Dekel09}
{Dekel}, A., {et~al.} 2009, \nat, 457, 451

\bibitem[{{Dickinson} {et~al.}(2003){Dickinson}, {Papovich}, {Ferguson}, \&
  {Budav{\'a}ri}}]{Dickinson03}
{Dickinson}, M., {Papovich}, C., {Ferguson}, H.~C., \& {Budav{\'a}ri}, T. 2003,
  \apj, 587, 25

\bibitem[{{Downes} \& {Solomon}(1998)}]{Downes98}
{Downes}, D., \& {Solomon}, P.~M. 1998, \apj, 507, 615

\bibitem[{{Dutton} {et~al.}(2010){Dutton}, {van den Bosch}, \&
  {Dekel}}]{Dutton10}
{Dutton}, A.~A., {van den Bosch}, F.~C., \& {Dekel}, A. 2010, \mnras, 405, 1690

\bibitem[{{Elbaz} {et~al.}(2011){Elbaz}, {Dickinson}, {Hwang},
  {D{\'{\i}}az-Santos}, {Magdis}, {Magnelli}, {Le Borgne}, {Galliano},
  {Pannella}, {Chanial}, {Armus}, {Charmandaris}, {Daddi}, {Aussel}, {Popesso},
  {Kartaltepe}, {Altieri}, {Valtchanov}, {Coia}, {Dannerbauer}, {Dasyra},
  {Leiton}, {Mazzarella}, {Alexander}, {Buat}, {Burgarella}, {Chary}, {Gilli},
  {Ivison}, {Juneau}, {Le Floc'h}, {Lutz}, {Morrison}, {Mullaney}, {Murphy},
  {Pope}, {Scott}, {Brodwin}, {Calzetti}, {Cesarsky}, {Charlot}, {Dole},
  {Eisenhardt}, {Ferguson}, {F{\"o}rster Schreiber}, {Frayer}, {Giavalisco},
  {Huynh}, {Koekemoer}, {Papovich}, {Reddy}, {Surace}, {Teplitz}, {Yun}, \&
  {Wilson}}]{Elbaz11}
{Elbaz}, D., {et~al.} 2011, \aap, 533, A119

\bibitem[{{Erb} {et~al.}(2012){Erb}, {Quider}, {Henry}, \& {Martin}}]{Erb12}
{Erb}, D.~K., {Quider}, A.~M., {Henry}, A.~L., \& {Martin}, C.~L. 2012, \apj,
  759, 26

\bibitem[{{Erb} {et~al.}(2006){Erb}, {Steidel}, {Shapley}, {Pettini}, {Reddy},
  \& {Adelberger}}]{Erb06}
{Erb}, D.~K., {Steidel}, C.~C., {Shapley}, A.~E., {Pettini}, M., {Reddy},
  N.~A., \& {Adelberger}, K.~L. 2006, \apj, 646, 107

\bibitem[{{Faber} {et~al.}(2007){Faber}, {Willmer}, {Wolf}, {Koo}, {Weiner},
  {Newman}, {Im}, {Coil}, {Conroy}, {Cooper}, {Davis}, {Finkbeiner}, {Gerke},
  {Gebhardt}, {Groth}, {Guhathakurta}, {Harker}, {Kaiser}, {Kassin},
  {Kleinheinrich}, {Konidaris}, {Kron}, {Lin}, {Luppino}, {Madgwick},
  {Meisenheimer}, {Noeske}, {Phillips}, {Sarajedini}, {Schiavon}, {Simard},
  {Szalay}, {Vogt}, \& {Yan}}]{Faber07}
{Faber}, S.~M., {et~al.} 2007, \apj, 665, 265

\bibitem[{{Fassbender} {et~al.}(2011){Fassbender}, {Nastasi}, {B{\"o}hringer},
  {{\v S}uhada}, {Santos}, {Rosati}, {Pierini}, {M{\"u}hlegger}, {Quintana},
  {Schwope}, {Lamer}, {de Hoon}, {Kohnert}, {Pratt}, \& {Mohr}}]{Fassbender11}
{Fassbender}, R., {et~al.} 2011, \aap, 527, L10

\bibitem[{{Fassbender} {et~al.}(2014){Fassbender}, {Nastasi}, {Santos},
  {Lidman}, {Verdugo}, {Koyama}, {Rosati}, {Pierini}, {Padilla}, {Romeo},
  {Menci}, {Bongiorno}, {Castellano}, {Cerulo}, {Fontana}, {Galametz},
  {Grazian}, {Lamastra}, {Pentericci}, {Sommariva}, {Strazzullo}, {{\v
  S}uhada}, \& {Tozzi}}]{Fassbender14}
---. 2014, \aap, 568, A5

\bibitem[{{Finn} {et~al.}(2010){Finn}, {Desai}, {Rudnick}, {Poggianti}, {Bell},
  {Hinz}, {Jablonka}, {Milvang-Jensen}, {Moustakas}, {Rines}, \&
  {Zaritsky}}]{Finn10}
{Finn}, R.~A., {et~al.} 2010, \apj, 720, 87

\bibitem[{{Fontana} {et~al.}(2003){Fontana}, {Donnarumma}, {Vanzella},
  {Giallongo}, {Menci}, {Nonino}, {Saracco}, {Cristiani}, {D'Odorico}, \&
  {Poli}}]{Fontana03}
{Fontana}, A., {et~al.} 2003, \apjl, 594, L9

\bibitem[{{Fontana} {et~al.}(2006){Fontana}, {Salimbeni}, {Grazian},
  {Giallongo}, {Pentericci}, {Nonino}, {Fontanot}, {Menci}, {Monaco},
  {Cristiani}, {Vanzella}, {de Santis}, \& {Gallozzi}}]{Fontana06}
---. 2006, \aap, 459, 745

\bibitem[{{F{\"o}rster Schreiber} {et~al.}(2011){F{\"o}rster Schreiber},
  {Shapley}, {Erb}, {Genzel}, {Steidel}, {Bouch{\'e}}, {Cresci}, \&
  {Davies}}]{ForsterSchreiber11}
{F{\"o}rster Schreiber}, N.~M., {Shapley}, A.~E., {Erb}, D.~K., {Genzel}, R.,
  {Steidel}, C.~C., {Bouch{\'e}}, N., {Cresci}, G., \& {Davies}, R. 2011, \apj,
  731, 65

\bibitem[{{F{\"o}rster Schreiber} {et~al.}(2006){F{\"o}rster Schreiber},
  {Genzel}, {Lehnert}, {Bouch{\'e}}, {Verma}, {Erb}, {Shapley}, {Steidel},
  {Davies}, {Lutz}, {Nesvadba}, {Tacconi}, {Eisenhauer}, {Abuter}, {Gilbert},
  {Gillessen}, \& {Sternberg}}]{ForsterSchreiber06}
{F{\"o}rster Schreiber}, N.~M., {et~al.} 2006, \apj, 645, 1062

\bibitem[{{F{\"o}rster Schreiber} {et~al.}(2009){F{\"o}rster Schreiber},
  {Genzel}, {Bouch{\'e}}, {Cresci}, {Davies}, {Buschkamp}, {Shapiro},
  {Tacconi}, {Hicks}, {Genel}, {Shapley}, {Erb}, {Steidel}, {Lutz},
  {Eisenhauer}, {Gillessen}, {Sternberg}, {Renzini}, {Cimatti}, {Daddi},
  {Kurk}, {Lilly}, {Kong}, {Lehnert}, {Nesvadba}, {Verma}, {McCracken},
  {Arimoto}, {Mignoli}, \& {Onodera}}]{ForsterSchreiber09}
---. 2009, \apj, 706, 1364

\bibitem[{{Genzel} {et~al.}(2010){Genzel}, {Tacconi}, {Gracia-Carpio},
  {Sternberg}, {Cooper}, {Shapiro}, {Bolatto}, {Bouch{\'e}}, {Bournaud},
  {Burkert}, {Combes}, {Comerford}, {Cox}, {Davis}, {Schreiber},
  {Garcia-Burillo}, {Lutz}, {Naab}, {Neri}, {Omont}, {Shapley}, \&
  {Weiner}}]{Genzel10}
{Genzel}, R., {et~al.} 2010, \mnras, 407, 2091

\bibitem[{{Genzel} {et~al.}(2013){Genzel}, {Tacconi}, {Kurk}, {Wuyts},
  {Combes}, {Freundlich}, {Bolatto}, {Cooper}, {Neri}, {Nordon}, {Bournaud},
  {Burkert}, {Comerford}, {Cox}, {Davis}, {F{\"o}rster Schreiber},
  {Garc{\'{\i}}a-Burillo}, {Gracia-Carpio}, {Lutz}, {Naab}, {Newman},
  {Saintonge}, {Shapiro Griffin}, {Shapley}, {Sternberg}, \&
  {Weiner}}]{Genzel13}
---. 2013, \apj, 773, 68

\bibitem[{{Genzel} {et~al.}(2015){Genzel}, {Tacconi}, {Lutz}, {Saintonge},
  {Berta}, {Magnelli}, {Combes}, {Garc{\'{\i}}a-Burillo}, {Neri}, {Bolatto},
  {Contini}, {Lilly}, {Boissier}, {Boone}, {Bouch{\'e}}, {Bournaud}, {Burkert},
  {Carollo}, {Colina}, {Cooper}, {Cox}, {Feruglio}, {F{\"o}rster Schreiber},
  {Freundlich}, {Gracia-Carpio}, {Juneau}, {Kovac}, {Lippa}, {Naab}, {Salome},
  {Renzini}, {Sternberg}, {Walter}, {Weiner}, {Weiss}, \& {Wuyts}}]{Genzel15}
---. 2015, \apj, 800, 20

\bibitem[{{Grogin} {et~al.}(2011){Grogin}, {Kocevski}, {Faber}, {Ferguson},
  {Koekemoer}, {Riess}, {Acquaviva}, {Alexander}, {Almaini}, {Ashby}, {Barden},
  {Bell}, {Bournaud}, {Brown}, {Caputi}, {Casertano}, {Cassata}, {Castellano},
  {Challis}, {Chary}, {Cheung}, {Cirasuolo}, {Conselice}, {Roshan Cooray},
  {Croton}, {Daddi}, {Dahlen}, {Dav{\'e}}, {de Mello}, {Dekel}, {Dickinson},
  {Dolch}, {Donley}, {Dunlop}, {Dutton}, {Elbaz}, {Fazio}, {Filippenko},
  {Finkelstein}, {Fontana}, {Gardner}, {Garnavich}, {Gawiser}, {Giavalisco},
  {Grazian}, {Guo}, {Hathi}, {H{\"a}ussler}, {Hopkins}, {Huang}, {Huang},
  {Jha}, {Kartaltepe}, {Kirshner}, {Koo}, {Lai}, {Lee}, {Li}, {Lotz}, {Lucas},
  {Madau}, {McCarthy}, {McGrath}, {McIntosh}, {McLure}, {Mobasher},
  {Moustakas}, {Mozena}, {Nandra}, {Newman}, {Niemi}, {Noeske}, {Papovich},
  {Pentericci}, {Pope}, {Primack}, {Rajan}, {Ravindranath}, {Reddy}, {Renzini},
  {Rix}, {Robaina}, {Rodney}, {Rosario}, {Rosati}, {Salimbeni}, {Scarlata},
  {Siana}, {Simard}, {Smidt}, {Somerville}, {Spinrad}, {Straughn}, {Strolger},
  {Telford}, {Teplitz}, {Trump}, {van der Wel}, {Villforth}, {Wechsler},
  {Weiner}, {Wiklind}, {Wild}, {Wilson}, {Wuyts}, {Yan}, \& {Yun}}]{Grogin11}
{Grogin}, N.~A., {et~al.} 2011, \apjs, 197, 35

\bibitem[Hatch et al.(2016)]{Hatch16} Hatch, N.~A., Muldrew, S.~I., Cooke, E.~A., et al.\ 2016, \mnras, 459, 387 

\bibitem[{{Hayward} \& {Smith}(2015)}]{Hayward15}
{Hayward}, C.~C., \& {Smith}, D.~J.~B. 2015, \mnras, 446, 1512

\bibitem[{{Hodge} {et~al.}(2013){Hodge}, {Carilli}, {Walter}, {Daddi}, \&
  {Riechers}}]{Hodge13}
{Hodge}, J.~A., {Carilli}, C.~L., {Walter}, F., {Daddi}, E., \& {Riechers}, D.
  2013, \apj, 776, 22

\bibitem[{{Hodge} {et~al.}(2012){Hodge}, {Carilli}, {Walter}, {de Blok},
  {Riechers}, {Daddi}, \& {Lentati}}]{Hodge12}
{Hodge}, J.~A., {Carilli}, C.~L., {Walter}, F., {de Blok}, W.~J.~G.,
  {Riechers}, D., {Daddi}, E., \& {Lentati}, L. 2012, \apj, 760, 11

\bibitem[{{Ilbert} {et~al.}(2010){Ilbert}, {Salvato}, {Le Floc'h}, {Aussel},
  {Capak}, {McCracken}, {Mobasher}, {Kartaltepe}, {Scoville}, {Sanders},
  {Arnouts}, {Bundy}, {Cassata}, {Kneib}, {Koekemoer}, {Le F{\`e}vre}, {Lilly},
  {Surace}, {Taniguchi}, {Tasca}, {Thompson}, {Tresse}, {Zamojski}, {Zamorani},
  \& {Zucca}}]{Ilbert10}
{Ilbert}, O., {et~al.} 2010, \apj, 709, 644

\bibitem[{{Ilbert} {et~al.}(2013){Ilbert}, {McCracken}, {Le F{\`e}vre},
  {Capak}, {Dunlop}, {Karim}, {Renzini}, {Caputi}, {Boissier}, {Arnouts},
  {Aussel}, {Comparat}, {Guo}, {Hudelot}, {Kartaltepe}, {Kneib}, {Krogager},
  {Le Floc'h}, {Lilly}, {Mellier}, {Milvang-Jensen}, {Moutard}, {Onodera},
  {Richard}, {Salvato}, {Sanders}, {Scoville}, {Silverman}, {Taniguchi},
  {Tasca}, {Thomas}, {Toft}, {Tresse}, {Vergani}, {Wolk}, \& {Zirm}}]{Ilbert13}
---. 2013, \aap, 556, A55

\bibitem[{{Karim} {et~al.}(2011){Karim}, {Schinnerer},
  {Mart{\'{\i}}nez-Sansigre}, {Sargent}, {van der Wel}, {Rix}, {Ilbert},
  {Smol{\v c}i{\'c}}, {Carilli}, {Pannella}, {Koekemoer}, {Bell}, \&
  {Salvato}}]{Karim11}
{Karim}, A., {et~al.} 2011, \apj, 730, 61

\bibitem[{{Kartaltepe} {et~al.}(2012){Kartaltepe}, {Dickinson}, {Alexander},
  {Bell}, {Dahlen}, {Elbaz}, {Faber}, {Lotz}, {McIntosh}, {Wiklind}, {Altieri},
  {Aussel}, {Bethermin}, {Bournaud}, {Charmandaris}, {Conselice}, {Cooray},
  {Dannerbauer}, {Dav{\'e}}, {Dunlop}, {Dekel}, {Ferguson}, {Grogin}, {Hwang},
  {Ivison}, {Kocevski}, {Koekemoer}, {Koo}, {Lai}, {Leiton}, {Lucas}, {Lutz},
  {Magdis}, {Magnelli}, {Morrison}, {Mozena}, {Mullaney}, {Newman}, {Pope},
  {Popesso}, {van der Wel}, {Weiner}, \& {Wuyts}}]{Kartaltepe12}
{Kartaltepe}, J.~S., {et~al.} 2012, \apj, 757, 23

\bibitem[{{Kauffmann} {et~al.}(2003){Kauffmann}, {Heckman}, {White}, {Charlot},
  {Tremonti}, {Brinchmann}, {Bruzual}, {Peng}, {Seibert}, {Bernardi},
  {Blanton}, {Brinkmann}, {Castander}, {Cs{\'a}bai}, {Fukugita}, {Ivezic},
  {Munn}, {Nichol}, {Padmanabhan}, {Thakar}, {Weinberg}, \&
  {York}}]{Kauffmann03}
{Kauffmann}, G., {et~al.} 2003, \mnras, 341, 33

\bibitem[{{Kawata} \& {Mulchaey}(2008)}]{Kawata08}
{Kawata}, D., \& {Mulchaey}, J.~S. 2008, \apjl, 672, L103

\bibitem[{{Kennicutt}(1998)}]{Kennicutt98b}
{Kennicutt}, Jr., R.~C. 1998, \apj, 498, 541

\bibitem[{{Kere{\v s}} {et~al.}(2009){Kere{\v s}}, {Katz}, {Fardal},
  {Dav{\'e}}, \& {Weinberg}}]{Keres09}
{Kere{\v s}}, D., {Katz}, N., {Fardal}, M., {Dav{\'e}}, R., \& {Weinberg},
  D.~H. 2009, \mnras, 395, 160

\bibitem[{{Koekemoer} {et~al.}(2011){Koekemoer}, {Faber}, {Ferguson}, {Grogin},
  {Kocevski}, {Koo}, {Lai}, {Lotz}, {Lucas}, {McGrath}, {Ogaz}, {Rajan},
  {Riess}, {Rodney}, {Strolger}, {Casertano}, {Castellano}, {Dahlen},
  {Dickinson}, {Dolch}, {Fontana}, {Giavalisco}, {Grazian}, {Guo}, {Hathi},
  {Huang}, {van der Wel}, {Yan}, {Acquaviva}, {Alexander}, {Almaini}, {Ashby},
  {Barden}, {Bell}, {Bournaud}, {Brown}, {Caputi}, {Cassata}, {Challis},
  {Chary}, {Cheung}, {Cirasuolo}, {Conselice}, {Roshan Cooray}, {Croton},
  {Daddi}, {Dav{\'e}}, {de Mello}, {de Ravel}, {Dekel}, {Donley}, {Dunlop},
  {Dutton}, {Elbaz}, {Fazio}, {Filippenko}, {Finkelstein}, {Frazer}, {Gardner},
  {Garnavich}, {Gawiser}, {Gruetzbauch}, {Hartley}, {H{\"a}ussler},
  {Herrington}, {Hopkins}, {Huang}, {Jha}, {Johnson}, {Kartaltepe},
  {Khostovan}, {Kirshner}, {Lani}, {Lee}, {Li}, {Madau}, {McCarthy},
  {McIntosh}, {McLure}, {McPartland}, {Mobasher}, {Moreira}, {Mortlock},
  {Moustakas}, {Mozena}, {Nandra}, {Newman}, {Nielsen}, {Niemi}, {Noeske},
  {Papovich}, {Pentericci}, {Pope}, {Primack}, {Ravindranath}, {Reddy},
  {Renzini}, {Rix}, {Robaina}, {Rosario}, {Rosati}, {Salimbeni}, {Scarlata},
  {Siana}, {Simard}, {Smidt}, {Snyder}, {Somerville}, {Spinrad}, {Straughn},
  {Telford}, {Teplitz}, {Trump}, {Vargas}, {Villforth}, {Wagner}, {Wandro},
  {Wechsler}, {Weiner}, {Wiklind}, {Wild}, {Wilson}, {Wuyts}, \&
  {Yun}}]{Koekemoer11}
{Koekemoer}, A.~M., {et~al.} 2011, \apjs, 197, 36

\bibitem[{{Larson} {et~al.}(1980){Larson}, {Tinsley}, \& {Caldwell}}]{Larson80}
{Larson}, R.~B., {Tinsley}, B.~M., \& {Caldwell}, C.~N. 1980, ApJ, 237, 692

\bibitem[{{Lee-Brown} {et~al.}(2017){Lee-Brown},{Rudnick},{Momcheva},{Papovich},{Lotz},{Tran},{Henke},{Willmer},{Brammer},{Brodwin},{Dunlop},{Farrah}}]{Lee-Brown17}Lee-Brown, D., Rudnick, G., Momcheva, I., Papovich, C., Lotz, J., Tran, K.-V., Henke, B., Willmer, C., Brammer, G., Brodwin, M., Dunlop, J., Farrah, D., \apj, 844, 43

\bibitem[{{Leroy} {et~al.}(2008){Leroy}, {Walter}, {Brinks}, {Bigiel}, {de
  Blok}, {Madore}, \& {Thornley}}]{Leroy08}
{Leroy}, A.~K., {Walter}, F., {Brinks}, E., {Bigiel}, F., {de Blok}, W.~J.~G.,
  {Madore}, B., \& {Thornley}, M.~D. 2008, \aj, 136, 2782

\bibitem[{{Lotz} {et~al.}(2013){Lotz}, {Papovich}, {Faber}, {Ferguson},
  {Grogin}, {Guo}, {Kocevski}, {Koekemoer}, {Lee}, {McIntosh}, {Momcheva},
  {Rudnick}, {Saintonge}, {Tran}, {van der Wel}, \& {Willmer}}]{Lotz13}
{Lotz}, J.~M., {et~al.} 2013, \apj, 773, 154

\bibitem[{{Ma} {et~al.}(2015){Ma}, {Smail}, {Swinbank}, {Simpson}, {Thomson},
  {Chen}, {Danielson}, {Hilton}, {Tadaki}, {Stott}, \& {Kodama}}]{Ma15}
{Ma}, C.-J., {et~al.} 2015, \apj, 806, 257

\bibitem[{{Magdis} {et~al.}(2012){Magdis}, {Daddi}, {B{\'e}thermin}, {Sargent},
  {Elbaz}, {Pannella}, {Dickinson}, {Dannerbauer}, {da Cunha}, {Walter},
  {Rigopoulou}, {Charmandaris}, {Hwang}, \& {Kartaltepe}}]{Magdis12}
{Magdis}, G.~E., {et~al.} 2012, \apj, 760, 6

\bibitem[{{Marchesini} {et~al.}(2009){Marchesini}, {van Dokkum}, {F{\"o}rster
  Schreiber}, {Franx}, {Labb{\'e}}, \& {Wuyts}}]{Marchesini09}
{Marchesini}, D., {van Dokkum}, P.~G., {F{\"o}rster Schreiber}, N.~M., {Franx},
  M., {Labb{\'e}}, I., \& {Wuyts}, S. 2009, \apj, 701, 1765

\bibitem[{{Martin} {et~al.}(2012){Martin}, {Shapley}, {Coil}, {Kornei},
  {Bundy}, {Weiner}, {Noeske}, \& {Schiminovich}}]{Martin12}
{Martin}, C.~L., {Shapley}, A.~E., {Coil}, A.~L., {Kornei}, K.~A., {Bundy}, K.,
  {Weiner}, B.~J., {Noeske}, K.~G., \& {Schiminovich}, D. 2012, \apj, 760, 127

\bibitem[{{McGee} {et~al.}(2014){McGee}, {Bower}, \& {Balogh}}]{McGee14}
{McGee}, S.~L., {Bower}, R.~G., \& {Balogh}, M.~L. 2014, \mnras, 442, L105

\bibitem[Momcheva et al.(2016)]{Momcheva16} Momcheva, I.~G., Brammer, G.~B., van Dokkum, P.~G., et al.\ 2016, \apjs, 225, 27 

\bibitem[Muldrew et al.(2015)]{Muldrew15} Muldrew, S.~I., Hatch, N.~A., \& Cooke, E.~A.\ 2015, \mnras, 452, 2528 

\bibitem[{{Muzzin} {et~al.}(2012){Muzzin}, {Wilson}, {Yee}, {Gilbank},
  {Hoekstra}, {Demarco}, {Balogh}, {van Dokkum}, {Franx}, {Ellingson}, {Hicks},
  {Nantais}, {Noble}, {Lacy}, {Lidman}, {Rettura}, {Surace}, \&
  {Webb}}]{Muzzin12}
{Muzzin}, A., {et~al.} 2012, \apj, 746, 188

\bibitem[{{Muzzin} {et~al.}(2013){Muzzin}, {Marchesini}, {Stefanon}, {Franx},
  {McCracken}, {Milvang-Jensen}, {Dunlop}, {Fynbo}, {Le Fevre}, {Brammer}, \&
  {Labbe}}]{Muzzin13}
---. 2013, \apj, 777, 18

\bibitem[{{Narayanan} {et~al.}(2011){Narayanan}, {Krumholz}, {Ostriker}, \&
  {Hernquist}}]{Narayanan11b}
{Narayanan}, D., {Krumholz}, M., {Ostriker}, E.~C., \& {Hernquist}, L. 2011,
  \mnras, 418, 664

\bibitem[{{Narayanan} \& {Krumholz}(2014)}]{Narayanan14}
{Narayanan}, D., \& {Krumholz}, M.~R. 2014, \mnras, 442, 1411

\bibitem[{{Narayanan} {et~al.}(2012){Narayanan}, {Krumholz}, {Ostriker}, \&
  {Hernquist}}]{Narayanan12}
{Narayanan}, D., {Krumholz}, M.~R., {Ostriker}, E.~C., \& {Hernquist}, L. 2012,
  \mnras, 421, 3127

\bibitem[{{Nicol} {et~al.}(2011){Nicol}, {Meisenheimer}, {Wolf}, \&
  {Tapken}}]{Nicol11}
{Nicol}, M.-H., {Meisenheimer}, K., {Wolf}, C., \& {Tapken}, C. 2011, \apj,
  727, 51

\bibitem[Noble et al.(2017)]{Noble17} Noble, A.~G., McDonald, M., Muzzin, A., et al.\ 2017, \apjl, 842, L21

\bibitem[{{Noeske} {et~al.}(2007){Noeske}, {Weiner}, {Faber}, {Papovich},
  {Koo}, {Somerville}, {Bundy}, {Conselice}, {Newman}, {Schiminovich}, {Le
  Floc'h}, {Coil}, {Rieke}, {Lotz}, {Primack}, {Barmby}, {Cooper}, {Davis},
  {Ellis}, {Fazio}, {Guhathakurta}, {Huang}, {Kassin}, {Martin}, {Phillips},
  {Rich}, {Small}, {Willmer}, \& {Wilson}}]{Noeske07}
{Noeske}, K.~G., {et~al.} 2007, \apjl, 660, L43

\bibitem[{{Pannella} {et~al.}(2009){Pannella}, {Carilli}, {Daddi}, {McCracken},
  {Owen}, {Renzini}, {Strazzullo}, {Civano}, {Koekemoer}, {Schinnerer},
  {Scoville}, {Smol{\v c}i{\'c}}, {Taniguchi}, {Aussel}, {Kneib}, {Ilbert},
  {Mellier}, {Salvato}, {Thompson}, \& {Willott}}]{Pannella09}
{Pannella}, M., {et~al.} 2009, \apjl, 698, L116

\bibitem[{{Papovich}(2008)}]{Papovich08}
{Papovich}, C. 2008, \apj, 676, 206

\bibitem[{{Papovich} {et~al.}(2007){Papovich}, {Rudnick}, {Le Floc'h}, {van
  Dokkum}, {Rieke}, {Taylor}, {Armus}, {Gawiser}, {Huang}, {Marcillac}, \&
  {Franx}}]{Papovich07}
{Papovich}, C., {et~al.} 2007, \apj, 668, 45

\bibitem[{{Papovich} {et~al.}(2010){Papovich}, {Momcheva}, {Willmer},
  {Finkelstein}, {Finkelstein}, {Tran}, {Brodwin}, {Dunlop}, {Farrah}, {Khan},
  {Lotz}, {McCarthy}, {McLure}, {Rieke}, {Rudnick}, {Sivanandam}, {Pacaud}, \&
  {Pierre}}]{Papovich10}
---. 2010, \apj, 716, 1503

\bibitem[{{Papovich} {et~al.}(2012){Papovich}, {Bassett}, {Lotz}, {van der
  Wel}, {Tran}, {Finkelstein}, {Bell}, {Conselice}, {Dekel}, {Dunlop}, {Guo},
  {Faber}, {Farrah}, {Ferguson}, {Finkelstein}, {H{\"a}ussler}, {Kocevski},
  {Koekemoer}, {Koo}, {McGrath}, {McLure}, {McIntosh}, {Momcheva}, {Newman},
  {Rudnick}, {Weiner}, {Willmer}, \& {Wuyts}}]{Papovich12}
---. 2012, \apj, 750, 93

\bibitem[{{Pierre} {et~al.}(2012){Pierre}, {Clerc}, {Maughan}, {Pacaud},
  {Papovich}, \& {Willmer}}]{Pierre12}
{Pierre}, M., {Clerc}, N., {Maughan}, B., {Pacaud}, F., {Papovich}, C., \&
  {Willmer}, C.~N.~A. 2012, \aap, 540, A4

\bibitem[{{Poggianti} {et~al.}(2006){Poggianti}, {von der Linden}, {De Lucia},
  {Desai}, {Simard}, {Halliday}, {Arag{\'o}n-Salamanca}, {Bower}, {Varela},
  {Best}, {Clowe}, {Dalcanton}, {Jablonka}, {Milvang-Jensen}, {Pello},
  {Rudnick}, {Saglia}, {White}, \& {Zaritsky}}]{Poggianti06}
{Poggianti}, B.~M., {et~al.} 2006, \apj, 642, 188

\bibitem[{{Pozzetti} {et~al.}(2007){Pozzetti}, {Bolzonella}, {Lamareille},
  {Zamorani}, {Franzetti}, {Le F{\`e}vre}, {Iovino}, {Temporin}, {Ilbert},
  {Arnouts}, {Charlot}, {Brinchmann}, {Zucca}, {Tresse}, {Scodeggio}, {Guzzo},
  {Bottini}, {Garilli}, {Le Brun}, {Maccagni}, {Picat}, {Scaramella},
  {Vettolani}, {Zanichelli}, {Adami}, {Bardelli}, {Cappi}, {Ciliegi},
  {Contini}, {Foucaud}, {Gavignaud}, {McCracken}, {Marano}, {Marinoni},
  {Mazure}, {Meneux}, {Merighi}, {Paltani}, {Pell{\`o}}, {Pollo}, {Radovich},
  {Bondi}, {Bongiorno}, {Cucciati}, {de la Torre}, {Gregorini}, {Mellier},
  {Merluzzi}, {Vergani}, \& {Walcher}}]{Pozzetti07}
{Pozzetti}, L., {et~al.} 2007, \aap, 474, 443

\bibitem[{{Riechers} {et~al.}(2010){Riechers}, {Capak}, {Carilli}, {Cox},
  {Neri}, {Scoville}, {Schinnerer}, {Bertoldi}, \& {Yan}}]{Riechers10}
{Riechers}, D.~A., {et~al.} 2010, \apjl, 720, L131

\bibitem[{{Rubin} {et~al.}(2014){Rubin}, {Prochaska}, {Koo}, {Phillips},
  {Martin}, \& {Winstrom}}]{Rubin14}
{Rubin}, K.~H.~R., {Prochaska}, J.~X., {Koo}, D.~C., {Phillips}, A.~C.,
  {Martin}, C.~L., \& {Winstrom}, L.~O. 2014, \apj, 794, 156

\bibitem[{{Rudnick} {et~al.}(2003){Rudnick}, {Rix}, {Franx}, {Labb{\' e}},
  {Blanton}, {Daddi}, {F{\" o}rster Schreiber}, {Moorwood}, {R{\" o}ttgering},
  {Trujillo}, {van de Wel}, {van der Werf}, {van Dokkum}, \& {van
  Starkenburg}}]{Rudnick03}
{Rudnick}, G., {et~al.} 2003, \apj, 599, 847

\bibitem[{{Rudnick} {et~al.}(2006){Rudnick}, {Labb{\'e}}, {F{\"o}rster
  Schreiber}, {Wuyts}, {Franx}, {Finlator}, {Kriek}, {Moorwood}, {Rix},
  {R{\"o}ttgering}, {Trujillo}, {van der Wel}, {van der Werf}, \& {van
  Dokkum}}]{Rudnick06}
---. 2006, \apj, 650, 624

\bibitem[{{Rudnick} {et~al.}(2009){Rudnick}, {von der Linden}, {Pell{\'o}},
  {Arag{\'o}n-Salamanca}, {Marchesini}, {Clowe}, {DeLucia}, {Halliday},
  {Jablonka}, {Milvang-Jensen}, {Poggianti}, {Saglia}, {Simard}, {White}, \&
  {Zaritsky}}]{Rudnick09}
---. 2009, \apj, 700, 1559

\bibitem[{{Rudnick} {et~al.}(2012){Rudnick}, {Tran}, {Papovich}, {Momcheva}, \&
  {Willmer}}]{Rudnick12}
{Rudnick}, G.~H., {Tran}, K.-V., {Papovich}, C., {Momcheva}, I., \& {Willmer},
  C. 2012, \apj, 755, 14

\bibitem[{{Saintonge} {et~al.}(2008){Saintonge}, {Tran}, \&
  {Holden}}]{Saintonge08}
{Saintonge}, A., {Tran}, K.-V.~H., \& {Holden}, B.~P. 2008, \apjl, 685, L113

\bibitem[{Saintonge {et~al.}(2011)Saintonge, Kauffmann, Kramer, Tacconi,
  Buchbender, Catinella, Fabello, Gracia-Carpio, Wang, Cortese, Fu, Genzel,
  Giovanelli, Guo, Haynes, Heckman, Krumholz, Lemonias, Li, Moran,
  Rodriguez-Fernandez, Schiminovich, Schuster, \& Sievers}]{Saintonge11a}
Saintonge, A., {et~al.} 2011, Monthly Notices of the Royal Astronomical
  Society, 415, 32

\bibitem[{{Sanders} \& {Mirabel}(1996)}]{Sanders96}
{Sanders}, D.~B., \& {Mirabel}, I.~F. 1996, \araa, 34, 749

\bibitem[{{Sandstrom} {et~al.}(2013){Sandstrom}, {Leroy}, {Walter}, {Bolatto},
  {Croxall}, {Draine}, {Wilson}, {Wolfire}, {Calzetti}, {Kennicutt}, {Aniano},
  {Donovan Meyer}, {Usero}, {Bigiel}, {Brinks}, {de Blok}, {Crocker}, {Dale},
  {Engelbracht}, {Galametz}, {Groves}, {Hunt}, {Koda}, {Kreckel}, {Linz},
  {Meidt}, {Pellegrini}, {Rix}, {Roussel}, {Schinnerer}, {Schruba}, {Schuster},
  {Skibba}, {van der Laan}, {Appleton}, {Armus}, {Brandl}, {Gordon}, {Hinz},
  {Krause}, {Montiel}, {Sauvage}, {Schmiedeke}, {Smith}, \&
  {Vigroux}}]{Sandstrom13}
{Sandstrom}, K.~M., {et~al.} 2013, \apj, 777, 5

\bibitem[{{Santos} {et~al.}(2014){Santos}, {Altieri}, {Tanaka}, {Valtchanov},
  {Saintonge}, {Dickinson}, {Foucaud}, {Kodama}, {Rawle}, \&
  {Tadaki}}]{Santos14}
{Santos}, J.~S., {et~al.} 2014, \mnras, 438, 2565

\bibitem[{{Santos} {et~al.}(2015){Santos}, {Altieri}, {Valtchanov}, {Nastasi},
  {B{\"o}hringer}, {Cresci}, {Elbaz}, {Fassbender}, {Rosati}, {Tozzi}, \&
  {Verdugo}}]{Santos15}
---. 2015, \mnras, 447, L65

\bibitem[{{Scoville} {et~al.}(1997){Scoville}, {Yun}, \& {Bryant}}]{Scoville97}
{Scoville}, N.~Z., {Yun}, M.~S., \& {Bryant}, P.~M. 1997, \apj, 484, 702

\bibitem[Scoville et al.(2016)]{Scoville16} Scoville, N., Sheth, K., Aussel, H., et al.\ 2016, \apj, 820, 83 

\bibitem[{{S\'{e}rsic}(1968)}]{Sersic68}
{S\'{e}rsic}, J.~L. 1968, {Atlas de galaxias australes}, Cordoba, Argentina: Observatorio Astronomico

\bibitem[{{Shapiro} {et~al.}(2008){Shapiro}, {Genzel}, {F{\"o}rster Schreiber},
  {Tacconi}, {Bouch{\'e}}, {Cresci}, {Davies}, {Eisenhauer}, {Johansson},
  {Krajnovi{\'c}}, {Lutz}, {Naab}, {Arimoto}, {Arribas}, {Cimatti}, {Colina},
  {Daddi}, {Daigle}, {Erb}, {Hernandez}, {Kong}, {Mignoli}, {Onodera},
  {Renzini}, {Shapley}, \& {Steidel}}]{Shapiro08}
{Shapiro}, K.~L., {et~al.} 2008, \apj, 682, 231

\bibitem[{{Skelton} {et~al.}(2014){Skelton}, {Whitaker}, {Momcheva}, {Brammer},
  {van Dokkum}, {Labb{\'e}}, {Franx}, {van der Wel}, {Bezanson}, {Da Cunha},
  {Fumagalli}, {F{\"o}rster Schreiber}, {Kriek}, {Leja}, {Lundgren}, {Magee},
  {Marchesini}, {Maseda}, {Nelson}, {Oesch}, {Pacifici}, {Patel}, {Price},
  {Rix}, {Tal}, {Wake}, \& {Wuyts}}]{Skelton14}
{Skelton}, R.~E., {et~al.} 2014, \apjs, 214, 24

\bibitem[Solomon \& Vanden Bout(2005)]{Solomon05} Solomon, P.~M., \& Vanden Bout, P.~A.\ 2005, \araa, 43, 677 

\bibitem[{{Stefanon} {et~al.}(2013){Stefanon}, {Marchesini}, {Rudnick},
  {Brammer}, \& {Whitaker}}]{Stefanon13}
{Stefanon}, M., {Marchesini}, D., {Rudnick}, G.~H., {Brammer}, G.~B., \&
  {Whitaker}, K.~E. 2013, \apj, 768, 92

\bibitem[{{Steidel} {et~al.}(2004){Steidel}, {Shapley}, {Pettini},
  {Adelberger}, {Erb}, {Reddy}, \& {Hunt}}]{Steidel04}
{Steidel}, C.~C., {Shapley}, A.~E., {Pettini}, M., {Adelberger}, K.~L., {Erb},
  D.~K., {Reddy}, N.~A., \& {Hunt}, M.~P. 2004, \apj, 604, 534

\bibitem[{{Strateva} {et~al.}(2001){Strateva}, {Ivezi{\' c}}, {Knapp},
  {Narayanan}, {Strauss}, {Gunn}, {Lupton}, {Schlegel}, {Bahcall}, {Brinkmann},
  {Brunner}, {Budav{\' a}ri}, {Csabai}, {Castander}, {Doi}, {Fukugita}, {Gy{\H
  o}ry}, {Hamabe}, {Hennessy}, {Ichikawa}, {Kunszt}, {Lamb}, {McKay},
  {Okamura}, {Racusin}, {Sekiguchi}, {Schneider}, {Shimasaku}, \&
  {York}}]{Strateva01}
{Strateva}, I., {et~al.} 2001, \aj, 122, 1861

\bibitem[{{Strazzullo} {et~al.}(2013){Strazzullo}, {Gobat}, {Daddi}, {Onodera},
  {Carollo}, {Dickinson}, {Renzini}, {Arimoto}, {Cimatti}, {Finoguenov}, \&
  {Chary}}]{Strazzullo13}
{Strazzullo}, V., {et~al.} 2013, \apj, 772, 118

\bibitem[{{Tacconi} {et~al.}(2010){Tacconi}, {Genzel}, {Neri}, {Cox}, {Cooper},
  {Shapiro}, {Bolatto}, {Bouch{\'e}}, {Bournaud}, {Burkert}, {Combes},
  {Comerford}, {Davis}, {Schreiber}, {Garcia-Burillo}, {Gracia-Carpio}, {Lutz},
  {Naab}, {Omont}, {Shapley}, {Sternberg}, \& {Weiner}}]{Tacconi10}
{Tacconi}, L.~J., {et~al.} 2010, \nat, 463, 781

\bibitem[{Tacconi {et~al.}(2013)Tacconi, Neri, Genzel, Combes, Bolatto, Cooper,
  Wuyts, Bournaud, Burkert, Comerford, Cox, Davis, Forster~Schreiber,
  Garcia-Burillo, Gracia-Carpio, Lutz, Naab, Newman, Omont, Saintonge,
  Shapiro~Griffin, Shapley, Sternberg, \& Weiner}]{Tacconi13}
Tacconi, L.~J., {et~al.} 2013, The Astrophysical Journal, 768, 74

\bibitem[{{Tanaka} {et~al.}(2010){Tanaka}, {Finoguenov}, \& {Ueda}}]{Tanaka10}
{Tanaka}, M., {Finoguenov}, A., \& {Ueda}, Y. 2010, \apjl, 716, L152

\bibitem[{{Tanaka} {et~al.}(2013){Tanaka}, {Toft}, {Marchesini}, {Zirm}, {De
  Breuck}, {Kodama}, {Koyama}, {Kurk}, \& {Tanaka}}]{Tanaka13b}
{Tanaka}, M., {et~al.} 2013, \apj, 772, 113

\bibitem[Toft et al.(2007)]{Toft07} Toft, S., van Dokkum, P., Franx, M., et al.\ 2007, \apj, 671, 285 

\bibitem[Tomczak et al.(2016)]{Tomczak16} Tomczak, A.~R., Quadri, R.~F., Tran, K.-V.~H., et al.\ 2016, \apj, 817, 118

\bibitem[{{Tran} {et~al.}(2010){Tran}, {Papovich}, {Saintonge}, {Brodwin},
  {Dunlop}, {Farrah}, {Finkelstein}, {Finkelstein}, {Lotz}, {McLure},
  {Momcheva}, \& {Willmer}}]{Tran10}
{Tran}, K.-V.~H., {et~al.} 2010, \apjl, 719, L126

\bibitem[{{Tran} {et~al.}(2015){Tran}, Nanayakkara, {Yuan}, Kacprzak,
  Glazebrook, Kewley, Momcheva, Papovich, Quadri, Rudnick, Saintonge, Spitler,
  Straatman, \& A.}]{Tran15}
---. 2015, \apj, 811, 28

\bibitem[{{Tremonti} {et~al.}(2007){Tremonti}, {Moustakas}, \&
  {Diamond-Stanic}}]{Tremonti07}
{Tremonti}, C.~A., {Moustakas}, J., \& {Diamond-Stanic}, A.~M. 2007, \apjl,
  663, L77

\bibitem[{{Tripp} {et~al.}(2011){Tripp}, {Meiring}, {Prochaska}, {Willmer},
  {Howk}, {Werk}, {Jenkins}, {Bowen}, {Lehner}, {Sembach}, {Thom}, \&
  {Tumlinson}}]{Tripp11}
{Tripp}, T.~M., {et~al.} 2011, Science, 334, 952


\bibitem[van de Voort et al.(2017)]{vandeVoort17} van de Voort, F., Bah{\'e}, Y.~M., Bower, R.~G., et al.\ 2017, \mnras, 466, 3460 

\bibitem[{{van der Burg} {et~al.}(2013){van der Burg}, {Muzzin}, {Hoekstra},
  {Lidman}, {Rettura}, {Wilson}, {Yee}, {Hildebrandt}, {Marchesini},
  {Stefanon}, {Demarco}, \& {Kuijken}}]{vanderburg13}
{van der Burg}, R.~F.~J., {et~al.} 2013, \aap, 557, A15

\bibitem[{{van der Wel} {et~al.}(2012){van der Wel}, {Bell}, {H{\"a}ussler},
  {McGrath}, {Chang}, {Guo}, {McIntosh}, {Rix}, {Barden}, {Cheung}, {Faber},
  {Ferguson}, {Galametz}, {Grogin}, {Hartley}, {Kartaltepe}, {Kocevski},
  {Koekemoer}, {Lotz}, {Mozena}, {Peth}, \& {Peng}}]{vanderwel12}
{van der Wel}, A., {et~al.} 2012, \apjs, 203, 24

\bibitem[{{van Dokkum} {et~al.}(2010){van Dokkum}, {Whitaker}, {Brammer},
  {Franx}, {Kriek}, {Labb{\'e}}, {Marchesini}, {Quadri}, {Bezanson},
  {Illingworth}, {Muzzin}, {Rudnick}, {Tal}, \& {Wake}}]{vandokkum10}
{van Dokkum}, P.~G., {et~al.} 2010, \apj, 709, 1018

\bibitem[{{Weiner} {et~al.}(2009){Weiner}, {Coil}, {Prochaska}, {Newman},
  {Cooper}, {Bundy}, {Conselice}, {Dutton}, {Faber}, {Koo}, {Lotz}, {Rieke}, \&
  {Rubin}}]{Weiner09}
{Weiner}, B.~J., {et~al.} 2009, \apj, 692, 187

\bibitem[{{Whitaker} {et~al.}(2012){Whitaker}, {van Dokkum}, {Brammer}, \&
  {Franx}}]{Whitaker12}
{Whitaker}, K.~E., {van Dokkum}, P.~G., {Brammer}, G., \& {Franx}, M. 2012,
  \apjl, 754, L29

\bibitem[{{Williams} {et~al.}(2009){Williams}, {Quadri}, {Franx}, {van Dokkum},
  \& {Labb{\'e}}}]{Williams09}
{Williams}, R.~J., {Quadri}, R.~F., {Franx}, M., {van Dokkum}, P., \&
  {Labb{\'e}}, I. 2009, \apj, 691, 1879

\bibitem[{{Wisnioski} {et~al.}(2015){Wisnioski}, {F{\"o}rster Schreiber},
  {Wuyts}, {Wuyts}, {Bandara}, {Wilman}, {Genzel}, {Bender}, {Davies},
  {Fossati}, {Lang}, {Mendel}, {Beifiori}, {Brammer}, {Chan}, {Fabricius},
  {Fudamoto}, {Kulkarni}, {Kurk}, {Lutz}, {Nelson}, {Momcheva}, {Rosario},
  {Saglia}, {Seitz}, {Tacconi}, \& {van Dokkum}}]{Wisnioski15}
{Wisnioski}, E., {et~al.} 2015, \apj, 799, 209

\bibitem[{{Wuyts} {et~al.}(2007){Wuyts}, {Labb{\'e}}, {Franx}, {Rudnick}, {van
  Dokkum}, {Fazio}, {F{\"o}rster Schreiber}, {Huang}, {Moorwood}, {Rix},
  {R{\"o}ttgering}, \& {van der Werf}}]{Wuyts07}
{Wuyts}, S., {et~al.} 2007, \apj, 655, 51

\bibitem[{{Wuyts} {et~al.}(2011){Wuyts}, {Forster Schreiber}, {Lutz}, {Nordon},
  {Berta}, {Altieri}, {Andreani}, {Aussel}, {Bongiovanni}, {Cepa}, {Cimatti},
  {Daddi}, {Elbaz}, {Genzel}, {Koekemoer}, {Magnelli}, {Maiolino}, {McGrath},
  {Perez Garcia}, {Poglitsch}, {Popesso}, {Pozzi}, {Sanchez-Portal}, {Sturm},
  {Tacconi}, \& {Valtchanov}}]{Wuyts11}
---. 2011, \apj, 738, 106

\end{thebibliography}

\end{document}